# Roadmap on Wavefront Shaping and deep imaging in complex media

Sylvain Gigan[4], Ori Katz[20], Hilton B. de Aguiar[4], Esben Ravn Andresen[14], Alexandre Aubry[8], Jacopo Bertolotti[10], Emmanuel Bossy[5] ,Dorian Bouchet[5], Joshua Brake[32] , Sophie Brasselet[9], Yaron Bromberg[29], Hui Cao[24], Thomas Chaigne[6], Zhongtao Cheng[33], Wonshik Choi[7], Tomáš Čižmár[11,23], Meng Cui[36],  Vincent R Curtis[16] , Hugo Defienne[30] , Matthias Hofer[9], Ryoichi Horisaki[22], Roarke Horstmeyer[34] ,Na Ji[2], Aaron K. LaViolette[1], Jerome Mertz[3], Christophe Moser[12] , Allard P. Mosk[27], Nicolas C. Pégard[16,18,19] , Rafael Piestun[31], Sebastien Popoff[8], David B. Phillips[10], Demetri Psaltis[13], Babak Rahmani[12], Hervé Rigneault[9], Stefan Rotter[28], Lei Tian[3,15], Ivo M. Vellekoop[21], Laura Waller[17] ,Lihong Wang[33,35] ,Timothy Weber[3], Sheng Xiao[3], Chris Xu[1], Alexey Yamilov[25], Changhuei Yang [35] Hasan Yılmaz[26]

[1]Cornell University, USA
[2]University of California Berkeley, USA
[3] Department of Biomedical Engineering, Boston University, USA
[4]Laboratoire Kastler Brossel, ENS-Université PSL, CNRS, Sorbonne Université, Collège de France, 24 rue Lhomond, 75005 Paris, France
[5]Univ. Grenoble Alpes, CNRS, LIPhy, 38000 Grenoble, France
[6]Aix Marseille Univ., CNRS, Centrale Marseille, Institut Fresnel, Marseille, France
[7]Institute for Basic Science, Department of Physics, Korea University, Korea
[8]Institut Langevin, ESPCI, PSL University, CNRS, Paris, France
[9]Aix Marseille Univ, CNRS, Centrale Marseille, Institut Fresnel, F-13013 Marseille, France
[10]School of Physics and Astronomy, University of Exeter, Exeter, EX4 4QL. UK.
[11]Leibniz Institute of Photonic Technology, Albert-Einstein-Straße 9, 07745 Jena, Germany.
[12]Laboratory of Applied Photonics Devices, School of Engineering, Ecole Polytechnique Fédérale de Lausanne, 1015 Lausanne, Switzerland
[13]Laboratory of Optics, School of Engineering, Ecole Polytechnique Fédérale de Lausanne, 1015 Lausanne, Switzerland
[14]Université de Lille, CNRS UMR 8523 – PhLAM – Laboratoire de Physique des Lasers, Atomes et Molécules, F-59000 Lille, France
[15]Department of Electrical and Computer Engineering, Boston University, Boston, MA 02215, USA
[16]Department of Applied Physical Sciences, UNC Chapel Hill, USA
[17] Department of Electrical Engineering and Computer Sciences, UC Berkeley, USA
[18] Department of Biomedical Engineering, UNC Chapel Hill, USA
[19] UNC Neuroscience Centre, UNC Chapel Hill, USA
[20]Department of Applied Physics, Hebrew University of Jerusalem, Jerusalem, Israel
[21] Department of Science and Technology, University of Twente, Enschede, The Netherlands
[22]The University of Tokyo, Japan
[23]Institute of Scientific Instruments of the Czech Academy of Sciences, Královopolská 147, 612 64 Brno, Czechia.
[24]Dept. of Applied Physics, Yale University, USA.
[25] Dept. of Physics, Missouri University of Science and Technology, USA
[26] National Nanotechnology Research Center (UNAM), Bilkent University, Turkey
[27]Debye Institute for Nanomaterials Science, Utrecht University, Utrecht, The Netherlands,
[28]Institute for Theoretical Physics, TU Wien, Vienna, Austria.
[29]Racah Institute of Physics, The Hebrew University of Jerusalem, Israel
[30]School of Physics and Astronomy, University of Glasgow, UK
[31]University of Colorado Boulder, USA
[32]Department of Engineering, Harvey Mudd College, Claremont, CA 91711, USA
[33]Caltech Optical Imaging Laboratory, Andrew and Peggy Cherng Department of Medical Engineering, California Institute of Technology, Pasadena, CA 91106, USA
[34] Department of Biomedical Engineering, Duke University, Durham, NC 27708, USA
[35] Department of Electrical Engineering, California Institute of Technology, Pasadena, CA 91106, USA
[36] Electrical and Computer Engineering and Biology department, Purdue University, USA

Email: sylvain.gigan@lkb.ens.fr , orik@mail.huji.ac.il

**Abstract**
The last decade has seen the development of a wide set of tools, such as wavefront shaping, computational or fundamental methods, that allow to understand and control light propagation in a complex medium, such as biological tissues or multimode fibers. A vibrant and diverse community is now working on this field, that has revolutionized the prospect of diffraction-limited imaging at depth in tissues. This roadmap highlights several key aspects of this fast developing field, and some of the challenges and opportunities ahead.

# Introduction

Sylvain Gigan[1], Ori Katz[2]

[1] Laboratoire Kastler-Brossel, ENS CNRS Sorbonne Université, College de France, France
[2] Department of Applied Physics, Hebrew University of Jerusalem, Jerusalem, Israel

Wavefront shaping in complex media, the high resolution manipulation of light waves in order to control light in disordered environments, is a relatively young field. Its commonly accepted inception can be traced back to 2007 when Vellekoop and Mosk demonstrated that via an iterative optimization algorithm, a diffraction-limited focus could be obtained through a visually-opaque strongly scattering medium by phase-control thousands of optical modes [1]. However, while being a relatively young field, it builds on decades of experimental works performed in several established fields, from adaptive optics for astronomy, through holography, time-reversal of ultrasound waves, to RADAR imaging. It also builds on decades of fundamental insight from mesoscopic physics.

Deep imaging in complex environment is a hugely important challenge, from noninvasive biomedical investigations to seeing through fog. However, due to the highly scattering nature of many real-world samples, direct imaging in complex samples, such as biological tissues, is conventionally limited to shallow depths of a few hundred microns. While other imaging modalities, such as ultrasound, MRI, and X-ray, can penetrate deeper into biological tissue, they are inferior to optical microscopy in terms of resolution (limited by the wavelength of light), variety of contrast mechanism (e.g. chemical sensitivity, and functional information), or its non-ionizing nature. Wavefront shaping offers a unique possibility: to achieve optical resolution focusing and imaging,  without being limited by the exponential decay of ballistic photons with depth. It is relatively safe to say that, to obtain micron or submicron resolution images deep in complex media, retrieving information from scattered light (either through physical wavefront control or through computation) are the only way to go.

One of the main reasons for the fast recent progress in the field of wavefront-shaping are the great technological advances in both digital modulators and detectors. Spatial light modulators based on a variety of technologies, from liquid crystals through MEMS to acousto-optic modulators, now allow to ramp up the number of controlled modes and speed. The progress in cameras and detectors technologies had also a great impact on the field due to several now widely available devices from multi-megapixel fast cameras, through ultrasensitive EMCCDs and sCMOS technologies, to single-photon detector arrays.

The great advancements in modulators and detectors comes hand-in-hand with the now available computational power, data bandwidth and memory, which are required to digitally transfer and process the huge amounts of scattered-light information. The field also naturally strongly benefits from the current advancements in signal processing and AI revolution, with the emergence of deep learning. In some instances, these advancements in signal-processing and machine-learning algorithms allow to simplify or even shortcut the stringent requirements of wavefront shaping, both in terms of number of measurements, and even in the need for a wavefront-shaping device.

Simultaneously, the field continuously revisits and makes use of fundamental concepts from mesoscopic physics to improve imaging and light delivery deep inside visually opaque samples. A first example is the transmission matrix, initially a theoretician concept, that became a versatile tool for imaging once it was demonstrated that it could be effectively recorded.  Another  notable examples for imaging is the optical "memory effect", a correlation of multiply scattered waves predicted in the 80s. It was proposed for imaging (and other tasks) as early as the 90s in a visionary work by Freund [2]. But it is only in the last decade that its practical potential was realized, and put to practice thanks to wavefront shaping. Very recently, fundamental concepts such as open-channels, or time-delay eigenstates, have been increasingly studied in the context of imaging and light control in complex media.

The applications of wavefront-shaping now span well beyond simple focusing and imaging inside complex media: it has been employed to allow looking around corners, for energy delivery through opaque samples, for trapping and optical manipulation,  near-field, plasmonics, spectroscopy, and ultrafast pulse shaping, virtually all aspects of photonics have been explored in combination with wavefront shaping. One specifically important extension of wavefront shaping is its application for light control through long multimode and multicore fibers, which has emerged as an extremely fruitful path. In particular, allowing the development of miniature lens-less endoscopes for deep imaging, now a very active subfield.

In terms of imaging modalities, wavefront shaping has been experimentally demonstrated in essentially every widespread optical modality (most often in proof of principle experiments): these include confocal imaging, multiphoton imaging, photo-acoustics, acousto-optics, phase-contrast, optical coherence tomography (OCT), fluorescence imaging, structured illumination, temporal and spectral control, Raman, and other spectroscopic techniques. Interestingly, optical super-resolution techniques, such as STED, still remain to date largely unaddressed.

By and large, the field is now relatively mature, but remains very active, and shows no sign of slowing down in terms of innovation: one can cite very recent progress such as fluorescence based incoherent transmission matrices, applying deep learning approaches to image reconstruction, the emergence of novel ultrafast spatial light modulators, to cite just a few salient results. It also focuses more and more on applications in real-world samples rather than basic proof of principle demonstrations.

In this roadmap we highlight, from multiple and different perspectives, many of these recent advances as well as the challenges and opportunities that the field may offer in the years to come.

## Section 1 – Three Dimensionally Resolved Fluorescence Microscopy in Deep Scattering Biological Tissue

Aaron K. LaViolette, Chris Xu, Cornell University

**Status**

One-photon (1P) confocal, two-photon (2P) and three-photon (3P) microscopy provide three dimensionally resolved, high spatial resolution fluorescence imaging in scattering biological tissues. 1P confocal microscopy is typically only applicable within shallow imaging depths, while 2P microscopy (2PM) and eventually 3P microscopy (3PM) are preferred as the imaging depth increases. Such a photon "upmanship" [1] for deep tissue fluorescence imaging can be understood by considering the effective attenuation coefficient ($\alpha_e$) of the tissues for ballistic photons, the maximum allowable power, multiphoton cross sections, out-of-focus background fluorescence generation, and availability of fluorophores.

$\alpha_e$ is a function of excitation wavelength and is the sum of the scattering coefficient ($\alpha_s$) and absorption coefficient ($\alpha_a$), i.e., $\alpha_e = \alpha_s + \alpha_a$. For in vivo imaging, the absorption of brain tissue is dominated by blood and water. Fig. 1 shows the calculated and experimentally measured $\alpha_e$ for mouse brain tissues in vivo [2–8]. Because the excitation power at the focus, $P(z)$, decreases exponentially as a function of imaging depth $z$, $P(z) = P_0 exp(-\alpha_e z)$, where $P_0$ is the power at the brain surface, the value of $\alpha_e$ is the most important consideration for deep imaging. Therefore, the preferred excitation wavelengths for deep tissue imaging reside within the long wavelength windows around 1300, 1700, and 2200 nm.

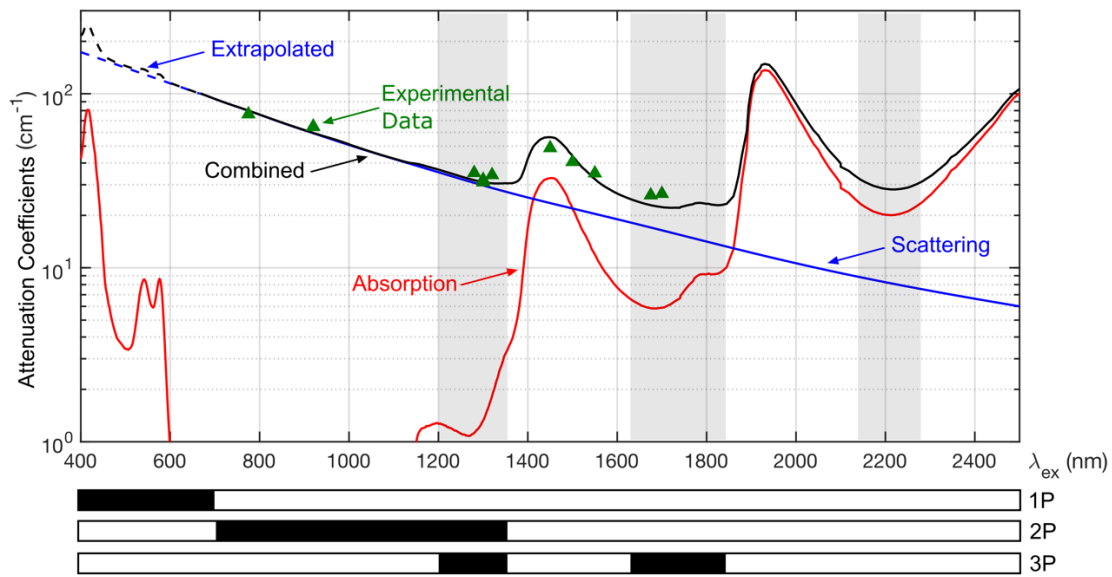

**Figure 1.** Absorption coefficient (red), scattering coefficient (blue) and the effective attenuation coefficient (black) are plotted as a function of wavelength. The scattering coefficient is calculated using Mie theory for a tissue-like phantom solution of polystyrene beads at a concentration of $5.4 \times 10^9$ mL$^{-1}$, which mimics scattering in the mouse cortex [2]. The absorption coefficient is the combined effect (sum) of water [7] and blood [8]. The blue dotted line is extrapolated for scattering below 667 nm by fitting an exponential model to the calculated scattering attenuation coefficient values between 667 and 1000 nm. The black dotted line reflects that extrapolated scattering data was used. Superimposed are green triangles which are experimental measurements of the mouse brain presented in [2–6]. The black bars below the graph indicate the ranges where 1P, 2P and 3P imaging are typical done, which are largely determined by the availability of fluorophores and the effective attenuation coefficient. The grey highlighted regions show the long wavelength windows for deep tissue imaging.

While $\alpha_e$ is the most important consideration for penetration depth, tissue absorption determines the maximum permissible power for imaging (i.e., the maximum value of $P_0$) [4]. For the three spectral windows with the lowest $\alpha_e$, the absorption increases at the longer wavelength (Fig. 1). Although the values of $\alpha_e$ around 1300 and 2200 nm are comparable, the 1300 nm window is preferred because of the higher allowable power due to the lower tissue absorption, while the low $\alpha_e$ around 1700 nm indicates that the 1700 nm window is the best for the deepest imaging.

Most existing fluorophores require excitation wavelengths within the visible to near infrared (IR) range, which limits 1P confocal imaging to excitations between about 350 and 700 nm, where $\alpha_e$ is large. Thus, 1P confocal microscopy is generally best in shallow regions. While 2P and 3P excitation makes the low attenuation spectral windows compatible with existing fluorophores, only red or near IR fluorophores can be excited in the 1300 nm window by 2P excitation, excluding the most commonly used green and yellow fluorophores for deep tissue 2PM. 3P imaging can be performed within the 1300 nm spectral window for blue, green, and yellow fluorophores, and the 1700 nm spectral window for red and near IR fluorophores. The spectral "gap" for 3P imaging is due to the large water absorption between 1400 and 1600 nm. The 2200 nm window, in addition to the high tissue absorption, is too long for nearly all existing fluorophores even with 3P excitation.

In addition to the practical considerations listed above, the imaging depth of three dimensionally resolved imaging is fundamentally limited by out-of-focus background fluorescence [9]. In general, a higher order nonlinear excitation will result in a higher excitation confinement and less out-of-focus background, resulting in a higher signal-to-background ratio (SBR). However, higher order nonlinear excitation is less efficient due to the small multiphoton excitation cross sections. Therefore, the improvement in SBR comes at the expense of the signal strength. The trade-off between SBR and signal strength indicates that there is a depth for a fluorophore under consideration beyond which 3P imaging outperforms 2P imaging [4]. This depth can be quantified based on a metric grounded in detection theory (e.g., for the detection of calcium-transients with the d' metric [4,10] or binary objectes with the binary detection factor metric [11]).

## Current and Future Challenges

The penetration depths of 1P confocal and 2P imaging are limited by the SBR (i.e., the depth limit). Neglecting the Stokes shift and the pinhole size, 1P confocal imaging has nearly the same point-spread function as 2PM. Therefore, the depth limit determined by the SBR is similar for 1P confocal and 2P imaging when the excitation wavelength is the same. Both theoretical and experimental studies showed that the depth limit for 1P confocal and 2P imaging is between 1.5 to 2 mm, when imaging mouse brain vasculature, in the long wavelength windows of 1300 and 1700 nm  [12–15]. The biggest challenge facing deep tissue long wavelength 1P confocal microscopy is the lack of IR dyes with excitation wavelengths >1200 nm. In addition, low-cost, low-noise, high-gain, high quantum efficiency detectors at the 1300 and 1700 nm windows must be developed. For long wavelength 2PM, the availability and performance of deep red and near IR functional indicators are major limitations for practical applications.

3P excitation best matches existing dyes with the low tissue attenuation windows, and 3P imaging depths of >1 and >2 mm have been demonstrated, respectively, for green/yellow fluorophores using 1300 nm excitation [16,17] and red fluorophores using 1700 nm excitation [18]. The higher order nonlinear excitation has also enabled 3PM to image deep in densely labelled samples or through a highly scattering layer (e.g., the mouse skull or corpus callosum [19]) where the imaging depth of long wavelength 2PM is severely limited by the SBR. Theoretical analysis and experimental studies in tissue phantoms showed that 3PM has the potential to image much deeper than what has been achieved so far. Indeed, the predicted maximum penetration depth limited by the SBR is about 3 to 4 mm for 3PM when imaging mouse brain vasculature, which is nearly twice the deepest imaging today. The biggest challenge in pushing the imaging depth of 3PM is the small 3P cross section. Together with the maximum allowable power, small 3P cross sections limit the 3P signal strength and currently set the practical imaging depth limit.

**Advances in Science and Technology to Meet Challenges**

The depth limits of long wavelength 1P confocal and 2P imaging have already been reached in the mouse brain. However, further improvements are required to transform them into valuable practical tools for biological research. For deep tissue 1P confocal microscopy, the biggest advancement would be creating a plethora of fluorophores, fluorescent proteins, and functional indicators with excitation wavelengths >1200 nm. However, significant effort has been devoted to finding long wavelength fluorophores for in vivo imaging in the last 10 to 20 years, which has proven to be challenging, particularly for fluorophores excited with wavelengths >800 nm. Quantum dots (QD) are probably the most promising path so far but making QDs into robust functional indicators may yet prove difficult. Additionally, the recent development of superconducting nanowire detectors (SND) is promising for 1P confocal imaging at 1300 and 1700 nm [15]. While still expensive, advancements in materials and manufacturing for SNDs could reduce the cost and make these detectors affordable for biological imaging. Noticeable progress has been made in deep red or near IR fluorescent proteins and functional indicators for 2PM around 1300 nm. Further improvement in their performance will greatly improve the practical utility of long wavelength deep tissue 2PM and could make long wavelength 2PM of deep red or near IR fluorophores an alternative to 3PM of green fluorophores for many applications.

The depth limit of 3PM has not been reached in any in vivo biological samples and increasing the imaging depth of 3PM will require improving the signal strength. Fluorophores with enhanced multiphoton cross sections can increase the imaging depth of 2PM and 3PM and lower the cost of the excitation source. While past attempts exploring the molecular structures of fluorophores have largely failed to create new ones with extraordinarily large 2P or 3P cross sections for in vivo imaging, exploring resonance enhanced 3P excitation appears to be promising. Indeed, already approximately 10 times enhanced 3P cross sections have been demonstrated [20]. Additionally, adaptive optics (AO), which is a well-established technique for improving the spatial resolution and increasing the signal generation for in vivo brain imaging [21,22], may be considered. AO has shown to have a larger impact in 3PM than in 2PM due to the higher order nonlinear excitation and deeper imaging depth [23], where a 5 to 10 times signal strength increase can be achieved for deep 3PM [24]. One of the AO challenges is the lack of a fast and direct wavefront sensing method in deep scattering tissue.

For both 2PM and 3PM, a promising way forward is the development of high pulse energy, low repetition rate femtosecond lasers for deep imaging such as optical parametric chirped pulse amplifiers. This is because the pulse energy required from the laser increases exponentially as a function of imaging depth, and so the repetition rate of the laser must be reduced exponentially due to the limit on the maximum allowable power. The ideal laser for deep tissue imaging should provide constant output average power and a user-defined, tuneable repetition rate. While lasers with high pulse energy and tuneable repetition rate have become available for 2PM and 3PM in the last 5 years, such sources cannot yet maintain a constant output power as the repetition rate is tuned. Furthermore, by illuminating the regions-of-interest (ROIs) only, pulse-on-demand systems (e.g., the adaptive excitation source [25]) can increase the signal strength without increasing the excitation power in the sample or requiring higher laser output. Such adaptive lasers can improve the performance of deep tissue 2PM (e.g., imaging speed) and are likely to prove essential for reaching the depth limit for 3PM.

**Concluding Remarks**

This roadmap aims to elucidate the challenges for high spatial resolution, deep tissue, three-dimensionally resolved fluorescence microscopy. The compatibility of the long wavelength windows and the availability of fluorophores, together with the trade-off between the SBR and the signal strength, form the basis for the current choices of 1P confocal microscopy, 2PM, and 3PM and the depth limit of each imaging modalities. High spatial resolution fluorescence imaging in deep scattering tissue is challenging because the "difficulty" grows exponentially as a function of imaging depth. While long wavelength multiphoton microscopy can image at >2 mm in the mouse brain, the imaging depth is still less than a quarter of an adult mouse brain in vivo. Future advancements in fluorophores, detectors, and lasers can perhaps push the imaging depth of three-dimensionally resolved fluorescence microscopy by another factor of two (e.g., 3 to 4 mm when imaging the mouse brain vasculature). Breakthrough innovations are needed to image much deeper than long wavelength multiphoton microscopy.

**Acknowledgements**

This work has been funded by a National Science Foundation (NSF) grant (DBI-1707312).

## Section 2 – Adaptive optical multiphoton fluorescence microscopy
Na Ji, University of California Berkeley

**Status**

Adaptive optics (AO) was originally developed to combat atmospheric aberrations that degrade image quality of astronomical objects. Here, wavefront distortion is measured directly using devices such as a Shack-Hartmann wavefront sensor. With the increasing applications of optical microscopy to imaging complex tissues, AO methods have been developed for microscopy to correct sample-induced aberrations in order to maintain optimal imaging performance[1]. Biological aberrations are distinct from their astronomical counterpart in that there is little to no temporal variation in the aberration profile but the samples are often optically opaque. Together, these characteristics have motivated the development of indirect wavefront sensing methods whose performance is not affected by light scattering.

Both direct and indirect wavefront sensing have been applied to multiphoton fluorescence microscopy (MPFM). The most popular and powerful method for imaging opaque samples is to measure the tissue-induced aberrations on the excitation light and then cancel them out by pre-shaping the excitation wavefront using a deformable mirror or spatial light modulator. To reduce scattering in direct wavefront sensing, far-red and near-infrared fluorophores were employed [2, 3]. Indirect wavefront sensing methods use serial evaluations of image metrics (e.g., brightness, spatial resolution, contrast, point spread function) while manipulating the excitation light to deduce the wavefront profile[4-8]. When implemented properly, both types of methods are capable of forming a diffraction-limited focus deep in tissue to excite fluorescence at diffraction-limited spatial resolution.

In the opaque mouse brain, AO has enabled MPFM to visualize subcellular structures such as dendrites and dendritic spines hundreds of microns below the brain surface (Fig. 1). It has also enabled biological discovery: using an AO-enabled two-photon fluorescence microscope, we characterized the input from visual thalamus in the mouse visual cortex and discovered previously unknown orientation selectivity of their synapses[9]. The rich repertoire of AO technologies and extensive demonstrations of their capabilities have firmly established AO to be essential for high-resolution MPFM investigations of complex tissues at depth.

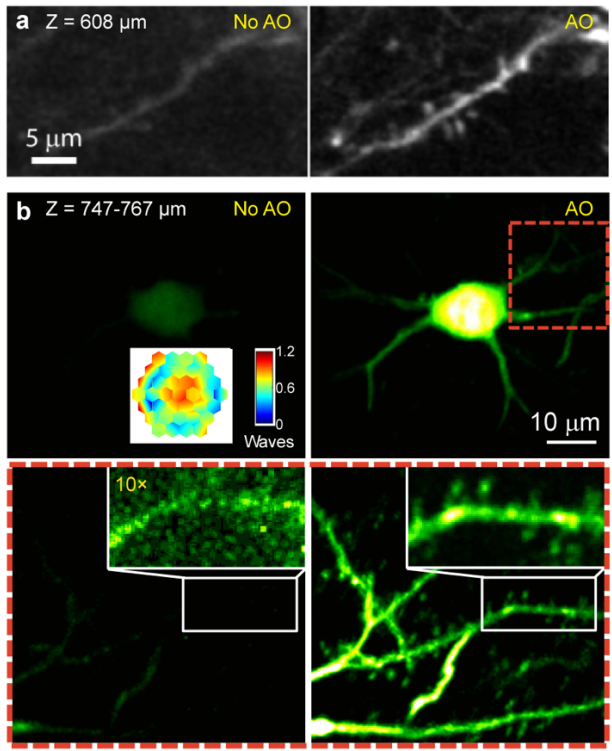

**Figure 1.** AO improves image quality of multiphoton fluorescence microscopy in the live mouse brain. (a) Two-photon fluorescence images of dendrites and dendritic spines before and after AO using direct wavefront sensing [3]. (b) Three-photon fluorescence images of a neuron, its dendrites, and dendritic spines before and after AO using indirect wavefront sensing [9].

## Current and Future Challenges

Because fluorophores emitting in the visible spectrum are most commonly used to probe biological processes, the requirement of introducing additional fluorophores with far-red and NIR emission for direct wavefront sensing complicates sample preparation. The easiest way to introduce these far-red/NIR fluorophores into brain tissues is by injecting them (typically chemical dyes) into the blood[3]. However, this approach may lead to corrections with smaller isoplanatic patch due to the high curvature of blood vessels and is not applicable to tissues devoid of vasculature. Indirect wavefront sensing methods can work with fluorophores in the visible spectrum, but the depth at which they can be applied in opaque tissues remains limited by scattering of the excitation light and the brightness of the fluorophores.

Currently, AO has largely remained the domain of physicists rather than biologists. One challenge, therefore, is how to maximize their impact on the biological fields where optical microscopy is routinely applied to enable discovery. AO systems developed for telescopes in large observatories have in-house staff ensure their optimal performance, allowing external users to benefit from the high resolution without requiring them to have optical expertise. However, there are no microscopy facilities that operate on a similarly large scale, with most biology laboratories having their own microscopes or relying on core facilities at their institutions. Given the lack of commercially available adaptive optical microscopes, implementation of AO in laboratories that pursue biological inquiry has been limited to a few groups straddling optics and biology.

**Advances in Science and Technology to Meet Challenges**

Because longer wavelength light is less scattered by tissue, using excitation light of longer wavelengths (three- versus two-photon, e.g., 1.3 µm versus 0.9 µm excitation for green fluorophores) can increase the imaging depth in tissue (see previous chapter). At such large imaging depths, AO remains essential in achieving high spatial resolution. Because tissues often absorb more at these longer wavelengths, by increasing the focal intensity, AO enables the reduction of average excitation power and reduces heating-induced tissue damage.

Effort has also been put into developing far-red and NIR fluorescence proteins, which target cell types, biomolecules, and biological processes with much higher specificity than chemical dyes. For example, recently a NIR protein was developed to sense intracellular calcium concentration[10]. Although still inferior to visible fluorescent proteins in terms of brightness and photostability, continued efforts in engineering better NIR proteins could eventually allow them to provide both structural and functional information, as well as act as guide stars for direct wavefront sensing, substantially reducing the demand on sample preparation.

With the lack of commercially available microscopy systems, to maximize the impact of AO technologies, it is essential to reduce the complexity of their implementation both in terms of hardware and software. Direct wavefront sensing requires a sensor and a modulator of the wavefront, both of which need to be carefully calibrated and aligned. Therefore, for labs to integrate AO into their existing microscopes, indirect wavefront sensing techniques that utilize a single wavefront modulator can be more easily incorporated into the microscopy beam path. Standalone software module that can be operated independently of microscope control program has also been developed[8], which should further lower the threshold of entry for biological laboratories.

**Concluding Remarks**

By cancelling out tissue-induced aberrations and recovering a diffraction-limited focus for multiphoton excitation, AO methods utilizing both direct and indirect wavefront sensing have led to drastic improvement of image quality of MPFM in complex tissues. Although their applications have yet to go much beyond demonstrations of physical principles, efforts have been made to improve the accessibility of these methods to non-experts. Together with the continued push in developing brighter fluorophores with longer wavelengths, AO would become an essential component in cutting-edge MPFMs in pushing the imaging depth for biological investigations at high spatial resolution.

**Acknowledgements**

*This work is supported by National Institutes of Health (U01NS118300).*

## Section 3 – Optical wavefront engineering for intravital fluorescence microscopy

Meng Cui, Purdue ECE and Biology

**Status**

Cellular resolution imaging in live biological systems holds great significance in biology and medicine[1]. Thanks to the rapid advance of genetic fluorescence function indicators, various cellular activities can be captured by optical measurement. However, a major challenge of applying optical measurement in live animals is the limited imaging depth as a result of the inhomogeneous refractive index of biological tissue[2-4]. Achieving *in vivo* large volume high-throughput 3D imaging remains a challenge in most animal models. Wavefront engineering has been explored to improve the performance of deep tissue imaging.  First, tissue-induced light scattering and aberration is a reversible process. Proper engineering of the optical wavefront can correct optical aberration and even suppress light scattering, which can improve the imaging spatial resolution and signal-to-noise ratio (SNR)[2-4]. Second, wavefront engineering can be employed to achieve 3D volumetric imaging[1]. Various devices can be employed to generate defocusing wavefront which leads to axial control of the imaging plane. Third, wavefront engineering can be coupled with miniature invasive imaging probes to access extremely large depth[5]. Miniature imaging devices often have inherent aberration, which greatly reduces their imaging performance including resolution, field-of-view, and imaging throughput. Wavefront engineering can help assist miniature probe imaging to improve the overall imaging performance. Further advances in all three directions are expected to enable new observations and knowledge in biomedical research.

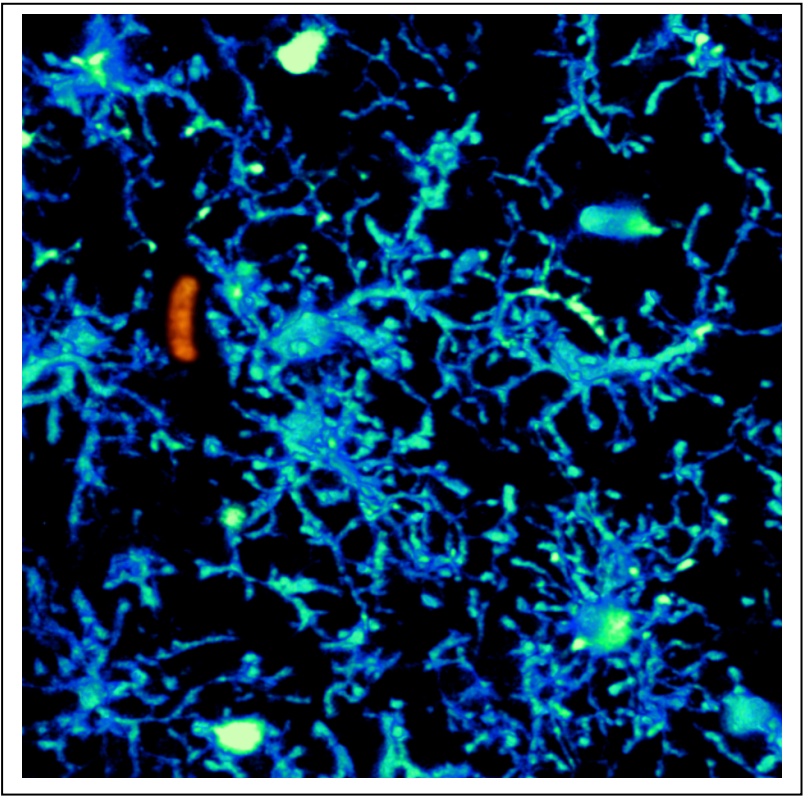

**Figure 1.** High-speed 3D volumetric imaging of mouse brain. Cyan, microglia; Orange, blood cell.

## Current and Future Challenges

Cellular resolution function imaging has been widely employed in biomedical research (Fig. 1). In immunology, intravital imaging has been used to track the motility and interactions of immune cells, which provides key information about the function and dynamics of various cell types[1]. In neuroscience, *in vivo* measurement is performed over a wide range of spatiotemporal scales. To study the neuronal plasticity related to development, learning, and memory, one needs to observe the subtle morphology change of neuronal structures at a sub-micron spatial resolution over days or weeks. To capture the neuronal activity during behavior, fluorescence images based on calcium or voltage indicators need to be recorded at 10-1000 Hz frame rate with cellular or subcellular resolutions. Spatial resolution, 3D signal confinement, and SNR are highly important to these measurements, which demand high-quality focus deep in biological tissue. Although the index of refraction of biological tissue is similar to that of water, the index of cellular components is slightly higher, which causes spatially varying wavefront distortion. Moreover, the movement of cellular structure and the trafficking of blood cells also cause temporal variation. To fully compensate for the wavefront distortion, we need to provide dynamic correction which also varies over space and time. As the cellular dynamics are inherently three-dimensional, we also need to capture the dynamics in 3D. For the applications on awake animals, the motion of the animal also causes image instability. As a result, slow recording will suffer from motion artifacts. With high-speed 3D recording techniques, we can eliminate such artifacts through post-measurement image registration. To access very deep regions (e.g. several millimeters or more), miniature invasive imaging probes are commonly employed. However, the design and the dimension limit of these miniature lenses cause inevitable aberrations, which are also field-position-dependent. Routine applications often suffer from reduced resolution, SNR, field-of-view, and throughput. Advanced wavefront correction is needed to enable high-quality large volume imaging through these miniature imaging devices.

**Advances in Science and Technology to Meet Challenges**

To provide *in vivo* wavefront measurement, sensor-based and sensorless methods have both been developed. For multiphoton excitation, the excitation and emission wavelengths are far apart. The short wavelength emission suffers much severe aberration and scattering. Therefore, sensorless methods that modulate the excitation wavefront profiles are often preferred[2]. High-speed methods can achieve microsecond level measurement time per spatial mode[6]. For correcting highly complex wavefront distortion, the Iterative Multi-Photon Adaptive Compensation Technique (IMPACT) has been developed[2], which leverages the iterative feedback and the inherent nonlinearity in multi-photon imaging to force the focus to converge to a diffraction-limited spot inside highly scattering biological tissue. In addition to high-resolution imaging inside the thick brain and lymph node tissue, IMPACT also enables high-resolution noninvasive transcranial imaging through intact mouse skulls[7]. For high-throughput large field-of-view imaging, the imaging system needs to provide simultaneous spatially varying aberration correction. Multi-Pupil Adaptive Optics (MPAO) has been developed to achieve high-speed high-resolution imaging[8]. Moreover, defocusing control can be applied to the desired region to achieve non-planar imaging such that the features of interest can be shifted to the same 2D recording plane for simultaneous fast recording. Towards fast 3D volumetric imaging, the optical phase-locked ultrasound lens has been explored to provide a microsecond scale defocusing wavefront control[1]. Such capabilities can convert existing 2D raster scanning microscopes for fast 3D volumetric recording. An important technique for 3D laser scanning imaging is the remote focusing method[9], which relays a defocusing wavefront through a pair of objective lenses to the desired focal plane for rapidly shifting the laser focus. The perfect operation of remote focusing demands perfect telecentric objective and relay lenses. However, perfect telecentricity is not the design goal of common objective lenses and is hardly achievable. To improve the imaging performance, an image plane adaptive correction method has been developed which can greatly extend the working range of remote focusing systems[10]. For imaging beyond several millimeters, miniature invasive probes are often used. The inherent aberration limits the resolution and accessible tissue volume. Recently, Clear Optically Matched Panoramic Access Channel Technique (COMPACT) has been developed, which provides two to three orders of magnitude increase in tissue access volume[5]. Combined with aberration correction, COMPACT can yield high-quality images over massive tissue volume.

**Concluding Remarks**

The ultimate goal of *in vivo* imaging is to noninvasively image deep inside live biological systems. Currently, the majority of the development is still to correct for the static slowly varying low-order aberrations. Although these developments lead to better resolution and SNR, the imaging depth advance is still moderate. Significant imaging depth increase can only happen if the wavefront correction can handle the high-order spatially varying dynamic wavefront distortion in live animals. Currently, none of the established methods is close to achieving this goal. Major innovations are needed to break the current limit. An important aspect of tool development is the broad and routine adoption by the users, which requires the developed technique to be highly robust and easy to use. Without such capabilities, the development will likely have negligible impacts.

**Acknowledgements**

*M.C. acknowledges the support by NIH grant 1U01NS094341, U01NS107689, RF1MH120005, RF1MH1246611, U01NS118302, 1R01NS118330, R21EY032382, Purdue University, and the scientific equipment from HHMI.*

## Section 4 – High-Speed Wavefront Shaping
Rafael Piestun, University of Colorado Boulder

**Status**

Spatially modulating light at high speed is critical for the success of multiple optical techniques and applications. Early use of spatial light modulators (SLM) in 3D holographic displays, optical signal processing, and pattern recognition were hampered by the lack of adequate SLM at the time. From initial liquid crystal displays with low space-bandwidth product and poor phase modulation to photorefractives, a plethora of modulation techniques would require decades to mature to the level required for current deep imaging needs. Adaptive optics became practical early on through the use of mechanically deformable mirrors which, despite having a relatively small number of actuators, adapt effectively to the task of compensating for optical aberrations.

Most biological applications of wavefront shaping (WFS) require fast wavefront modulation, regardless of the specific technique used to calculate the compensating wavefront. In techniques that use iterative optimization, optical systems utilize feedback over multiple iterations to attain a wavefront that satisfies a target performance. Alternatively, in transmission matrix calculations, a large set of wavefronts needs to be projected and the respective outputs measured, ideally fast enough before the medium changes again. Similarly, in direct digital phase conjugation, the spatial modulation is calculated directly from direct measurements of the wavefront but latencies and dynamic changes of the medium are still an issue. Fast spatial modulation is critical in high speed scanning through or inside a complex medium regardless of how the SLM pattern is obtained, including techniques based on digital phase conjugation, transmission matrix, or iterative optimization.

The need for fast modulation techniques goes beyond biology, being a requirement in dynamic imaging, sensing and focusing, with implications in optical communications as well as quantum and nonlinearity control.

Recent demonstrations using micro-electro-mechanical systems (MEMS) and acousto-optics have shown a path from the early modulation rates in the 10s of Hz to 100s of KHz. These experiments help motivate the development of larger and better SLM arrays, as well as the investigation into novel physical modulation mechanisms. In effect, current SLM constraints imply that speed is typically achieved at the expense of a lower number of degrees of freedom or a reduction in efficiency, if not both.

This chapter reviews recent progress to achieve high-speed WFS, requiring the adoption of new modulation mechanisms, as well as optical, electronic, and computation optimization.

**Current and Future Challenges**

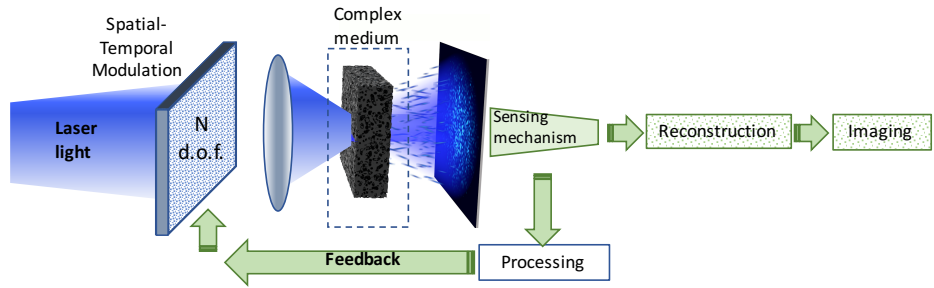

**Figure 1.** A general wavefront shaping problem: Coherent light is modulated by N independent degrees of freedom of a spatial light modulator (SLM) to attain a target metric that helps imaging through/within scattering media. A sensing mechanism (located on either side of the scatterer) provides feedback based on measurements that inform the state of the SLM. Reconstruction algorithms deliver the answer to the target imaging question.

A general WFS problem, as depicted in Figure 1, seeks to image through or inside a complex medium. These are defined as highly inhomogeneous media that generate multiple scattering events for any light propagating in their interior, hence scrambling the information to a degree that has been traditionally approached via statistical methods. This is the result of having a huge number of scatterers, with unknown locations and optical properties. Additionally, complicating the problem even further, these scatterers and the whole medium are dynamically changing.

The use of coherent laser light provides high intensity sources that somehow mitigate the need to consider chromatic dispersion while enabling sensitive phase measurements. Notwithstanding, multispectral or short pulse generalizations have been considered.

Measurements are provided via a sensing mechanism such as photodetectors collecting excited fluorescence or acoustic transducers detecting photoacoustic signals. These measurements are used to inform and update the SLM state, typically multiple times as the complex medium changes or as part of a sequence of measurements to characterize the medium. The process continues with a series of measurements followed by a matched reconstruction algorithm to attain the target imaging task.

The principle of phase conjugation for the correction of distortions by inhomogeneous objects was recognized shortly after the invention of holography[1]. Nowadays, the concepts of holography, optical, digital, or computer-generated, provide a useful framework for understanding and devising techniques for imaging through highly inhomogeneous media.

The traditional optical holographic process is slow for most dynamic imaging situations. The advent of fast detector arrays, interfacing electronics, SLMs, and computers in combination with new understanding of complex media and insights in computational imaging have opened opportunities for previously unthinkable imaging performance.

The ideal SLM, in particular, has a large number of pixels, number of phase levels, phase stroke, and diffraction efficiency; while simultaneously achieving fast response time, switching frequency and low latency. While progress in SLM technologies has been significant, the currently available spatial and temporal space bandwidths are far from what is needed when compared to the degrees of freedom of most complex media. Hence, managing tradeoffs is critical to advance imaging applications.

**Advances in Science and Technology to Meet Challenges**

One approach to overcome current limitations is to use fast SLMs with lower number of pixels and/or lower number of states (e.g. binary), while compensating using holographic encoding. For instance, it is possible to use a binary-amplitude SLM such as the deformable mirror device in conjunction with computer generated holography concepts to achieve millisecond scale transmission matrix measurements[2,3]. The electronic implementation via dedicated hardware is critical to reduce latencies due to communication and computation[3].

Another possibility is to use the dimensionality transformation enabled by scattering media to take advantage of existing fast 1D modulators[4]. In effect, the actual arrangement of the input modes is almost irrelevant as far as the coupling to the output modes is strong enough. In this case, the scattering medium converts the light from each pixel of the 1D SLM into a 2D speckle pattern. Adjusting the phase of each pixel provides a means to exert control of the output light distribution in two or even three dimensions. Hence, 1D modulation (at 100s of kHz) provides opportunities to control light about three orders of magnitude faster than with liquid crystal SLMs[4].

A little-explored area is the control of nonlinear phenomena using WFS[5]. Nonlinear WFS imaging techniques, such multi-photon excitation, Raman scattering, or second-harmonic generation provide new means to interrogate different materials. High-speed modulation could also benefit emerging hybrid imaging modalities [6-7].

The functional imaging of the brain, an imaging grand challenge, could help understand the neural pathways that trigger human brain function. The use of WFS in multimode fibers has enabled the thinnest endoscopes to functionally image the activity of live neurons[8]. Real-time imaging was critical to follow the neuron action potential patterns (Figure 2), accessing the brain with cellular resolution at depths where non-invasive techniques cannot reach.

Emerging techniques also include the use of acousto-optic devices to control the phase via programmable RF signals that encode beams in the medium. This is followed by measurement of the phases of the scattered beams with a fast single-pixel detector. The acousto-optic deflector phase conjugates the beams and creates a spatio-temporal focus possibly scanned at high speed[9].

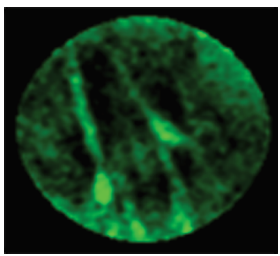
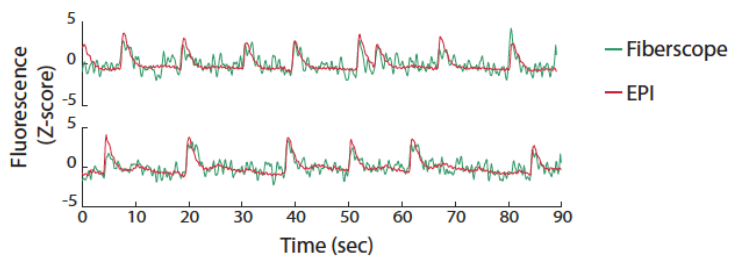

Figure 2.  Structural and in-vivo functional imaging with an ultrathin endoscope based on a multimode fiber and wavefront shaping. (Left) Imaging baby hamster kidney cells expressing eGFP. (Right) Hippocampal neuronal tissue culture expressing GCaMP6f, plots show the time signal of two neuronal cells compared with simultaneous measurements using traditional epi-fluorescence microscopy (from [8]).

## Concluding Remarks

Progress in high-speed WFS is accelerating to such an extent that the description of all advances exceeds the scope of this article. While some ideas might take years to become practical, others are already having an impact by managing tradeoffs such as speed vs degrees of freedom or speed vs system complexity. Novel concepts are still needed, for instance with the adoption of tools such as machine learning or novel physical mechanisms for modulation.  An intriguing development to watch is based on the use of the so-called active metasurfaces [10]. An outgrowth of subwavelength diffractive optics, they Implement planar nano-structured arrays whose optical response can be dynamically tuned. The control is achieved by different physical mechanisms and material systems, including mechanical deformation, phase-change media, free-carrier density modulation, thermo-optic and electro-optics effects.

## Acknowledgements

The author acknowledges support from the National Science Foundation (Award 1548924) and the Colorado Office of Economic Development and International Trade.

## Section 5 – Imaging in complex media by scattered back-illumination

Jerome Mertz, Timothy Weber, Sheng Xiao, Boston University

**Status**

When imaging in complex media one must first be clear on what exactly one is interested in imaging. For example, in fluorescence imaging, one is interested in the spatial distribution of fluorescent markers. Complexity of the medium in this case can introduce aberrations that degrade image contrast and resolution. On the other hand, when performing label-free imaging, contrast generally arises from scattering, enabling structures of interest to be distinguished by their size. For example, one might wish to image small, point-like scatterers embedded in an aberrating medium of spatially varying refractive index (RI). Alternatively, one might be interested in reconstructing the RI variations themselves, whereupon point-like scatterers only undermine image quality by introducing multiple scattering. This last case can be of particular interest since a retrieval of the RI distribution can enable the application of adaptive optics to improve image quality.

However, obtaining a 3D image of the RI distribution in a complex medium is no easy task, particularly if the medium is so thick that it can only be accessed from one side. A difficulty comes from the fact that only sharp structures in the medium, such as point-like reflectors or interfaces, provide sufficiently high spatial frequencies to scatter light in the backward direction so that it can be detected. More slowly varying structures scatter light only in the forward direction, making them invisible to conventional epi-detection devices. Such is the case with optical coherence tomography (OCT), which can only reveal sharp sample structures, producing highly granular images. The retrieval of slowly varying structures must be inferred indirectly in this case, for example by considering their effects on the image granularity and solving an inverse problem to simultaneously reconstruct both high and low frequency structure [1][2]. An alternative simpler approach, which is the subject of this roadmap, is to exploit the multiple scattering from deeper layers within the sample and use this to back-illuminate regions of interest in shallower layers. In this manner, contrast becomes based on light transmission rather than reflection, providing access to low-frequency sample structure. Such is the principle of oblique back-illumination microscopy (OBM) [3], depicted in Fig. 1, along with its scanning analogue based on oblique back-detection (formally equivalent by virtue of Helmholtz reciprocity [4]). OBM has recently been shown to enable quantitative RI reconstruction in 3D [5].

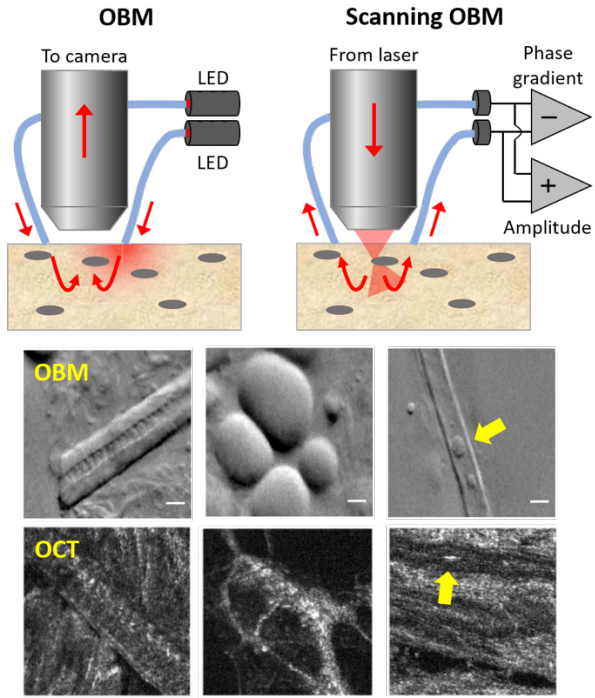

**Figure 1.** Top: Schematics of OBM and scanning OBM (adapted from [4]). Bottom: Comparison of sub-surface OBM and en-face OCT images in mouse skin. Left: Hair shafts. Middle: Fat globules. Right: Blood vessels (arrows indicate red blood cells). Scale bar: 10μm.

## Current and Future Challenges

OBM is simple to implement, basically as an add-on to any conventional microscope, however it does suffer from drawbacks. For one, while it allows imaging within arbitrarily thick, complex samples (indeed, complexity is required to obtain backscattering in the first place), it does not provide particularly deep imaging. For example, it does not attain the same penetration depths as OCT, despite the fact that both modalities are based on scattering. One reason for this is that OCT makes use of coherent illumination which, in turn, allows the possibility of interferometric time gating (or coherence gating) to reject background and improve contrast. Another reason is that OCT is based on ballistic light illumination and detection, which is easy to control (e.g. by spatial filtering). In contrast, OBM relies on diffuse backscattering, which is much more difficult to control.

Nevertheless, diffuse backscattering is not impossible to control. In particular, a remarkable phenomenon called the memory effect [6] prescribes that diffuse light obeys ballistic transmission or reflection laws even in thick complex media, but only within a very narrow angular range dependent on basic media properties (thickness in transmission; transport scattering length in reflection, see section 13). A hint as to how the memory effect can be exploited to improve OBM comes from the field of acoustical imaging [7], as shown in Fig. 2. Here, diffuse back-insonification, even though it is spatially incoherent, is controlled by the memory effect to enable essentially a coherent version of OBM based on measurements of differential phase rather than differential intensity. Because the back-insonification is in the transmission direction, it can reveal weak structures within the medium that are completely invisible to standard ultrasound imaging based on pulse-echo sonography (the acoustic equivalent of OCT). In addition, what can be loosely thought of as an acoustic version of scanning OBM has also been demonstrated, though with a more involved inversion-based image reconstruction algorithm where the memory effect is more implicit [8].

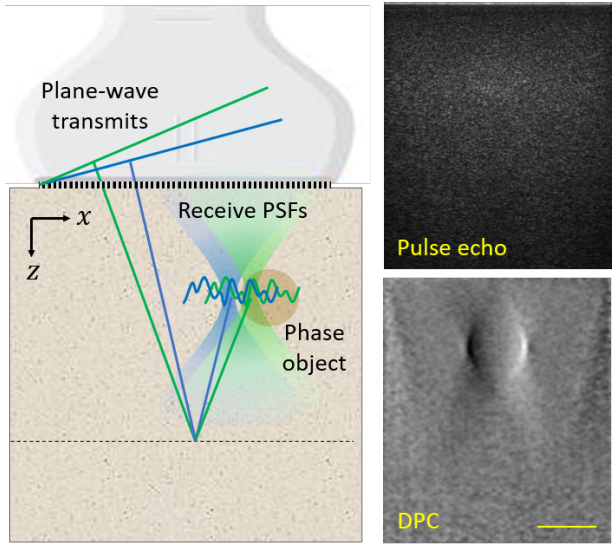

Figure 2. Left: Principle of ultrasound differential phase contrast (DPC). A linear-array probe transmits pairs of plane-wave pulses of different tilt angles. Memory effect leads to a controlled translational shift of the scattered back-insonification from deeper layers, allowing differential phase imaging at shallower layers. Right: a phase inclusion invisible to conventional pulse-echo sonography becomes apparent with DPC. Scale bar: 10mm. Adapted from [7].

**Advances in Science and Technology to Meet Challenges**
A feature of acoustic imaging compared to optical imaging is that the wave frequencies involved are orders of magnitude smaller (typically ~8). As such, actuators are readily available that allow the coherent receive of acoustic waves where both amplitude and phase can be directly resolved, leading to the possibility of sub-cycle time gating. The same cannot be said of optical detectors, which are currently too slow to directly resolve optical phase. This becomes important in applications involving the memory effect, since the range of this effect is known to increase when time gating becomes more refined [9]. It is for this reason that the memory effect could be easily exploited with acoustic OBM, enabling coherent differential phase detection, while it could not with optical OBM, which to date has only been demonstrated with incoherent differential intensity detection.

Time resolved imaging is, of course, possible and routinely performed with light, but attaining better than picosecond time resolution requires some kind of trick typically involving interferometry. For example, ultrafast pulsed lasers are readily available, enabling sub-picosecond time gating by interference with a pulsed reference wave. This principle is exploited in time-domain OCT. Alternatively, broadband spatially coherent sources are also readily available that allow the Fourier synthesis of a time gate, as exploited in frequency-domain OCT (and also in [9]). The application of such tricks to optical OBM may be envisaged, though perhaps more conceivably in its scanning configuration which involves only single-element detectors. However, the problem of engineering appropriate reference waves still remains. Much more straightforward would be a method of time gating that does not require interferometry, perhaps making use of a Kerr gate or, better still, by direct detection with an ultrafast detector. Advances in single-photon avalanche detectors (SPADs), either single element or in array form, may lead the way here. Indeed, the development of higher speed devices enabling the direct coherent detection of light is one of the key technological advances that the optical imaging community is eagerly awaiting.

**Concluding Remarks**
The purpose of this section is to highlight the possibility of indirectly rather than directly imaging structures within complex media by way of scattered back-illumination, leading to the possibility of transmission-based imaging with its attendant benefits, and allowing structures to be revealed that would normally be invisible. Scattered back-illumination can be controlled to a surprising degree of precision by way of the memory effect (at least in acoustics). Remarkably, this memory effect is far more general than utilized here [10], extending beyond spatial degrees of freedom to even polarization and spectral degrees. Such a generality opens the door to far richer contrast modalities than simple phase imaging as shown here, and promises to span a wide range of applications including biomedical, imaging around corners, lidar, sonar, seismology, and many more.

**Acknowledgements**
National Institutes of Health: R01CA182939, R21GM134216.

# Section 6 – Guidestar-assisted wavefront shaping

Joshua Brake 1†, Zhongtao Cheng 2†, Roarke Horstmeyer 3, Lihong Wang 2,4*, Changhuei Yang 4*

1 Department of Engineering, Harvey Mudd College, Claremont, CA 91711, USA
2 Caltech Optical Imaging Laboratory, Andrew and Peggy Cherng Department of Medical Engineering, California Institute of Technology, Pasadena, CA 91106, USA
3 Department of Biomedical Engineering, Duke University, Durham, NC 27708, USA
4 Department of Electrical Engineering, California Institute of Technology, Pasadena, CA 91106, USA
† Joshua Brake and Zhongtao Cheng contributed equally to this work.
* email: lihong@caltech.edu; chyang@caltech.edu

## Status

Focusing light efficiently into complex scattering media is significant for many applications, including optical imaging, manipulation, therapy, and stimulation. However, scattering media randomize the wavefront of an incident optical field, preventing the light from being easily focused, as it would in free space. To overcome this challenge, guidestar-assisted wavefront shaping methods are being actively developed. The general principle of guidestar-assisted wavefront shaping is illustrated in Fig. 1(a). The guidestar, which is typically located at the desired focus spot inside a scattering medium, interacts with the scattered photons and encodes its location in the scattered light. The measurement system detects the exiting scattered light and identifies the components that originate from the guidestar's location. The system then determines a wavefront modulation map to present on a spatial light modulator (SLM) to tailor the wavefront of the incident laser. Various information from the guidestar can be used such as the scattered wavefront itself [Fig. 1(b)][1], wavefront variation induced by the guidestar [Fig. 1(c)][2], the scattering point spread function [Fig. 1(d)][3], or the total signal intensity from the guidestar [Fig. 1(e)]. According to the specific measured quantity, the wavefront modulation map is updated iteratively to maximize the desired optical pattern (e.g., a focal spot) inside the scattering media, or directly determined by phase conjugation, to refocus light to the guidestar location.

One example of a popular guidestar mechanism is to generate new photons with a different frequency. Examples of this class of guidestar include nanoparticles which generate a nonlinear second-harmonic signal (SHG)[4], focused ultrasound which generates frequency-shifted photons using the acousto-optic effect[1], and fluorescent materials[3,5]. Among these examples, focused ultrasound provides a non-invasive and freely addressable approach for focusing light into scattering media [Fig. 1(b)], which is more promising for general wavefront shaping applications. A related guidestar is the photo-acoustic guidestar [Fig. 1(e)], which is detailed in section 7. A guidestar can also encode its location information by inducing wavefront variation, which is termed dynamic guidestar. Dynamic guidestars can be physical, such as magnetic particles[6] and microbubbles[7], or virtual, such as the adapted-perturbation in samples[2] or perturbations induced by an ultrasound field[8]. Through detecting and conjugating the differential field of collected scattered light at two different states of a dynamic guidestar, an optical focus can be realized at the position where the wavefront variation originates inside the scattering medium (i.e., the dynamic guidestar location) [Fig. 1(c)]. Since fluorescence is an important contrast mechanism in optical imaging, the fluorescence-based guidestar has intrinsic advantages of adoption in fluorescence microscopy. This type of approach has been previously demonstrated for adaptive optics correction with standard one-photon fluorescence[9] as well as two-photon (2P) excitation microscopy. As 2P excitation-based detection facilitates clear image formation deeper within tissue than standard fluorescence, the intrinsic guidestar approach has been integrated into several 2P systems as a promising means to extend imaging depths to 400 μm or more within biological tissue[10]. Existing approaches include an iterative SLM update technique[11] (IMPACT), a scanning-based approach for point-spread function estimation[3] (F-SHARP, Fig. 1(d)), and a holographic phase stepping approach for rapid correction[12] (DASH). Such techniques provide a promising means to jointly focus and rapidly scan out images at improved tissue depths with minimal additional hardware and in a non-invasive manner.

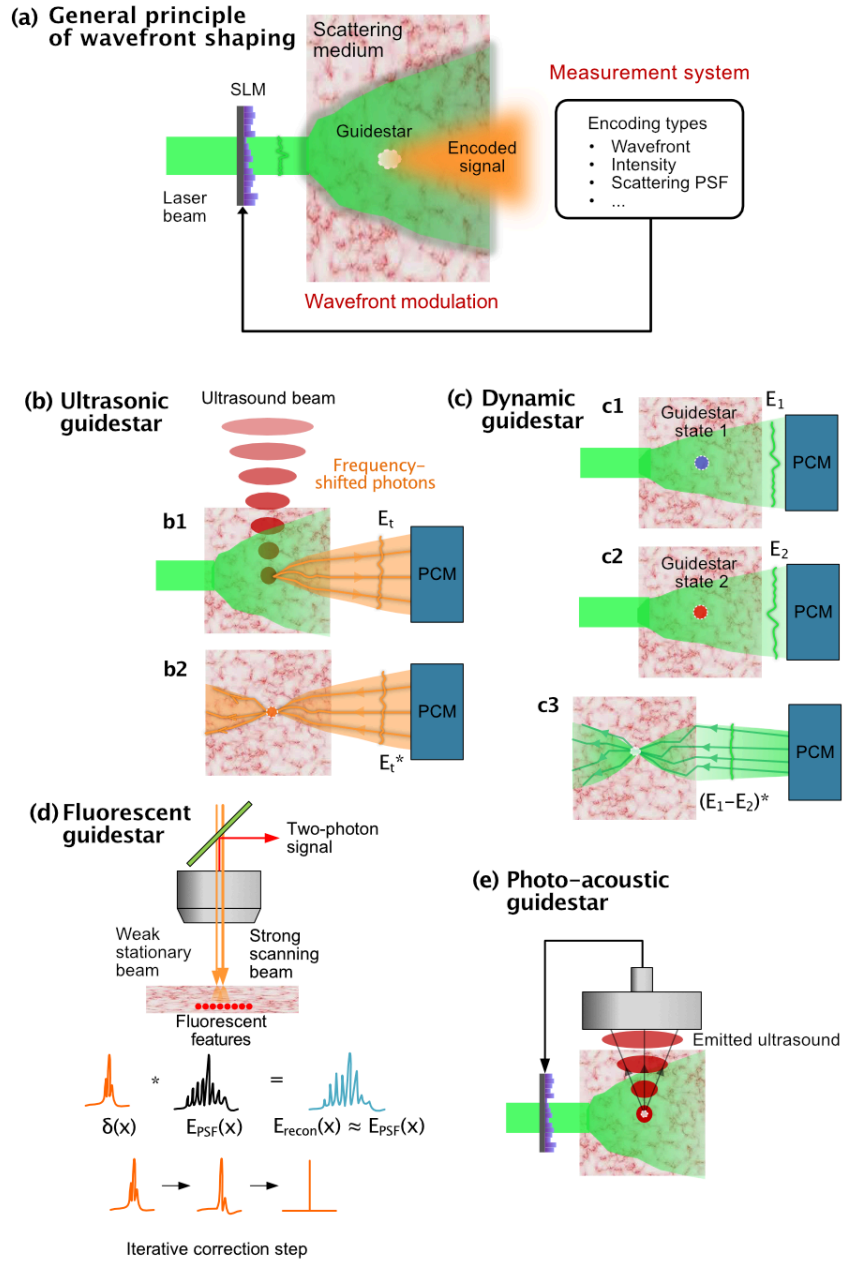

**Figure 1**. The state-of-the-art techniques for guidestar-assisted wavefront shaping. (a) The general principle of guidestar-assisted wavefront shaping. (b) Ultrasonic guidestar: (b1) wavefront recording of the frequency-shifted photons from the ultrasound modulation through a phase conjugation mirror (PCM); (b2) generating the conjugated wavefront for optical focusing. (c) Dynamic guidestar: (c1)-(c2) wavefront recording of the scattered fields at two different states of the guidestar; (c3) generating the conjugated differential field for optical focusing. (d) Fluorescent guidestar. This panel shows the iterative measurement process for focusing on features of interest using the scattering point spread function in two-photon fluorescence microscopy with F-SHARP. (e) Photo-acoustic guidestar. Emitted ultrasonic waves from a feature of interest are monitored to provide feedback for the wavefront shaping system.

## Current and Future Challenges

The major areas of development that can improve guidestar-assisted wavefront shaping fall into two main categories: (1) improvement of system latency to enable focusing deeper in dynamic scattering media and (2) development of new guidestar mechanisms to improve focusing performance and enable increased adoption. Solving these challenges will enable wavefront shaping to advance beyond experiments with carefully controlled and designed samples to more practical applications throughout biomedicine.

The latency between recording and playback in a wavefront shaping system is one of the most critical specifications for practical biomedical applications. The wavefront which focuses light to a desired location is valid only for a limited time due to biological motion such as blood flow. This time scale can be as long as several seconds in acute brain slices in vitro[13], but the decorrelation time at the same depth within *in vivo* specimens drops by three orders of magnitude to around 1 ms. Such fast decorrelation requires that the latency of a wavefront shaping system be on the timescale of milliseconds or less, which is challenging for most current techniques. In addition, for effective wavefront shaping into thicker samples, the control of additional degrees of freedom (i.e., additional pixels on the wavefront shaping device) is desirable, which places additional demands on system latency that must be considered.

The second main challenge is related to the guidestar mechanisms themselves. While many different guidestar mechanisms (such as those discussed previously) have been developed, they each have specific applications for which they are best suited. For example, one of the main advantages of the ultrasonic guidestar is that it is freely addressable and can be easily moved to target a desired focal location. However, it has a low modulation efficiency (defined as the percentage of light interacting with the guidestar location and subsequently "tagged" with a different frequency), it is difficult to use in a reflection geometry due to its non-isotropic tagging behavior, and it is impacted by acoustic absorption, especially at higher ultrasound frequencies which are optimal for achieving high resolution foci. As this one example indicates, further development of additional guidestar techniques or improvements to existing guidestars is necessary to achieve the goals of a freely-addressable, high-resolution, efficient guidestar which is compatible with a reflection geometry and can access depths up to the optical absorption limit.

**Advances in Science and Technology to Meet Challenges**
Addressing the challenges of lower latency wavefront shaping systems and better guidestars requires a combination of technological development and the invention of new ways to leverage intrinsic or extrinsic signals within a sample for focusing.

Improving the latency of a wavefront shaping system can be accomplished by developing faster techniques and technologies for the various stages of the recording and playback process. These primary areas for improvement are (1) spatial light modulator response, (2) data transfer, (3) phase map calculation, and (4) guidestar integration.

The most widely-used spatial light modulator (SLM) technology is based on liquid crystal (LC) technology, which can be configured to control either the amplitude or phase of the wavefront. Digital micromirror devices are another popular technology that are several orders of magnitude faster than LC-based SLMs, but in their simplest configuration offer only binary amplitude control of a wavefront. While other novel spatial light modulator technologies based on acousto-optics or liquid light valves exist and offer even faster modulation, these methods are normally limited in terms of the number of modes they can control. Thus, they are limited in terms of their practical capabilities. State-of-the-art wavefront shaping systems leverage fast spatial light modulators (e.g., digital micromirror devices) and optimized electronics (e.g., field programmable gate arrays, FPGAs) to maximize the system throughput and minimize the latency between recording and playback.

When controlling many modes simultaneously, the data transfer requirements also quickly become non-trivial. A typical scientific CMOS camera contains $5x10^6$ 16-bit pixels, meaning that a single raw frame is on the order of 10 MB. Even using the fastest commercially available data links, maximum frames rates typically top out at 100 frames per second. The amount of data also impacts the time required to calculate the appropriate phase mask for playback. Using a phase shifting approach requires a few addition operations and a division per pixel, whereas more complicated measurement schemes such as those based on off-axis interferometers may require more involved computations such as Fourier transforms. Fortunately, many of these calculations can be highly optimized and parallelized using specialized hardware such as graphics processing units (GPUs) or custom digital hardware on FPGAs. Besides advances in hardware, computational methods are also emerging for facilitating wavefront shaping via reduced data acquisition, such as the single-shot ultrasound-assisted optical focusing[14].

A promising approach toward improving the system latency is an integrated sensor and wavefront-shaping architecture that combines the functionality of an image sensor and an SLM into a single device with some basic computation capabilities provided in each pixel. In this device, each pixel can capture and compute its playback phase in parallel, thus eliminating the need to transfer data. Using a micromirror-based architecture for the wavefront-shaping elements will provide low-latency shaping capabilities. Furthermore, this architecture would solve the alignment challenges of time-reversal-based wavefront shaping systems by co-locating the sensing and shaping elements

To deal with the low signal-to-noise ratio of the ultrasonic guidestar, multiple iterative measurements of the phase map can be made to improve the focus peak-to-background ratio. However, these iterations each require a recording and playback cycle and thus can significantly increase the effective wavefront shaping system latency. Other guidestars, such as magnetic particles that can provide higher wavefront modulation efficiency, are better in this respect since they can provide higher SNRs at the same focusing depth. In summary, each guidestar technology comes with its own tradeoffs and benefits, and balancing these tradeoffs for a particular application is critical.

**Concluding Remarks**

Guidestar-assisted wavefront shaping is an attractive technique for deep imaging and efficient delivery of light energy beyond the diffusion limit. Although the system latency and guidestar mechanism of current wavefront shaping technologies are still the main limiting factors for more practical biomedical applications, advances in science and technology are showing great potential to gradually resolve or alleviate these limitations. The rapid evolution of semiconductor and liquid crystal technologies for optical sensors and modulators will enable high-speed and high-resolution wavefront sensing and controlling in the future. In the meantime, advanced computational methods are being actively developed for more efficient information extraction to accelerate wavefront shaping. The continued development of these techniques offers the potential for guidestar-assisted wavefront shaping to expand the optical imaging capabilities of scientists both in the laboratory and clinic.

**Acknowledgements**

## Section 7 – Imaging at depth with linear feedback

Hilton B. de Aguiar, Sylvain Gigan

Laboratoire Kastler Brossel, ENS-Université PSL, CNRS, Sorbonne Université, Collège de France, 24 rue Lhomond, 75005 Paris, France

**Status**

High-resolution non-invasive imaging at depth is currently performed mostly by wavefront shaping to compensate for scattering, be it by optimization, phase-conjugation, or transmission matrices approaches [1] (see also section 8-10). To be non-invasive, they require a "guide-star" mechanism. Besides acoustic methods, all methods require a non-linear feedback mechanism, for instance SHG, or multiphoton fluorescence in order to converge into a single-grain focus, typically diffraction-limited [2] (see also section 5). Even acoustic methods benefit from non-linearities, allowing to improve the resolution close to the optical diffraction limit. Linear feedback is conventionally considered not to be feasible, because of the impossibility to converge to a focus from an extended object. In parallel, algorithmic tools have been introduced and became very popular to retrieve hidden fluorescent objects thanks to the memory effect. Nevertheless, they remain limited in size and complexity of the object. However, linear contrasts, such as fluorescence and Raman, are extremely important in biomedical imaging: they are the easiest and cheapest to implement, they provide an unprecedented level of signal (compared to their non-linear counterparts), and are extremely widespread in life science and medicine. Reaching deep imaging using such linear incoherent feedback is therefore an important goal for the field. In this chapter, we want to cover a few computational strategies, allowing retrieving objects with linear incoherent contrast, such as fluorescence and spontaneous Raman scattering.

**Current and Future Challenges**

Deep imaging has mostly been done exploiting nonlinear feedback for wavefront shaping experiments. Quickly after the seminal work from Mosk's group, it was realized that a feedback signal for single-grain focusing within the scattering medium was far from simple. Hence, the field has drastically focused on nonlinear signals as a feedback mechanism due to its ability to converge to a single focus in wavefront shaping experiments, as seen for instance very recently [3]. However, nonlinear processes are less popular in science and engineering due to the cost associated with the hardware (mostly the laser source).

With the emergence of the Big Data era, computational microscopy has brought new directions to the wavefront shaping toolbox. In particular, computational tools exploit other properties of scattering – that may allow for unique single-grain focusing in wavefront shaping experiments or retrieve an image without resorting to wavefront shaping – using linear optical process as a readout, in particular fluorescence. A first set of computational tools have recently been put forward using variance-based methods using linear fluorescence [4,5]: variance calculations can be seen as a "nonlinear metric", therefore allowing for focusing convergence in wavefront shaping experiments. However, imaging was only possible with focus-scanning techniques, hence time consuming. More recently, [6] has solved this issue by combining wavefront shaping with computational modelling and algorithm tools allowing for imaging without focus-scanning methodologies. Indeed, most recent results have shown that costly wavefront shapers are actually not needed: reading out linear fluorescence excited by a set of speckles excitation, combined with computational microscopy tools, allows for image recovery at depth [7]. Along the same lines, one may achieve super-resolution capabilities within scattering media exploiting computational tools without resorting to wavefront shaping hardware [8]. These examples highlight the wealth of information for deep imaging the incoherent properties of linear fluorescence allows for.

While conventional wisdom forces us to think that actual imaging is necessary, the randomness embedded in the speckle allows for imageless retrieval of in-depth information. For instance, in [9] demixing algorithms were used to retrieve the temporal activity of neurons (fluctuations of fluorescence), without the need to modulate the illumination.

*Figure 1: an example of Fluorescence demixing in scattering media, here for functional*

*activity. a DMD is used to excite temporally fluorescent beads to mimick neurons firing. The resulting low contrast fluorescent patterns are collected through a skull bone (highly scattering) and can be demixing thanks to a NNMF algorithm, and the activity (but not the location of the neuron) can be retrieved (adapted from [9]).*

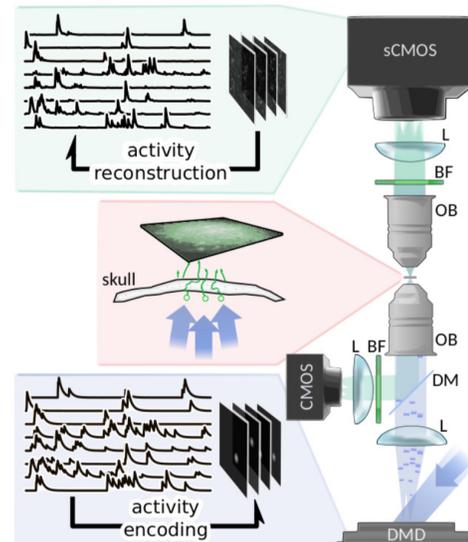

**Advances in Science and Technology to Meet Challenges**

These recent outcomes exploiting computational tools highlight a fertile field ahead. Several of these successful demonstrations rely on the fact that fluorescence is an incoherent effect and often sparse. While these aspects are embedded in many off-the-shelf algorithms, going beyond the incoherent assumption and low sparsity object in computational tools are certainly new challenges to tackle.

As we have learned [1], correlations in scattering helps in convergence in the image retrieval process. Up to now, mostly spatial correlations have been used (aka memory effect). Other correlation properties are known, such as spectral (see below) and polarization [10], and could also be exploited in future computational tools.

Despite the fact that fluorescence has been the major incoherent contrast mechanism in wavefront shaping methods, there is still much to be done with Raman scattering, another popular contrast method with molecular selectivity. Different from fluorescence, the main challenge in Raman scattering is due to its weak signals that deteriorate the performance of any method. Nevertheless, there is richness in the Raman spectrum that is still to be unveiled as no method today exploits the spectral information embedded in the Raman signal readout. To date, only one report has shown non-invasive focusing using highly sparse samples (therefore focusing convergence is guaranteed) [11], highlighting the need to develop tools for deep Raman imaging. Nevertheless, recently, speckles have been used for enabling super-resolution in Raman-based processes [12], opening another research venue for exploiting computational tools in deep imaging.

## Concluding Remarks

To conclude, linear feedback mechanisms have originally been seen as no-go for single-grain focusing in wavefront shaping experiments. Nevertheless, they have been recently been re-analysed, with the advances of computational tools, showing that there is still lots to be exploited from these well-established and popular microscopy contrasts, therefore opening important perspectives for deep imaging. Beyond incoherent contrasts discussed here, coherent mechanisms (in particular Raman) would also open important new applications. The computational tools developed so far have shown to be very useful, but modern machine learning tools, in particular physics-informed neural networks (see section 11 and 16) may also prove extremely powerful and versatile.

## Acknowledgements

This research has been funded by the FET-Open (Dynamic-863203) and European Research Council ERC Consolidator (SMARTIES-724473) grants.

# Section 8 - Photoacoustic-guided optical wavefront shaping

Emmanuel Bossy, Univ. Grenoble Alpes, CNRS, LIPhy, 38000 Grenoble, France
Thomas Chaigne, Aix Marseille Univ., CNRS, Centrale Marseille, Institut Fresnel, Marseille, France

## Status

In the ongoing effort to develop optical imaging techniques that can reach large depths into scattering biological tissue, photoacoustic imaging stands out by its unique capabilities. Relying on the emission of an ultrasonic wave upon the absorption of a pulsed illumination, this modality can be used in a variety of optical excitation and acoustic detection schemes. Depending on the acoustic frequency content of the detected signal, the geometry of the detector and the optical illumination system, various regimes can be explored, with a typical depth to resolution ratio of about 200.

The imaging resolution depends on the depth and scattering regime, and one can distinguish two main categories: below a few scattering mean free paths, it is possible to focus the optical illumination beam, and thus to achieve optical resolution. Beyond this limit, the illumination light is multiply scattered and diffuse, and, without scattering compensation, the resolution is set by the acoustic detection bandwidth, higher frequencies providing a better resolution. The depth-resolution trade-off is then ultimately set by the increasing attenuation of ultrasound with frequency (typically ~1dB/cm/MHz for biological tissue) and the maximum optical energy that biological tissue can receive without damage. Typically, photoacoustic imaging of biological tissue can provides maps of light absorption with a resolution of about 100 µm at depth around 2 cm.

Because the amount of light absorption depends on both the light intensity and the absorption coefficient, photoacoustics has been proposed about a decade ago as a guidestar mechanism for wavefront shaping, as it enables to probe the light intensity impinging on an optically absorbing structure lying deep inside soft tissue. Various schemes involving optimisation procedures or the measurement of a so-called photoacoustic transmission matrix have been proposed [1]–[10]. Any of these methods basically comes down to the following: when a diffuse coherent light field (a speckle pattern) impinges on an optically absorbing object imaged using a photoacoustic system, a modulation of the speckle pattern translates into a modulation of the photoacoustic signal. This photoacoustic signal can serve as a feedback signal to implement optical wavefront shaping methods, in particular to focus light through scattering media. The size of the optical focus is generally dictated by the ultrasound resolution [1], [11], but focusing down to the optical diffraction limit has also been demonstrated [6], [8], [10] (see Fig.1).

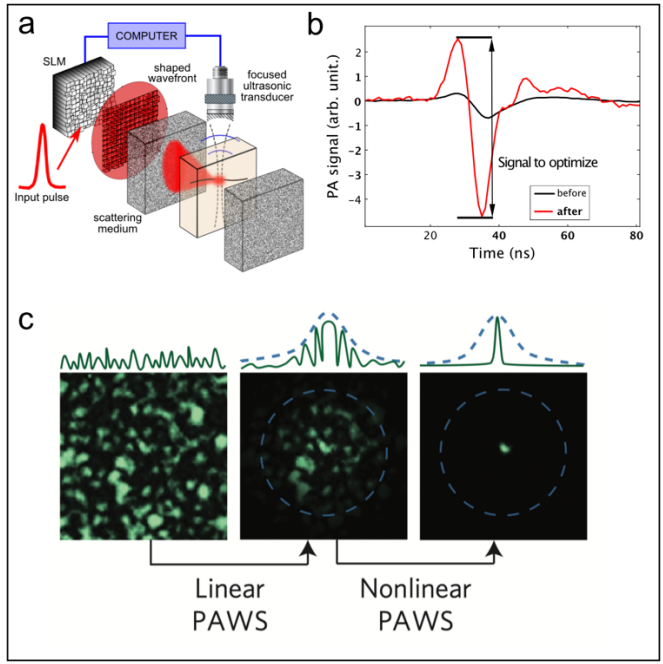

**Figure 1.** Illustration of photoacoustic-guided optical wavefront shaping through a scattering medium. (a) A spatial light modulator shapes the wavefront of a nanosecond pulsed laser source. The wavefront then propagates through a scattering medium and illuminates an absorbing object located within the focal region of a single element ultrasonic transducer. (b) Photoacoustic signal before and after optimisation of the peak-to-peak amplitude of the photoacoustic signal. (c) Speckle patterns (left) before the optimisation, (middle) after the optimisation of the photoacoustic signal amplitude, and (right) after optimisation of non-linear photoacoustic signal (based on temperature dependance of the Grüneisen factor). The dashed line represents the width of the focal region of the transducer. Sources: a,b adapted from [4]; c adapted from [6] with permission from NPG.

## Current and Future Challenges

To date, focusing through a scattering medium with photoacoustic feedback has been demonstrated only in a quite biologically irrelevant experimental configuration: all past experiments required speckle illumination with large enough speckle grains, obtained via free space propagation between a scattering medium and the absorbing target (see Fig.2a). This configuration provides an enhanced relative modulation of the photoacoustic feedback signal, as the feedback region contains only a limited number of speckle grains. The larger the number of speckle grains inside the photoacoustic feedback region, the lower is the relative modulation. Currently, a modulation of photoacoustic signals from optical absorbers embedded inside a multiply scattering medium, i.e. illuminated with diffraction-limited optical speckle patterns, have never been measured, and remains a hurdle towards photoacoustic-guided wavefront shaping at depth. This is due to the extremely weak amplitude modulation as compared to the mean photoacoustic signal, inherently caused by the mismatch between the ultrasound resolution and the optical resolution, as illustrated in Fig.2d-e. In other words, the current fundamental limitation to perform photoacoustic-guided wavefront shaping at depth in tissue is fundamentally a limitation from signal-to-noise and dynamic range considerations. Increasing the ultrasound detection bandwidth to increase the ultrasound resolution may increase the relative amplitude modulation on the one hand, but the subsequent increase of the ultrasound attenuation will decrease the SNR on the other hand.

Very importantly, *focusing* through or inside a complex media via wavefront shaping does not mean *imaging*, as in this context imaging usually requires the ability to scan a focused spot. Currently, there has never been any realistic demonstration of imaging based on photoacoustic-guided optical focusing: imaging was only demonstrated by scanning an object behind a *fixed* diffuser and *fixed* focal spot (see [8] for instance), whereas imaging a complex medium requires scanning a focal spot through scattering media, at best fixed, but in most practical scenarios dynamically varying within milliseconds. The optical memory effect through scattering media provides a mean to scan a focal spot through or inside a scattering media (see section by Pr. Ori Katz), but it is limited to very shallow depths, while photoacoustic imaging aims at depths beyond several millimeters.

We note that performing photoacoustic-guided wavefront shaping at depth *in vivo* also has the same limitations common to all optical wavefront shaping techniques (see section by Pr. Changheui Yang and Pr. Lihong Wang), i.e. the need to measure and control an enormous numbers of modes within times shorter than decorrelation times induced from various motion (flow, breathing, Brownian motion) in living tissue.

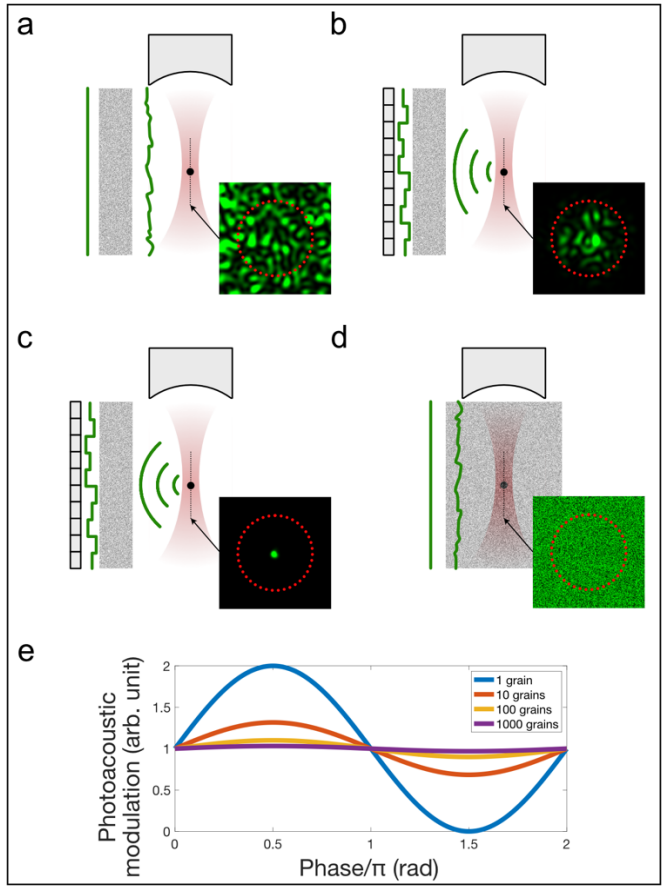

**Figure 2.** Green: optical wavefront; Red: sensitivity profile of the transducer. (a-c) Model configuration used in all past demonstration of photoacoustic-guided optical wavefront shaping: large speckle grains are obtained via free-space propagation *behind* a scattering medium, and allow focusing to the acoustic focus by linear signal optimization (b) or to the optical speckle grain (which is large compared to the optical wavelength), using nonlinear signal optimization. (d) Realistic configuration, illustrating the huge mismatch between the acoustic resolution and the size of optical speckle grains *inside* a scattering medium. (e) Modulation of the photoacoustic amplitude for various mismatches between the optical and acoustic wavelengths. These illustrate two main challenges of the approach: the number of mode required for (d) is extremely large, and the modulation of the photoacoustic signal is very low, as it varies as the square root of the number of speckle grains within the acoustically probed region.

## Advances in Science and Technology to Meet Challenges

As discussed above, advancing photoacoustic-assisted optical imaging requires both performing wavefront shaping extremely fast and detecting very weak photoacoustic signals and associated modulations. Because the average power density that a tissue can safely withstand is limited, going fast with high repetition-rate lasers turns into weak photoacoustic signals per pulse. At the end, SNR and dynamic range are always the fundamental limitations of photoacoustic sensing, whether to perform photoacoustic imaging or to guide optical wavefront shaping. Because the amount of light is limited by tissue safety limits rather than by laser sources, the only possible direction to meet the requirement of photoacoustic-assisted optical imaging seems to be the development of ultra sensitive and large-bandwidth acoustic detectors. Very recent technological breakthroughs in this domain, with the advent of transducers having sensitivities enhanced by orders of magnitude as compared to conventional piezo-electric sensing or optical sensing, are very promising and could address the challenges of photoacoustic-assisted optical wavefront shaping and imaging [12].

While we have discussed the use of the photoacoustic effect as a feedback mechanism for wavefront shaping, wavefront shaping has also been proposed as a way to improve photoacoustic imaging [13]. Controlling light propagation through tissue could help minimizing the intensity attenuation caused by multiple scattering and improve the SNR of photoacoustic detection while maintaining the same total power delivered on the tissue, therefore increasing the resolution (by extending the exploitable acoustic bandwidth) and/or the imaging depth of photoacoustic imaging. We also note that optical wavefront shaping is also exploited for photoacoustic imaging in the context of minimally invasive photoacoustic endomicroscopy through multi-mode fibers [14].

All these applications require nanosecond laser sources with coherence lengths of several centimeters, with high repetition rates (> 1 kHz), tunable wavelength, short pulse duration (~ 1ns), and reasonable power (~1W). Sources meeting these requirements in full are yet to be developed.

**Concluding Remarks**

Several proof-of-concepts experiments have showcased how photoacoustics and optical wavefront shaping can be exploited synergistically, either to assist optical wavefront shaping or to improve photoacoustic imaging. However, the relevance of the proposed approaches for practical applications remains to be demonstrated. Novel ultrasensitive ultrasound detectors could hopefully push this field a significant step forward in a close future.

**Acknowledgements**

This research has been funded by the European Research Council (ERC-COHERENCE-681514).

## Section 9 – Deep optical imaging based on a reflection matrix

Wonshik Choi,
Institute for Basic Science
Department of Physics, Korea University

**Status**

Optical imaging is an action of probing a sample with a light wave and finding object information from the wave momentum change induced by a target object of interest. One of the most widely used imaging configurations in biology and medicine is a confocal detection scheme, which is to scan a focused beam and collect backscattered wave at a point conjugate to the illumination spot. This detection scheme, which is in fact equivalent to tracking the momentum change, works reasonably well up to a shallow depth where the returning wave forms a sharp focus. However, the increase in imaging depth gives rise to a sample-induced aberrations distorting the point-spread-function (PSF) and multiple scattering noise. Distinguishing the distorted PSF from multiple scattering noise is an ill-posed problem in confocal imaging due to the insufficient data acquisition.

Reflection matrix approach has come to the rescue [1]–[4]. Unlike confocal detection, it records electric-field maps of the backscattered waves arriving at non-confocal points as well as those at the confocal points. In other words, both the obscured PSFs and multiple scattering noise are recorded in full (Fig. 1). Essentially, the reflection matrix constructed by a set of these electric field maps deterministically characterizes the input–output response of the sample to the best possible degree. A unique algorithm termed closed-loop accumulation of single scattering (CLASS) was developed to process the reflection matrix for separating the distorted PSF from multiple scattering noise without need for guide stars. Furthermore, it enables us to find out the one-way wavefront aberration from the roundtrip aberrations where input and output distortions are convolved [5]. With the addition of time-gated detection for rejecting a majority of multiple scattering noise, the depth limit to which the obscured PSFs can be corrected has been pushed close to 10 times the scattering mean free path in tissue. The degree of wave distortion that the reflection matrix approach can deal with is much higher than the conventional adaptive optics (AO) approaches relying on limited resources. For instance, ideal diffraction-limit could be recovered in imaging underneath an intact mouse skull presenting an extreme form of aberrations [6]. All these exceptional capabilities come from the recording the of full input-output response, which in turn slows down the image acquisition and processing. While this limits its applicability in a short term, proper addressing of these issues is likely to make reflection matrix approach an essential tool for life science and medicine.

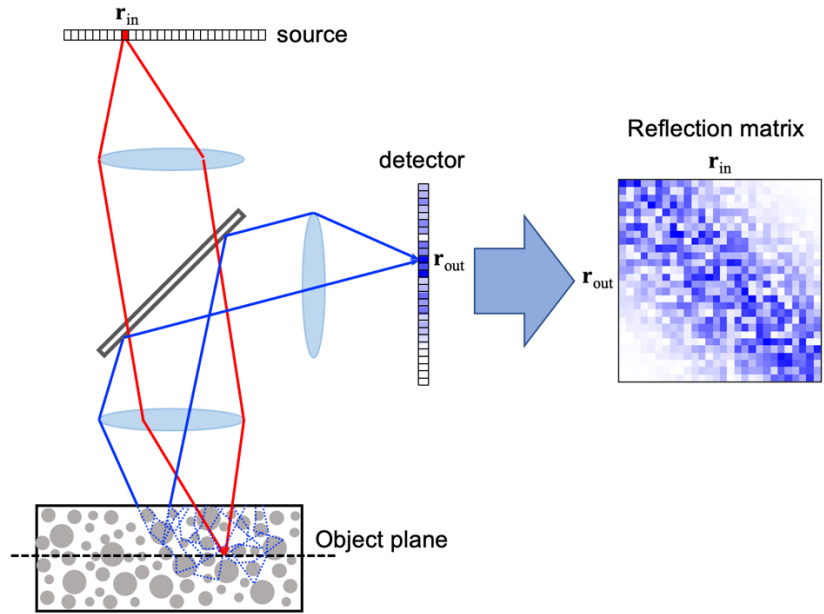

**Figure 1.** Schematic of recording a reflection matrix. Light wave emanating from each source channel $r_{in}$ (red) illuminates a sample, and the reflected waves (blue) arriving at various detector channels $r_{out}$ are recorded in their phases and amplitudes. Each pair constitutes matrix element of a reflection matrix.

## Current and Future Challenges

**Data acquisition and image processing speed:** Recording of a reflection matrix is intrinsically slower than confocal detection as it requires wide-field interferometric image for various illumination modes as opposed to single-point detection. It used to take a few minutes to record a single matrix at the early studies [2], [5], but ingenuous experimental configurations employing scanning mirrors made it possible to reduce the matrix recording time well below 1 second [6], [7]. This allows in vivo imaging of nervous systems in small animals such as zebrafishes and through-skull imaging of a living mouse. However, the covered field of view is still too narrow for comprehensive biological studies. Strategies for optimal downsampling will be necessary depending on the types of samples. New approaches of handling the matrix, such as forming a time-reversal matrix [8], can be adopted to allow sparse sampling without sacrificing too much of performance. Another challenge arises from the processing time for a recorded reflection matrix containing tens of millions of elements. Rapid advances in graphics processing unit (GPU) technology in terms of processing speed and memory capacity along with the optimization of the algorithms will put the real-time reconstruction and visualization forward.

**Outreach to other imaging modalities:** In the context of adaptive optics, reflection matrix approach can be considered a software-based AO as it can computationally reconstruct aberration-free images from the recorded matrix itself. However, it can also serve as a wavefront sensing AO as the measured aberration map can be transferred to a wavefront shaping device to physically correct the wavefront distortion. In comparison with other wavefront sensing AOs, the key benefit is its capability to retrieve tissue aberration map without resorting to guide stars. In fact, image reconstruction algorithm enables us to use a target object itself, any types of targets including structures generating speckle-like reflections, as effective guide stars. By applying the aberration correction map to a spatial light modulator in the excitation beam path of multi-photon fluorescence microscopy, through-skull imaging of dendric spines was realized with near-diffraction-limited resolution (Fig. 2) [6]. This strategy can be extended to other fluorescence imaging modalities such as single-molecule localization microscopy, coherent Raman microscopy, STED microscopy and so on to extend their working depths. To expedite this outreach, shortening the data acquisition and processing time will be crucial again. Outreach to lensless fiber endoscopes is another interesting possibility. Phase retardations due to fiber bending and twisting require the calibration of fiber transmission matrix, precluding the realization of flexible endoscopes. Fiber transmission matrix could be retrieved from reflection matrix in the case of a fiber-bundle endoscope [9], which leads to calibration-free flexible endoscopic microscope. Further investigation may extend this approach to endoscopes using a single multimode optical fiber.

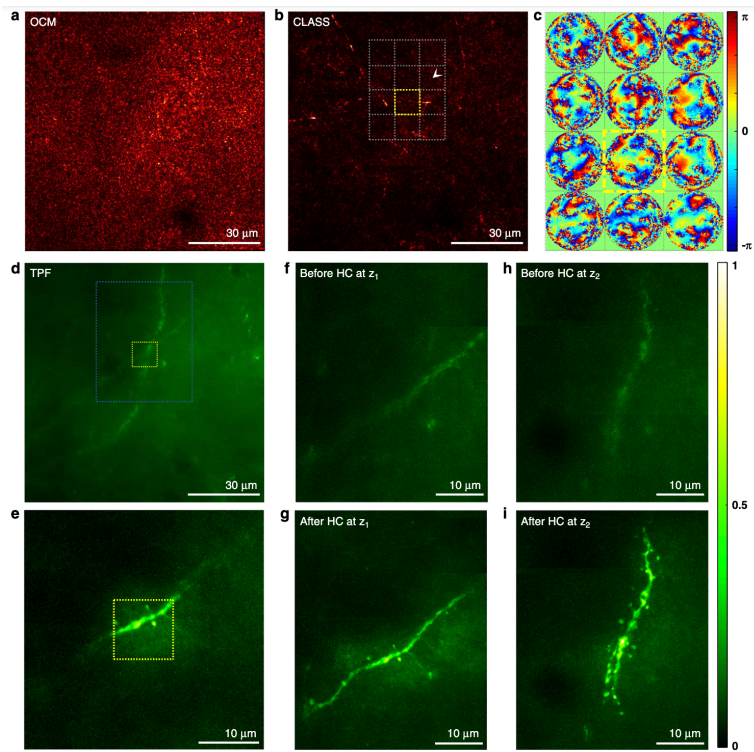

Figure 2. Exemplary application of reflection matrix microscopy for label-free imaging and two-photon fluorescence (TPF) imaging through intact skull. **a,** Conventional OCM image under an intact mouse skull before aberration correction. The thickness of the skull was about 85 μm, and the focal plane was set to a depth of 125 μm from the upper surface of the skull. **b,** Aberration-corrected image by the reflection matrix microscopy. 15 × 15 subregions were individually processed to find local aberrations. **c,** Aberration maps of subregions at 125 μm depth indicated by the gray dotted boxes in **b**. The size of the subregion is 10 × 10 μm², and each phase map contains 9880 angular modes. Color bar, phase in radians. **d,** TPF images at the same position as **a** before hardware aberration correction. **e,** TPF image after physical aberration correction for the subregion indicated by the yellow box in **b**. Yellow boxes in **d** and **e** correspond to the same yellow box area in **b**. **f** and **g,** TPF images at the depth of 113 ± 1.5 μm before and after aberration correction, respectively, for the area, indicated as a white dashed box in **d**. **h** and **i,** Same as **f** and **g**, respectively, for the depth of 122 ± 1.5 μm. Color bar, intensity normalized by the maximum intensity in **i**. Scale bars indicate 30 μm in **a**, **b** and **d**, and 10 μm in **e**–**i**. This figure was adopted from Ref. [6].

## Advances in Science and Technology to Meet Challenges

Can we increase imaging depth further? Immediate approach is to follow a recent trend and employ light sources with wavelengths of 1.3 mm and 1.7 mm. Although there are additional difficulties in implementing matrix recording system at these wavelengths in comparison with multi-photon imaging due to the requirement for wide-field interferometric imaging, they will surely be addressed in a short term.

Can there be more fundamental and yet practically useful advances? Achievable imaging depth will ultimately be determined by how much of multiple scattering to be used for image reconstruction [4]. Currently used CLASS algorithm identifies ballistic waves in the reflection matrix that do not alter their propagation directions in the scattering medium surrounding a target object and find their one-way phase retardations causing the distortion of shift-invariant PSFs. In this respect, it only exploits ballistic waves, not the multiply scattered waves. How can we extend this algorithm such that multiple scattering can be incorporated into the image reconstruction? One possible strategy is to extend its capability to deal with local aberrations. As shown in the through-skull imaging (Fig. 2), reflection matrix can be processed to find wavefront distortion in each subregion whose size is as small as $10 \times 10 \ \mu m^2$. This means that multiple scattering responsible for translationally variant PSFs at the length scale of $10 \ \mu m$ was identified and used for image reconstruction. Question remains how to extend this concept to the extremely short-range local aberrations. Another potential strategy is to solve high-order inverse scattering problem based on various types of forward scattering models incorporating multiple scattering. The problem is severely underdetermined, and it will be challenging to selectively train multiply scattered waves that interact with the object of interest while excluding the majority of those from bulk scattering medium. Novel computational tools including deep neural networks may help to expedite the progress.

## Concluding Remarks

Reflection matrix approach has made it possible to extract translationally invariant or slowly varying distorted PSFs in the epi-detection geometry in the presence of strong multiple scattering noise. Full characterisation of the input-output response of a sample along with a unique algorithm provides robust solution to this problem that used to be an ill-posed in a confocal detection scheme. Technical advances in matrix recording and processing speeds have realised in vivo biological studies and expedited its dissemination to other powerful imaging modalities. Extracting a short-range translationally variant PSFs hidden in a recorded reflection matrix of a thick scattering sample will be essential to further extend the imaging depth. This task is likely to demand new physical insights and the active use of computational resources.

Aside from the technical advances, it will also be important to find killer applications of the reflection matrix approach for its long-term establishment. Label-free detection is one of its major strengths, but it can only be meaningful in medicine where administration of exogenous labelling agents is not available. This means that long-term extensive collaborations with medical doctors needs to be formulated. In the meantime, the expansion of its approach as a wavefront sensing AO for other fluorescence imaging modalities can find its use in the near term for the immediate use in biological studies using tissue slices and animals.

**Acknowledgements**

*The author acknowledges Seokchan Yoon and Sungsam Kang for helpful discussion. This work is supported by the Institute for Basic Science (IBS-R023-D1).*

# Section 10 – Broadband Reflection Matrix: Deterministic and Learning-based Approaches to Deep Imaging

Alexandre Aubry and Sebastien Popoff
Institut Langevin, ESPCI, PSL University, CNRS, Paris, France

**Status**

In wave imaging, one aims at characterizing an unknown environment by actively probing it and then recording the waves reflected by the medium. It is, for example, the principle of optical coherence tomography for light. However, wave propagation from the sensors to the focal plane is often degraded by the heterogeneities of the medium itself. They can induce wave-front distortions (aberrations) and multiple scattering events that can strongly degrade the image resolution and contrast (Fig. 1). However, the emergence of high-resolution sensors array and recent advances in data science paves a way towards the breaking of these fundamental limits for optical deep imaging.

To that aim, a matrix formalism is the perfect tool to capture the input-output correlations of the light scattered by the medium. The Holy Grail for imaging is indeed to have access to the transmission matrix $\mathbf{T}$ that connects any point inside the medium to a sensor array outside (Fig. 1h). The experimental access to the $\mathbf{T}$-matrix has allowed experimentalists to take advantage of multiple scattering for optimal light focusing and communication across a diffusive layer or a multimode fiber [1]. However, a transmission configuration is not adapted to non-invasive and/or *in-vivo* imaging of biological tissues. The reflection matrix $\mathbf{R}$ that links the incoming and outgoing waves recorded by the sensor array should then be considered. To retrieve $\mathbf{T}$ from $\mathbf{R}$, both deterministic and learning based approaches have been proposed.

Inspired by adaptive optics, deterministic approaches take advantage of the correlations exhibited by scattered light over coherence volumes called isoplanatic patches [2,3]. Relying on the optical memory effect (see Chapter by Katz), such correlations can be exploited to compensate for low-order aberrations induced by forward scattering. Unfortunately, beyond one transport mean free path ($\ell_t \sim 1$ mm in biological tissues), waves lose the memory of their initial direction and start to follow a random walk that requires a precise knowledge of the microscopic properties of the medium to be harnessed. This is where a learning-based approach can be fruitful. Nevertheless, it requires a large training set and experimental results were so far only obtained for thin diffusers [4].

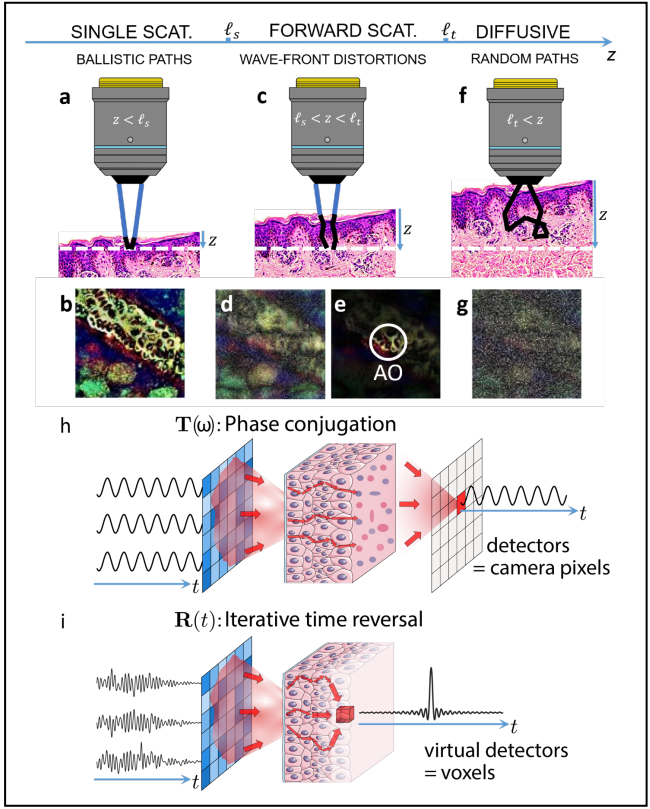

**Figure 1.** Deep imaging of biological tissues. (a) For a penetration depth $z < l_s$ (scattering mean free path), single scattering predominates and (b) optical microscopy provides a diffraction-limited image of the sample reflectivity. (c) For $l_s < z < l_t$, forward multiple scattering induces (d) a loss of resolution and contrast of the image. (e) Adaptive optics (AO) can compensate for these phase distortions but is only effective over an isoplanatic patch. (f) For $z > l_t$, the multiple scattering paths follow a random walk and (g) conventional imaging methods fail. (h) The transmission matrix approach can help to design the incident wave-fronts that make the multiple scattering paths interfere constructively at any point behind the scattering layer. (i) An iterative time reversal approach applied to a broadband reflection matrix shall give access, in principle, to the time-resolved transmission matrix between sensors outside and voxels inside the medium.

## Current and Future Challenges

Physical insights, used in deterministic approaches, and measured data, used as training sets in learning-based approaches, both provide information about the propagation of light in inhomogeneous media. However, they are mostly exploited separately in current imaging techniques, and both approaches have their limits.

Until now, matrix imaging has relied on singly-scattered and forward multiply-scattered photons, the number of which decaying exponentially with the penetration depth. Those photons exhibit a deterministic time of flight and are currently discriminated from the diffuse background by: (*i*) a time gating process; (*ii*) a spatial compensation of their phase distortions. Nevertheless, in the future, one will have to play with both spatial and temporal degrees of freedom in order to harness multiply-scattered waves. The measurement of a broadband **R**-matrix and a spatio-frequency analysis of its correlations should be coupled to learning based methods in order to retrieve a time-dependent **T**-matrix that will allow using the medium heterogeneities as a scattering lens and extend the penetration depth of matrix imaging beyond the transport mean free path (Fig. 1i). Such an approach can be rewarding since scattering can increase the effective numerical aperture of the imaging system and lead to super-resolution [5].

With regards to learning based approaches, models such as deep learning ones were developed initially for problems where the underlying model is not known, e.g. we do not know a mathematical formula that allows identifying a cat from a dog in a picture. This lack of information is replaced by a large number of parameters that has to be trained using large data sets. When the complexity of the task increases, so does the risk of not converging to a solution that generalizes well for unknown configurations (overfitting). It likely explains why deep learning approaches were so far not successful to predict the transmission properties of an inhomogeneous medium in the multiple scattering regime. Finding a way to efficiently combine physical insights and learning approaches may be the key to retrieve transmission information from the $\mathbf{R}$-matrix in multiple scattering media.

## Advances in Science and Technology to Meet Challenges

To address the multiple scattering limit in optics, the challenges to be met are both experimental and computational.

From an experimental point-of-view, the challenge is to record the $\mathbf{R}$-matrix over a broad bandwidth in an acquisition time ideally smaller than the decorrelation time of the medium. This can be a drastic barrier for living tissues which exhibit a decorrelation time ranging from 50 ms to 2.5 s depending on the level of immobilization [6]. Two strategies can be followed to circumvent that key issue: (i) reduce the number of input illuminations and use physical insights to retrieve the complete information thanks to compressed sensing or model-based approaches; (ii) develop a dynamic matrix approach of optical imaging by considering e.g the generalized Wigner Smith operator [7] in order to discriminate scattering paths as a function of their decorrelation time and address them independently. Interestingly, dynamic scattering can give access to a large number of speckle realizations for each voxel which can be used, in return, to extract the $\mathbf{T}$-matrix without relying on isoplanicity.

From a computational point-of-view, the training of numerical models designed to incorporate physical insights shows promising results, e.g demonstrating the compensation of aberration in optical measurements [8] or predicting the transmission properties of thin diffusers [9]. Exploiting those insights has two interesting consequences: (i) by drastically restricting the space of solutions, it limits overfitting and reduces the amount of information required for the training process; (ii) accessing the model parameters after training could allow predicting various properties of the physical system.

Using deep learning frameworks one can create physics inspired models for light propagation inside inhomogeneous media. The propagation equation, its invariants and/or its statistical properties can be incorporated into numerical models. For instance, for biological tissues, where the scattering is anisotropic, light propagation is well modeled by a series of diffraction events by thin diffusers with free space propagation in between [10] (Fig.2). One could then envision models that mimic these effects using layers with trainable parameters. Once trained on a measured **R**-matrix, it can be used to predict the associated **T**-matrix.

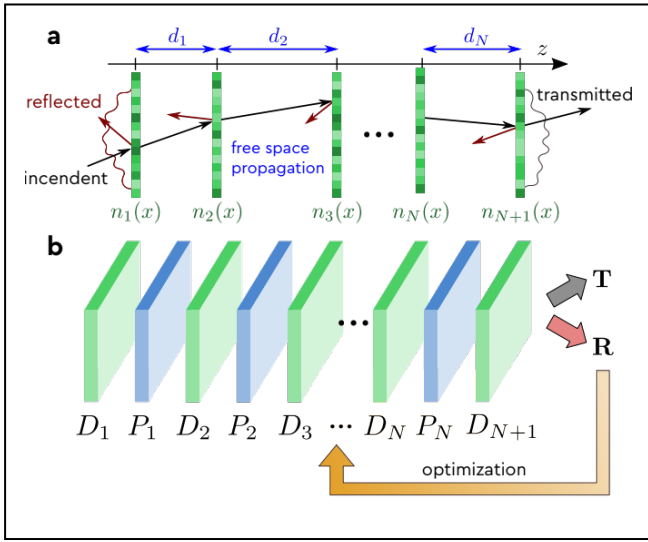

Figure 2. Proposition of a numerical model architecture (b) that mimics physical phenomena (a). The propagation of light in scattering media can be broken down into a series of events of scattering and propagation in free space. Deep learning frameworks can be used to build physics-aware models in which each layer simulate scattering by a thin diffuser ($D_i$) or free-space propagation in a slab ($P_i$). Trainable parameters, e.g. the index profiles of the thin diffusers $n_i(\mathbf{r})$ and the thickness of the slabs $d_i$, can then be optimized to match with the measured reflection matrix **R** using deep learning optimization tools. Once trained, the system could predict the transmission properties of the system, allowing imaging through or inside the medium.

**Concluding Remarks**

Predicting transmission properties from non-invasive and label-free reflection measurements is a subject of paramount importance with a wide range of applications. For biomedical imaging, the effort was originally oriented towards the ballistic and forward-scattered photons, which limits the penetration depth to one transport mean free path. To reach greater depths, it is necessary to harness multiply-scattered photons, whose trajectories cannot be *a priori* predicted. To that aim, a spatio-temporal control of light is required. On the experimental side, the acquisition of a broadband **R**-matrix can provide a post-processing solution to this challenging problem. On the computational side, the correlation properties of this matrix can feed a physical model whose numerous parameters can be adjusted through learning-based methods. The combination of deterministic and data-driven approaches constitutes a promising route towards the realization of an old dream researchers in optics are chasing for ages: Seeing in or through the fog.

**Acknowledgements**

A. A. acknowledges funding from the European Research Council under the European Union's Horizon 2020 Research and Innovation Program Grant n° 819261 (REMINISCENCE: REflection Matrix ImagiNg In wave SCiENCE). S. M. P. and A. A. acknowledge funding from the Labex WIFI (ANR-10-LABX-24, ANR-10-IDEX-0001-02 PSL*).

# Section 11 - Use of the linear scattering matrix for nonlinear coherent optical imaging at large depths in biological tissues

Sophie Brasselet, Matthias Hofer

Aix Marseille Univ, CNRS, Centrale Marseille, Institut Fresnel, F-13013 Marseille, France

**Status**

Scanning nonlinear microscopy is a widely used approach in biological imaging, with the advantage of benefiting from label-free optical contrasts that are specific to intrinsic tissue properties. Nonlinear optical imaging has first exploited two-photon excitation contrasts, through either incoherent (two-photon fluorescence) or coherent (second harmonic generation, SHG) processes, then reached higher order contrasts thanks to progresses made by pulsed laser sources. Today, three photon fluorescence, third harmonic generation (THG) and four wave mixing (FWM) processes based on the coherent mixing of three wavelengths are current imaging tools applied to neuro-imaging as well as in researches on pathologies such as in cancer or immunology. The requirements of nonlinear imaging are however stringent, since it necessitates the spatio-temporal coherent superposition of focussed near-infra red laser pulses, possibly at different frequencies, that are scanned over the sample volume with sub-micrometric resolution. This quality is rapidly lost when imaging at depth larger than a few hundreds of micrometers in biological tissues, e.g. a few scattering mean free paths. Preserving a high focus quality in space and time is at the center of strategies based on optimization and adaptive optics, which rely on the natural selection of coherent constructive optical paths by a nonlinear feedback such as two/three-photon fluorescence signals, or SHG. In parallel, the development of strategies to refocus incident beams based on the manipulation of the scattering matrix either in transmission (see section 6) or reflection (see section 8 and 9) has seen very fast technological and conceptual progresses. Learning the medium's scattering matrix by experimentally measuring the relation between incoming and outgoing fields is particularly interesting to master wavefronts' reversal processes in a deterministic way, but also to investigate how field degradations of input focused beams relate to the spatial, temporal and polarization correlation properties of the medium. Accessing such information is also crucial to determine a decipher strategies to correct not only for smooth aberrations but also for more complex propagation perturbations due to scattering. This section describes recent strategies based on linear photons manipulation to generate nonlinear signals through and inside biological tissues, as well as the challenges still to overcome. Here we describe the context of pure optical control, methods using other contrasts such as ultrasounds can be found in other sections.

**Current and Future Challenges**

The coherent manipulation of linear photons opens interesting routes for the excitation of nonlinear mechanisms, however the necessity to preserve short pulses (fs-ps range) synchronized in space and time for scanning nonlinear imaging brings additional challenges as compared to linear optics. The randomization of optical propagation in a scattering medium breaks the coherence propagation of pulses and distorts the pulse profiles into spatio-temporal speckles, also randomizing their spectral phase. Linear feedback control is nevertheless intrinsically limited by the correlation lengths of the medium along all dimensions (space, time/spectrum but also polarization). Elaborating on this property, a complete spatio-temporal control of a focus was recently demonstrated in the Gigan lab by the measurement and manipulation of a multi-spectral transmission matrix (TM) of a medium made of a stack of monochromatic, initially uncorrelated TMs, in a scheme that is also transposable in the time domain via time-gated TMs [1]. Focusing all components in the same output spatial position with accurate spectral phase/time control, enabled deterministic control of the output pulse time profile. More recently, it was found that by measuring this matrix with a broadband pulse, a transform-limited refocus naturally forms [2]. The self-reference interferometry used to measure the matrix creates an intrinsic coherent selection of low propagation paths photons in the medium, which has been shown to lead to strong nonlinear SHG signals after refocussing through a thick scattering biological medium as well as a natural polarization recovery [3]. Using a similar configuration in frequency mixing processes could open to the yet not much explored FWM manipulation in scattering media, for which a particularly interesting contrast is coherent anti-Stokes Raman scattering (CARS). This process necessitates two distant wavelengths from short pulses to be overlapped coherently, which frequency difference fits with a chemical vibration of the medium. Thanks to the manipulation of spectrally correlated TMs for which two ~150 fs pulses were manipulated at about 100 nm wavelengths distances, the recovering of a CARS signal was made possible through a thick scattering medium [4]. In this application, the knowledge of the spectral correlation properties as well as spectral-spatial coupling in the medium is crucial [5]. Yet, while these demonstrations make it clear that manipulating the linear incident photons leads to impressive recoveries of nonlinear polarized optical signal at large depths in transmission, it needs to be transposed to a reflection geometry to be pertinent for microscopy imaging in vivo.

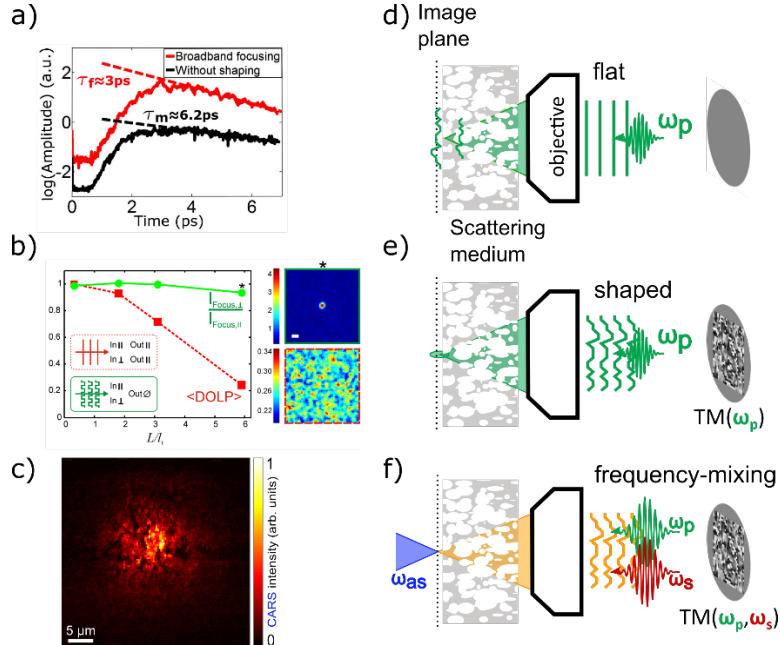

*Figure 1: (a) Broadband TM allows short path photons selection (adapted from [2]). (b) Polarization recovery using a broadband TM (adapted from [3]). c) CARS imaging by broadband TM manipulation (adapted from [4]). Wave propagation through a scattering medium, without shaping d), with broadband focusing e) and broadband shaping of spectrally correlated frequencies able to generate a CARS signal f).*

## Advances in Science and Technology to Meet Challenges

Strategies to transpose the control of complex waves in reflection, in the linear optical regime, hold great promises for nonlinear coherent optical imaging. A first set of strategies is based on an in-situ characterization of the scattered optical wavefront inside the medium by linear interferometry in presence of nonlinear signals generation inside the medium. Monitoring in-situ wavefronts has been achieved by focus scanning holographic aberration probing (F-SHARP) followed by wavefront conjugation [6]. More recently dynamic adaptive scattering compensation holography (DASH) has exploited similar principles with iterative wavefront front implementations [7]. Those approaches have allowed two-photon fluorescence imaging with high resolution over extended fields of views at a few hundreds of micrometers depth inside a mouse brain. A second set of strategies consists in measuring the reflection matrix measurement and refocusing light by phase front reversal. Constructing a distortion matrix to connect scanning focusses with reflected wavefronts, its singular value decomposition has allowed to correct for high-order aberrations and forward multiple scattering over isoplanatic patches at depth of ten scattering mean free paths in the turbid cornea [8]. Using a related approach, laser scanning reflection-matrix microscopy has allowed correcting for the matrix aberration components and generating reflectance and 2-photon images of axons underneath an intact mouse skull [9]. At last, an emerging strategy aims at modelling the medium and computationally correcting for propagation distortions to retrieve an image [10]. Capitalizing on the possibility to intrinsically monitor optical propagation properties in a complex sample, most of these strategies are able to correct for both smooth aberrations and scattering distortions and overcome the limiting memory range of the medium.

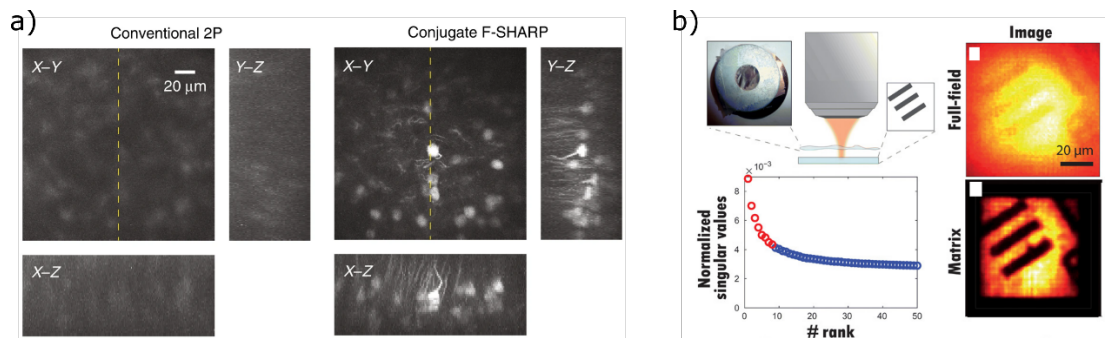

*Figure 2: Strategies of full characterization of a medium in the reflection mode. a) Comparison of conventional 2-photon fluorescence (TPF) through a thinned skull of a mouse with labelled neurons and dynamic F-SHARP based on TPF signals as feedback (adapted from [6]). b) Distortion matrix measured in reflection through interferometry of linear photons which allows for aberration correction of the imaged reflexion object (USAF target) (adapted from [8]).*

## Concluding Remarks

Challenges remain to accommodate the complexity of explored media and adapt nonlinear imaging to requirements in nanosciences, biology and biomedical optics. Interestingly, novel emerging schemes that exploit reflection-wavefront correction are developing fast and in parallel with novel nonlinear imaging tools which are compatible, such as spectral focussing or light sheet microscopy. Among the directions followed in the field of linear photons' manipulation, large progresses on both technological and computational tools will most probably allow future steps towards dynamic, high resolution and large field of view tissue imaging in depth.

## Acknowledgements

This work is supported by ANR-15-CE19-0018-01 (MyDeepCARS) and ANR-10-INBS-04-01 (France-BioImaging).

# Section 12 - Deep learning for imaging in complex media

Lei Tian, Department of Electrical and Computer Engineering, Boston University, Boston, MA 02215, USA

**Status**

Deep learning (DL) has shown tremendous success in solving ill-posed computational imaging problems. Interested readers can refer to a comprehensive review on this topic in [1]. Here we focus on DL techniques for imaging in complex media.

Recovering the object information from scattering measurements can be treated as an inverse problem. Significant progress has been made using the physical model-based techniques, such as those based on the memory effect and transmission matrix theory, which are detailed in other sections in this article. Due to the underlying assumptions in these models, the recovery is often limited by the field of view (FOV), the system's calibration requirement and stability. Instead of relying on a physical model, DL instead takes a data-driven approach, in which an implicit model is learned from a large training dataset. A main insight from recent works is that by intelligently engineering different training conditions, a deep neural network (DNN) based model can increase the FOV, relax the calibration requirement, and improve the system's stability. In the following, we highlight a few notable advances.

The first successful demonstration of using DL to overcome the FOV limitation imposed by the memory-effect (discussed in section 13) is shown in [2]. The DNN is trained to recover the object directly from a speckle pattern. By training the DNN using a diverse image dataset captured on a fixed diffuser, it can generalize over different objects and achieves diffraction-limited resolution across a FOV well beyond the isoplanatic region of the system. However, this network is susceptible to changes of the scattering medium. To overcome this limitation, a different training strategy is proposed in [3]. By incorporating variations induced by changes in scattering media itself in the training, the DNN is able to learn "hidden" correlation information from multiple realizations of a random medium. Specifically, the authors show that, by the training a DNN on multiple diffusers with the same macroscopic property, it can make high-quality predictions through different unseen diffusers (**Fig. 1**). Recently, unsupervised dimension reduction technique is used to provide additional insights on this learning process [4]. A similar strategy has been successfully applied to imaging through nonstatic thick turbid media [5].

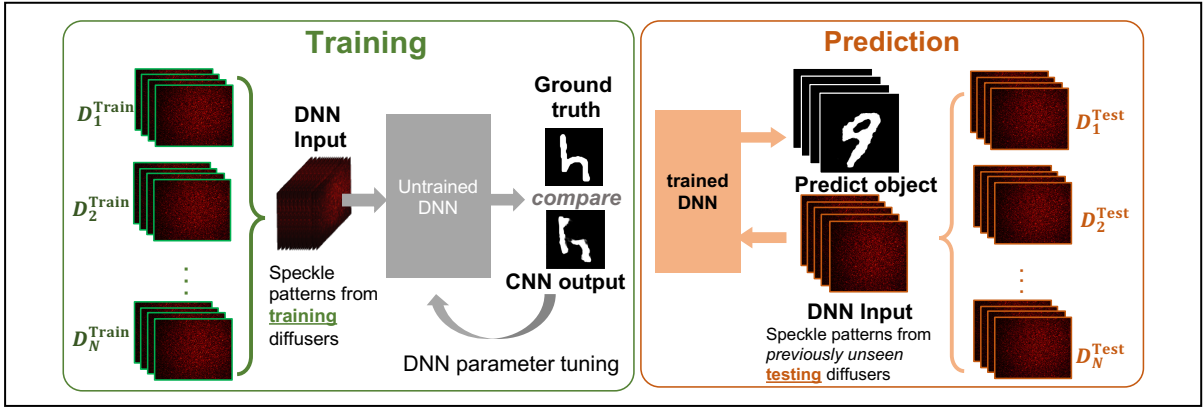

**Figure 1.** A deep learning framework to make the DNN robust to changes of the scattering media. Left: the DNN is trained by speckle patterns collected through multiple diffusers. Right: the DNN can generalize well to novel unseen diffusers.

12N

## Current and Future Challenges

Despite recent advances, there are multiple challenges remain to be solved to make the DL techniques applicable in broader imaging scenarios, as outlined below:

**1) Need for a large-scale and diverse paired training dataset.** The supervised learning framework used in the existing techniques dictates that paired input and output images are required to train the DNN. In general, both the scale and the diversity of the dataset required for training a generalizable DL model can be challenging in many practical applications, such as biomedical microscopy. We can further dissect this issue into several scenarios.

**1a) Volumetric imaging.** In all the existing work reviewed here, the training input images are generated by a spatial light modulator. This imposes several assumptions about the object, including a) it must be 2D planar, b) no other object sources are present before or after the 2D plane of interest. It will be challenging to directly apply the existing training strategies and DNN architectures to volumetric 3D imaging applications.

**1b) Imaging objects inside a scattering medium.** Another limitation comes from the requirement for the direct access to both the input and output planes during the training process. This becomes challenging when the objects are buried inside a continuously distributed scattering medium.

**1c) Imaging in reflection geometry.** A related challenge is when imaging objects inside a scattering medium, while the measurements can only be made outside the medium.

**2) Generalization to different scattering conditions.** Although advances have been made to make the DL model to be generalizable to scattering media having the macroscopic property, i.e. the same scattering condition, the model is still susceptible to changes to the scattering conditions, such as changes in the scattering density and scattering mean free path. Innovations in both training strategies and DNN architectures are needed to make the DL model robust over a wide range of scattering conditions.

**Advances in Science and Technology to Meet Challenges**

Here we outline several promising directions to pursue to overcome the above challenges and further push the fundamental limit of imaging in complex media using DL.

**1) Physics-informed learning.** The large training dataset requirement of the supervised learning techniques stem from its pure data-driven framework without using any knowledge of the physics. However, many physical insights and models are available and can be utilized to describe the scattering measurements. As a result, it is conceivable that physics-informed learning approaches, which synergistically combines physical models and DL, can overcome many of the existing challenges. For example, Metzler et al. showed that a DNN robust to low signal-to-noise ratio (SNR) in non-line of sight imaging can be trained by taking the autocorrelation of the speckle pattern as the input [6]. This data preprocessing step is directly drawn physical insights from the classical correlography theory.

**2) Multiple-scattering simulator-based training.** To alleviate the need to physically acquire experimental training data, another possibility is to use a physics simulator to generate the training data. To simulate a large-scale scattering measurement dataset, the simulator needs to be both accurate and computationally efficient. Additional benefit of this simulator-based training is that it is not limited to any specific imaging geometry. For example, Matlock et al. showed that a DNN that is generalizable to experimental measurements on 3D samples can be trained using a multiple-scattering simulator based on an accurate and efficient split-step non-paraxial beam-propagation model [7].

**3) Exploiting correlations between transmission and reflection measurements.** To address the challenge associated with the reflection imaging, another promising direction is to exploit the speckle correlations between the transmission and reflection measurements. For example, Skarsoulis et al. recently showed that a DNN can be trained to predict the speckle patterns through a scattering medium purely based on the pattern measured in the reflection [8].

**4) Adaptive learning framework.** To build a DL model that is robust over a wide range of scattering conditions, adaptive DNN architecture is a promising direction to investigate. For example, Tahir et al. recently proposed a new dynamic synthesis network architecture that can dynamically adjust the DNN's model weights and adapt to different scattering conditions [9].

**Concluding Remarks**

Recent advances in DL have shown potentials to push the fundamental limit for imaging in complex media. By combining new physical insights in scattering physics and novel learning framework, we expect novel imaging and sensing techniques will continue to emerge and make their ways to practical applications in many impactful areas, such as biomedical microscopy, metrology, and material science.

**Acknowledgements**

*The author acknowledges funding from National Science Foundation (1813848, 1846784).*

## Section 13 - Model-based wavefront shaping

Ivo M. Vellekoop, Department of Science and Technology, University of Twente, Enschede, The Netherlands

**Status**

Most approaches for imaging through scattering rely on the optical memory effect in one way or the other. One approach is to use wavefront shaping (WFS) with feedback from a guide star or a detector to form a focus (e. g. [1], c. f. Section 5), and then use the memory effect to raster-scan this focus for microscopy [2]–[4]. Sometimes, we can even use the memory effect by itself for computational imaging (see Section 13). Unfortunately, the optical memory effect only has a significant range for objects placed far behind a thin scattering layer [5], or inside forward scattering materials, such as biological tissue [6].

Without the optical memory effect, the possibilities for deep imaging are limited: if we can only focus onto the guide star itself, we will get a 'single-pixel image'. An interesting solution is to use ultrasound tagging to define a 'movable' guide star (Section 5). This solution does, however, have the drawback of severely decreasing the resolution and contrast of the focus.

*Model-based wavefront shaping (MB-WFS)* represents a completely different that neither requires guide stars nor the optical memory effect. The idea is simple: if we have an accurate refractive index model of the scattering structure, we can *compute* how to form a focus at any arbitrary point inside the sample.

Of course, one typically does not have an exact model of the structure. In many cases, however, *a priori* knowledge about the sample can be combined with additional measurements to construct the model. A rudimentary demonstration of this concept was given in [7], where the structure was a flat piece of glass with known thickness and refractive index. From this model the wavefront corrections for aberration-free focusing at any depth could easily be computed. A more advanced example is the construction of a model of a multimode fibre [8]. By fitting a 12-parameter model to calibration measurements, the fibre could be digitally modelled to compute wavefronts for focusing at arbitrary points through it.

The use of *MB-WFS* to see through highly scattering structures was first demonstrated in[9], see Figure 1. Here, a rough scattering surface was imaged from the outside, and these images were converted to a 3-dimensional refractive index model. *MB-WFS* resulted in high-quality wavefront corrections even at depths where guidestar-based techniques fail due to the low signal-to-noise ratio of the feedback signal.

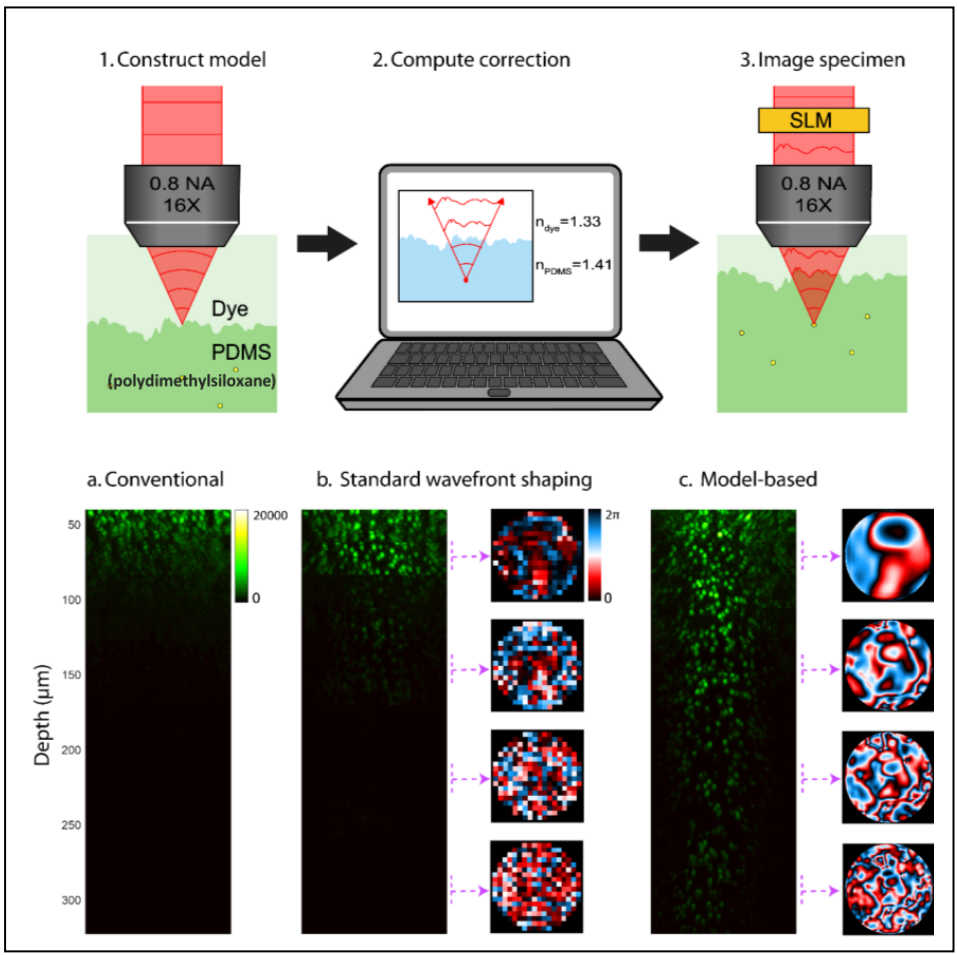

**Figure 1. Model-based wavefront shaping microscopy [9].** 1) In this proof-of-concept experiment, a 2-photon microscope was used to image the 3-D profile of a scattering interface between water (n=1.33) and polydimethylsiloxane (n=1.41). This height map was converted to a computer model, which was used to compute wavefront corrections for focusing through the layer (2). Finally, these corrections were applied on a spatial light modulator to enable deep imaging (3). Results: without WFS, imaging was possible till 75 μm behind the diffuse layer (a). With feedback-based WFS, a signal improvement was realized. However at depths exceeding 75 μm the feedback signal was too weak to be usable, causing noisy wavefronts (right column) (b). With model-based wavefront shaping, high quality wavefront corrections could be computed till a depth of over 300 μm, with a signal decrease only caused by volumetric scattering in the PDMS (c).=

## Current and Future Challenges

These promising new approaches introduce a class of challenges that are new to the field of wavefront shaping:

**1) Refractive index model reconstruction.** First of all, it is essential to have a sufficiently accurate model of the refractive index. A challenge here is that most tomographic techniques are developed in the weak scattering regime where the Rytov approximation, first-order Born approximation, or geometrical optics are applicable. When scattering becomes more dominant, unfortunately, these methods no longer produce correct results, so alternatives are needed. The alternative of directly mapping the surface (Figure 1) does work for micrometre-scale irregularities. However, the challenge here is to extend the concept to multi-layer or truly volumetric samples.

**2) Fast light propagation simulations.** A practical challenge is to make these computations fast enough for real-time imaging. Especially when the memory effect is very small or absent, a new wavefront needs to be computed for every point in the image. These computations currently take seconds to minutes, making them the bottleneck for model-based imaging.

**3) Coordinate mapping.** An experimental challenge is to achieve a true 1:1 mapping of the 'virtual' coordinates used in the computer model, and 'physical' coordinate spaces spanned by SLM pixels, camera pixels, galvo scan angles, etc.

**Advances in Science and Technology to Meet Challenges**
Rapid progress is being made to meet these challenges. Some of the highlights are:

**1) Refractive index reconstruction algorithms and iterative measurements.**
The need for refractive index reconstruction inside scattering materials calls for the development of reconstruction algorithms that are robust against multiple scattering. Promising developments in this direction are the development of optical coherence refraction tomography [10], neural-network based methods [11] and reconstruction algorithms specifically designed for strong scattering [12].

**2) Light propagation algorithms and parallel computation.**
Recently developed solvers can solve Maxwell's equations in 3-D media of $\sim 10^4$ cubic wavelengths in a matter of seconds on a single GPU [13]. Still, when the scattering structure is millimetre-sized ($10^{10}$ cubic wavelengths), faster, less accurate, algorithms are more appropriate, such as angular spectrum methods [9], [11], [12] or ray tracing [10].

**3) Automated alignment and calibration**
These coordinate mapping challenges are very similar to the challenges encountered in digital optical phase conjugation (DOPC). An interesting development in this field is the development of fully automated calibration protocols, that are even capable of digitally correcting alignment imperfections [14].

**Concluding Remarks**
After feedback-based wavefront shaping and phase conjugation, *MB-WFS* provides a third route to focusing light inside scattering materials. This new approach does not rely on guide stars or on the optical memory effect. Rather, it uses a variety of algorithms for refractive index mapping, light propagation computations, and automated alignment, combined with *a priori* knowledge about the structure. Interestingly, this new route shifts the problem of deep imaging from optics to a multi-disciplinary research field incorporating inverse problem methods, numerical mathematics, cloud computing, and artificial intelligence.

**Acknowledgements**

*The author acknowledges funding from the Dutch Research Council (14879) and the European Research Council (678919).*

## Section 14 - Memory-effect based imaging in complex media

Ori Katz

Department of Applied Physics, Hebrew University of Jerusalem, Jerusalem, Israel

### Status

Scattering induced distortions in complex media are one of the major hurdles for imaging in many applications in optics, from astronomical observations through the turbulent atmosphere, through imaging in foggy conditions, to deep-tissue imaging. Interestingly, this challenge appears in many important domains beyond optics, from acoustic imaging to geophysics. While wavefront-shaping allows focusing to a diffraction-limited bright spot even through multiply scattering complex samples, *imaging* requires the ability to focus to *multiple* points in the desired field-of-view (FoV). In general, one has thus to find and apply a large number of different wavefront corrections that correspond to all points in the FoV, making *imaging* in complex media a much more difficult challenge than single-point *focusing*.

However, what may alleviate the problem at many instances, is if a single wavefront-correction can be effective over more than a single point in the FoV. Such 'shift-invariance' of the wavefront correction is termed "isoplanatism" in the fields of adaptive-optics and acoustics, and it is in essence the shift- (or more precisely tilt-) invariance of the scattering point-spread-function (PSF), or Green function. Surprisingly enough, even thick, multiply-scattering diffusive samples can scatter light in an isoplanatic fashion [1], meaning that waves that illuminate the medium at slightly different angles, scatter to nearly-identical speckle patterns that propagate at corresponding relative angles (Fig.1a). This effect, termed the angular optical 'memory-effect' was first discovered and measured in the 80s [1]. Strikingly, the memory-effect exists even at depths well beyond the transport mean free path, $l_t$, where the propagation direction has been totally scrambled by multiple scattering.

A direct implication of the 'memory-effect' isoplanatism is that a single wavefront-correction can be used to scan a wavefront-shaped focus within the memory-effect angular range (also termed the 'isoplanatic patch') to produce an image [2] (Fig.2a). Thanks to Helmholtz reciprocity, a widefield single-shot variant of this approach can be performed [3] (Fig.2b), since light from adjacent point-sources is scattered to highly correlated speckle patterns [1], they can all be corrected *simultaneously* by the same static wavefront corection [3] (Fig.2b). These approaches for imaging based on a single-point wavefront-correction [2,3] determine the required wavefront correction invasively, i.e. by accessing a point at the target plane for measuring the wavefront-correction. However, this limitation can be overcome by noninvasively optimizing a nonlinear signal [4], or an image sharpness metric [5] (Fig.2b). Interestingly, the angular memory-effect is present also in back-scattering from complex samples such as white painted walls, and in light propagation through multi-core fibers (Fig.1c), opening the path to looking 'around corners' [1,3], and to lensless diffraction-limited endoscopy [6] (see section 18).

As was first suggested by Freund three decades ago [1], 'memory-effect' based imaging can be possible even without wavefront-shaping correction. The first realization in multiply-scattering media was demonstrated by scanning unknown (but correlated) speckle patterns over a fluorescence target (Fig.1a), and computationally reconstructing the image via phase-retrieval [7].

A widefield single-shot approach, inspired by Labeyrie's stellar speckle interferometry [9] was then put forward [8], where the diffraction-limited image is reconstructed from a single high pixel-count image of the scattered light [8] (Fig.2c), bringing Labeyrie's approach from astronomy to complex media.

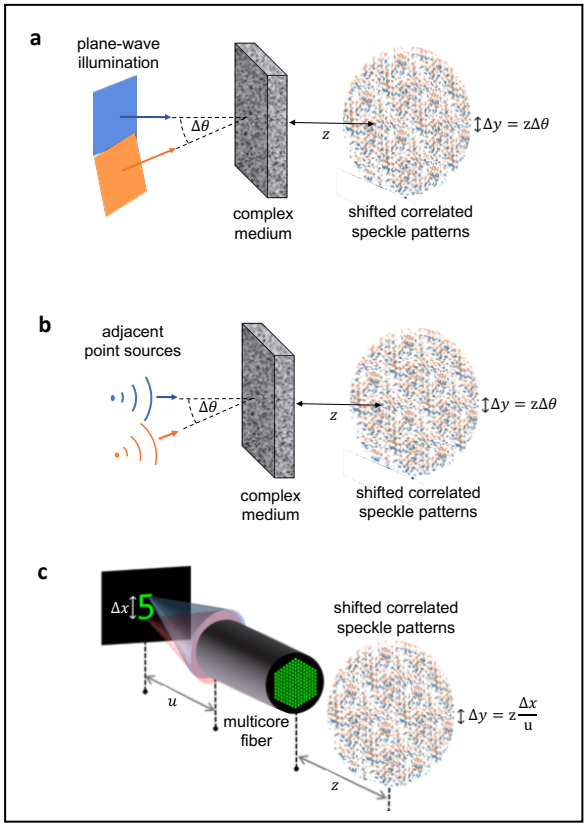

**Figure 1.** The angular 'memory-effect' is the isoplanatic scattering of light by complex samples. It is manifested as: **(a)** scattering of plane waves that illuminate the medium at different angles, to correlated speckle patterns that propagate at corresponding relative angles; **(b)** scattering of light from nearby point sources to correlated speckle patterns in complex media (b) and through multicore fibers **(c)**. A similar effect is present also in back-scattering from complex samples.

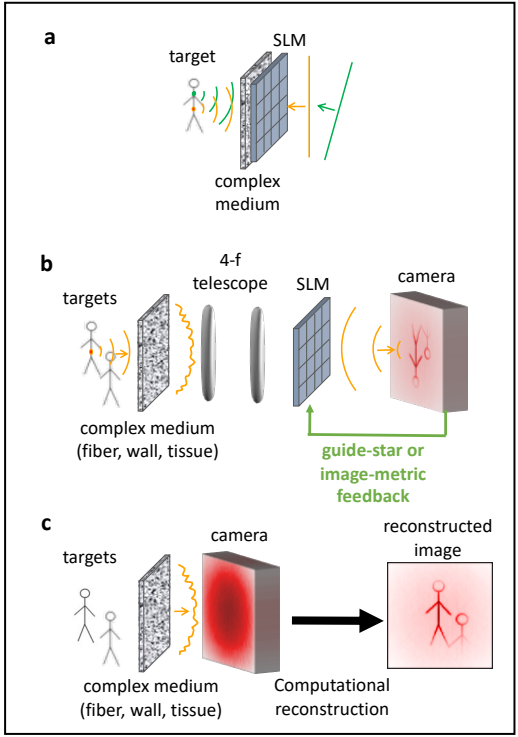

**Figure 2.** Using the angular 'memory-effect' for imaging is possible by: **(a)** scanning a wavefront-shaped focus on the target; **(b)** using a single wavefront-correction to correct scattering from all points within the memory-effect field-of-view; **(c)** computationally reconstruct the target from correlations in a single high-resolution scattered-light image, without wavefront-shaping.

## Current and Future Challenges

The major limitations of using the memory-effect for imaging are its small FoV and limited imaging depth. While the angular memory-effect is a universal property of multiple-scattering samples, its angular range is inversely proportional to the sample thickness, $L$. In the diffusive regime, i.e. at $L \gg l_t$, the memory effect angular range is $\Delta\theta_{FoV} \approx \frac{\lambda}{\pi L}$. The imaging FoV. which is calculated by multiplying this angle by the imaging depth, is thus unfortunately of the order of a wavelength - too small for most applications. Nonetheless, the FoV can be larger when imaging at a large standoff distance through a thin scattering layer (e.g. in an 'eggshell' geometry), or when imaging 'around-corners' (where the angular memory-effect range in reflection is $\Delta\theta_{FoV} \approx \frac{\lambda}{\pi l_t}$). The imaging FoV can also be significantly larger than a wavelength when imaging through biological tissues at depths smaller than $l_t$ [10]. In addition, the large anisotropy of the scattering in soft tissues give rise to speckle correlations also for transverse translations of the incident wavefront [10], which may also be exploited for imaging.

The isoplanatic patch size limitation can be overcome by mosaicking multiple isoplanatic patches into a single large-FoV image. This can be done by separating the different isoplanatic patches by decomposition of the medium's reflection matrix (see section 8), or of the matrix containing scattered fluorescence patterns (see section 6). Another approach to enlarge the FoV is to physically limit the sample probing using a probe that is smaler than the isoplanatic patch. This was recently realized via localized acousto-optic tagging followed by a ptychographic reconstruction [11].

The advanced computational reconstructions of the matrix-based approaches or ptychographic techniques do not only allow wider FoV imaging, but also address the convergence instability of the iterative phase-retrieval algorithms used in the first memory-effect works [6-8]. Another possibility to overcome the need for phase-retrieval is through the calculation of the scattered light bispectrum, as first developed for astronomical observations [12].

Another fundamental challenge is in three-dimensional imaging, which is addressed to some extent by time-gating in either reflection-matrix measurements or incoherent compounding [13]. Time-gating may also give rise to increase FoV [14].

Finally, a major challenge is in performing the large number of required measurements in an acquisition time that is shorter than the sample decorrelation time. One potentially interesting way around this is to exploit the sample dynamics to retrieve additional information vfrom multiple speckle realizations [11], rather than to try to overcome it.

## Advances in Science and Technology to Meet Challenges

The depth, FoV, and speed limitations of memory-effect based imaging require significant advances in both technology and scientific approaches. A straightforward step to extend the FoV and imaging depth is to utilize longer infrared wavelengths for imaging, possibly utilizing high-resolution InGaAs cameras for detection. Another step forward would be in boosting the acquisition speed by using high-speed cameras, fast spatial light modulators, and parallelized acquisition schemes, potentially exploiting spectral information.

The combination of memory-effect based imaging with acoustic tagging carries interesting potential for combining the benefits of both modalities: the high resolution of light and the large penetration depth of ultrasound. However, the current implementations [e.g. 11] are too slow for most applications. Faster acquisition and parallelization or some innovative approaches are thus critically required.

Deep-learning based approaches carry a huge potential, as they can in principle address all current challenges (namely the number of measurements, parallelization, reconstruction stability, FoV, and sample dynamics). Deep-learning has been recently shown to generalize correlations-based reconstruction beyond angular or translational correlations, and to include model-based physical insights (see chapter by L.Tian). As in all deep-learning works, the limited interpretability ('black-box'-iness), and the limited generalizability may present drawbacks, but their full promising potential is yet to be realized.

Finally, a fundamental limitation for deep imaging is the limited photon-budget: the deeper one tries to image the more speckle grains (modes) each photon is scattered to. This necessitates higher resolution wavefront shaping, highly sensitive detectors, and more noise-robust computational approaches.

## Concluding Remarks

The fact that multiply scattered diffuse light has inherent correlations is at first glance surprising, and indeed give rise to some counterintuitive results, such as imaging around corners. However, the intuition for the existence of angular correlations emerges from the fact that illuminating a point on a sample facet results in a bright halo only around the illumination point. The angular correlations are simply the manifestation of this effect in the Fourier domain [1]. Thus, one should not be surprised that some imaging information is contained in scattered light. How to distil this information in the fastest and most efficient manner is left to be found. Combining spatial and temporal gating, matrix based decomposition, generalized corrleations and deep-learning reconstruction are a potentially interesting path towards this goal.

**Acknowledgements**

We acknowledge funding from the European Research Council under the European Union's Horizon 2020 Research and Innovation Program Grant n° 677909, and the Israel Science Foundation (1361/18).

# Section 15 - Information from correlations

Jacopo Bertolotti, University of Exeter

## Status

Multiple scattering of light in biological tissues, clouds, etc is so complex and dependent from so many tiny details, that it is tempting to pretend that the whole process is random, and only deal with the relative simplicity of the diffusion approximation. The price one pays for that simplicity is big, as true randomness means that information is forever lost, and therefore there is only so much we can do if we want to image an object through a scattering layer. Despite its complexity, multiple scattering is still a perfectly deterministic process, and for the typical powers involved in imaging, it is also often a completely linear one. As a result, all the information contained in the signal before it was scattered, must still be present after it has been scattered. In this sense, multiple scattering effectively performs a rotation in a very highly dimensional space [1]. If we knew this rotation (the scattering matrix) we could invert it [2], but otherwise the information is now spread out and only visible in the form of correlations between the intensity at different points.

Speckle correlations come in many forms, each carrying a bit of the desired information. The most commonly used speckle correlation in imaging is the optical memory effect, i.e. the fact that, by tilting the incident beam, the transmitted speckle will tilt by the same angle, as long as the angle is not too large (isoplanatism) [3]. This correlation is a useful tool to image through a scattering layer, because it gives you information about what is happening on the hidden side of the layer using only information measured on the accessible one, thus allowing non-invasive imaging [4] (see section 13).

The success of methods based on the memory effect suggests that, among the many possible correlations that we can find in speckle patterns, the easiest to exploit are those that contain mutual information between the region where the hidden object is, and the region we can freely measure.

## Current and Future Challenges

The main limitations to the use of correlations for imaging are that they are often weak, as most correlations decrease with the "dimensionless conductance" $g$, which is very large for most scattering media [5]. Furthermore, most correlations are probabilistic in nature, requiring some form of averaging over disorder to extract an image. In this respect the optical memory effect is an outlier, being a perfect ($C$=1) and deterministic correlation, as long as the angles involved are small enough.
Another complication is that there are many ways that mutual information between one part of the scattered field and another can manifest. 2-points correlations in a single speckle pattern have an elegant classification in terms of how the correlation decreases with the distance between the two points, but other form of correlation, e.g. 3-points correlations (bispectrum) [6], are difficult to classify using that framework. This is, of course, both a challenge and an opportunity, as it is possible to find useful mutual information in many places. For instance, linear scattering preserves spatial coherence, so one can estimate the extension of a source hidden in a turbid medium by looking at the signal spatial coherence [7] (see section 6).

The presence of yet undiscovered or unexploited speckle correlation is also why machine learning-based technique can work on systems they were never trained on. Although we are not aware of them, a properly trained neural network can find and exploit correlations that are not sample-specific, thus enabling it to recover the desired image even when the scattering system is different from the one(s) used for the training [8] (See section 11). The challenge here is that, while the success of machine learning tells us that the correlations are there, there is no obvious way to find out which correlation are being exploited, and thus use them for other imaging techniques.

**Advances in Science and Technology to Meet Challenges**

Starting with the work of Feng *et al.* in 1988 [9], where the first speckle correlations were described and classified, there have been a lot of interest in the properties of speckle, and in recent years a number of new correlations has been characterized. From the "tilt-tilt" memory effect [10], to the "chromato-axial" memory effect [11], to the "transmission-reflection" correlations [12], and other. All these new correlations allow to push the boundaries of what can be done for imaging through scattering media, but there have been no coordinated effort to make the search for new correlation more systematic yet.

Measuring speckle correlations often entails measuring small signals over large backgrounds, and/or having to average over a large ensemble, which requires fast detectors with large dynamic ranges. At the same time, fast measurements almost invariably means less light per time bin, which exacerbate the dynamic range problem, as the signal one wants to measure can be on the limits of sensitivity. This conundrum has no easy solution, and improvements will require a performance increases on all fronts for cameras

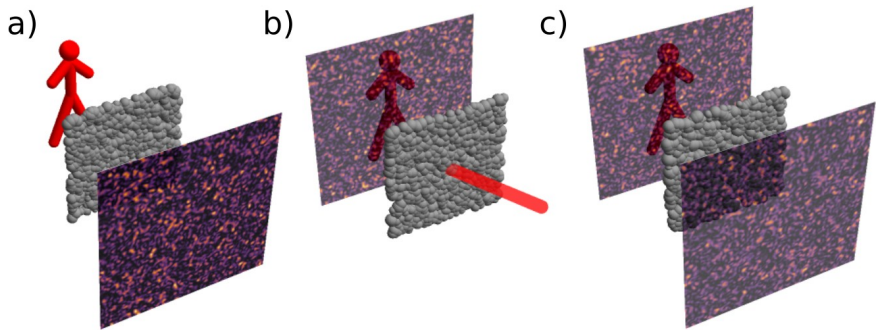

Figure 1 Different kind of correlations give us different kind of information. (a) When the light coming from an object is scrambled by a scattering layer, the resulting speckle will depend on the properties of the object, e.g. its spatial coherence. (b) If we can control the input light, the pattern illuminating the object will depend on the properties of the scattering medium, e.g. the optical memory effect range will depend mostly on the layer thickness. (c) The scattered light that goes back in our direction is correlated with the scattered light illuminating the hidden object, so we can use that to retrieve an image.

The number of modes over which the information is spread out is geometry- dependent, but in most cases it grows approximately quadratically with the thickness-to-wavelength ratio, which means that correlations tend to become weaker, and thus harder to measure and exploit, for thick scattering media. Currently, this is one of the main roadblocks to the real-world application of any correlation-based imaging technique (see section 13). Using longer wavelengths ameliorate the problem, and so does using media that are mostly forward scattering [13], but this the most important problem to solve in the near-medium future. A possible way forward is to use several correlations at the same time, and combine the small amount of information that can be gathered by each to obtain a more complete picture.

**Concluding Remarks**

Linear multiple scattering does not change the information content of a wavefront, but scramble it. In principle this scrambling can be reversed, but this requires a complete characterization of the system. Whenever this is not possible, we need to extract as much information as possible from the apparently random speckle patterns. This is possible because speckle patterns are not really random, and the information we seek is now encoded in their correlations. The large variety of correlations present is both a challenge and an opportunity, and with only a handful of correlations that have been studied and understood, this field is still in its infancy.

**Acknowledgements**

We acknowledge funding from EPSRC (UK, Grants EP/S026630/1 and EP/T00097X/1).

# Section 16- Computational imaging with randomness

Ryoichi Horisaki, The University of Tokyo

## Status

Recent advances in information science, such as compressive sensing and machine learning, have contributed to various fields, imaging being a typical example. In particular, computational imaging, which is a powerful framework for developing innovative photography and display systems by combining optical and computational processes, is an attractive field in which to apply state-of-the-art tools in information science to optics [1]. Simplification of optical hardware, enhancement of imaging performance, nonconventional imaging modalities and applications have been studied in the field of computational imaging. Established examples of computational imaging include computed tomography, digital holography, and computer-generated holography. The advances in information science mentioned above and the rapidly growing power of computer hardware, such as graphics processing units (GPUs), are driving forces in the field of computational imaging.

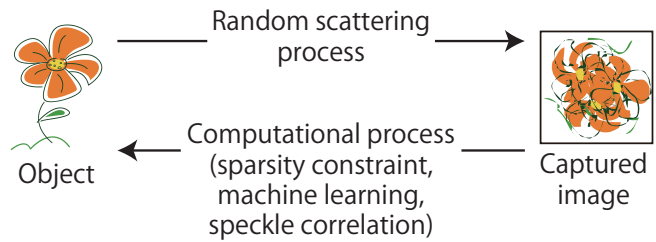

Fig. 1. Computational imaging with randomness.

Optical modulation with randomness has been used in computational imaging, as shown in Fig. 1. Random processes have some interesting features, for example, broad frequency spectra and low cross-correlation. There are two examples that exploit these features: 1. Pinhole cameras are the simplest optical modality in lensless imaging. To enhance the spatial resolution based on geometrical optics, the pinhole must be small, but this decreases the light efficiency, and vice versa. Thus, pinhole cameras have a serious tradeoff between spatial resolution and light efficiency. Coded aperture imaging employs random pinhole arrays and a computational deconvolution process to overcome this tradeoff in lensless imaging [2]. 2. Optical information is multidimensional and includes, for example, three-dimensional spatial position, wavelength, time, etc., but image sensors are two dimensional. To compensate for this dimensional gap between the object and the sensor, the resolution along a certain dimension, such as space or time, is generally compromised in conventional multidimensional imaging. Compressive sensing is an innovative sampling framework for capturing object information with fewer measurements compared with the sampling theorem, thus overcoming the above compromise [3]. It is based on dimensionality reduction with random projection and computational reconstruction with a sparsity constraint. Single-shot multidimensional imaging, such as single-shot depth imaging and single-shot spectral imaging, have been realized by using random processes implemented with scattering media [4].

## Current and Future Challenges

The transmission matrix represents the linear input/output relationship through scattering processes, and it has been used for imaging and focusing through scattering media. One advantage of the transmission-matrix-based approach is the ability to achieve single-shot imaging and focusing through scattering media after a calibration process for observing the transmission matrix [5]. However, this approach requires costly and careful calibration processes, such as an interferometrical optical setup with no stray light. To address this issue, we presented a machine learning approach for estimating the input/output relationship through a scattering process. This approach is applicable for observing not only linear relationships but also nonlinear ones, and it alleviate the requirements for the calibration process. Imaging and focusing based on this machine learning approach have been demonstrated with a simple non-interferometric setup [6, 7]. One challenge with this learning-based method is the tradeoff between the generalization capability and imaging performance.

Speckle-correlation imaging is also an approach for imaging through scattering media [8]. An advantage of speckle-correlation imaging over other methods, including the approach based on the transmission matrix, is non-invasiveness. In speckle-correlation imaging, by assuming shift-invariance of the scattering process, it is not necessary to access the region inside or behind the scattering media for the calibration mentioned above. The shift-invariance is called the memory effect, and it enables us to approximately identify the autocorrelation of the object and that of the captured speckle image. The lateral memory effect has realized two-dimensional speckle-correlation imaging. We have extended two-dimensional speckle-correlation imaging to three-dimensional cases and have demonstrated single-shot depth imaging through scattering media with an axial memory effect, where the speckle is laterally scaled when the object is axially shifted [9]. The object is reconstructed from a single speckle image with a three-dimensional correlation process and a three-dimensional phase retrieval process. Similarly, we have also presented single-shot spectral imaging through scattering media with a spectral memory effect [10]. These methods have realized calibration-free multidimensional imaging. However, the memory effect is a serious limitation in these methods and restricts the applications of speckle-correlation imaging.

**Advances in Science and Technology to Meet Challenges**
To address the issues in imaging through scattering media mentioned above, state-of-the-art technologies in information science, such as deep learning and unsupervised learning, can be important. Also, cross-disciplinary approaches might contribute to overcoming these issues. Computational imaging is one such approach in which optics and information science are combined. Further crossover, including biomedicine, chemistry, and so on, is necessary to make the current approaches for imaging through scattering media more practical and more general.

**Concluding Remarks**
I presented the current situation, future issues, and possibilities in computational imaging with randomness. Computational imaging has contributed to imaging through scattering media, and various promising methods have been reported, such as compressive sensing, transmission-matrix based imaging, and speckle-correlation imaging. However, crucial issues still remain. Further inter-disciplinary approaches, not only optics and computer science, may contribute to this field.

## Section 17 – Sculpted Illumination for deep Tissue Imaging and Interrogation

Vincent R Curtis[1], Laura Waller[2], and Nicolas C. Pégard[1,3,4]

[1]Department of Applied Physical Sciences, UNC Chapel Hill, USA

[2] Department of Electrical Engineering and Computer Sciences, UC Berkeley, USA

[3] Department of Biomedical Engineering, UNC Chapel Hill, USA

[4] UNC Neuroscience Centre, UNC Chapel Hill, USA

**Status**

Scattering and aberrations in dense biological tissue are a major barrier for imaging beyond superficial depths. To observe deeper layers, a popular strategy is to illuminate samples with engineered light to capture optically encoded information that static illumination cannot. Scanning techniques such as multiphoton and confocal microscopy concentrate light to isolate signal from noise and scattered photons, but they introduce trade-offs between imaging speed, spatial resolution, and light exposure. More efficient imaging strategies, such as compressed Hadamard imaging [1] and Fourier ptychographic microscopy [2], leverage custom illumination and computation to enhance diversity in the recorded data, followed by the reconstruction of deep images.

The success of many of these imaging strategies currently depends on advanced light sculpting techniques such as Computer-Generated Holography (CGH), where algorithms control spatial light modulators (SLMs), either to rectify aberrations [3] (Fig.1a) or to focus light deep into tissue [4]. These wavefront engineering methods are routinely used in ground telescopes to undo atmospheric distortions, but they are inadequate for deep tissue imaging because dense aggregations of living cells disturb the free propagation of light with far more degrees-of-freedom (DoF) than there are pixels on the SLM. Hence, new illumination techniques that can sculpt light with many more degrees-of-control to synthesize incoherent distributions [5] or light fields [6] are critically needed for next-generation deep tissue imaging.

Another critical aspect to the performance of deep imaging techniques is managing the joint operation of image acquisition and illumination hardware operating simultaneously in closed-loop systems [1], [4], [7]. Since the number of possible sampling modalities is too large to be explored exhaustively, the integration of advanced optical hardware in future deep imaging methods must be met with equally advanced computational methods and smart sampling strategies to collect as much optically encoded information as possible within the available observation time window.

**Current and Future Challenges**

Perhaps the most important current challenge in the development of high-performance optical instrumentation is the need to innovate with commercially available equipment. SLMs and digital micromirror devices are only commonly available today because they are mass-produced for video projection. Likewise, the video game industry drives the development of the Graphics Processing Units (GPUs), which are at the heart of scientific computation. As a result, while many research groups have conceptualized innovative technologies to push the limits of deep imaging beyond the state-of-the-art, experimental implementations are routinely biased towards readily available mass-produced hardware. While gaps between accessible technology and experimental needs can sometimes be filled with additional computational resources, the additional data processing steps that are required eventually affect imaging performance.

Significant work remains to be done to improve light sculpting technology and compensate for dense scattering in tissue. Existing light sculpting techniques operating with SLMs can only modulate the phase or the amplitude of a single coherent wavefront. For each square millimeter of tissue surface (Fig 1a) they enable, at best, a few million pixels of control. Conversely, deep tissue imaging at depths as short as 200 microns requires the compensation of hundreds of millions of aberrating features per square millimeter. Dense layers of cells disturb the free propagation of light, both spatially and in the angular domain, and repeatedly. The 3D accumulated effects yield complex, incoherent distributions of light (Fig. 1b) that cannot be approximated accurately by a single engineered 2D coherent wave. As a result of this dimensional discrepancy, even the most advanced adaptive wavefront shaping techniques do not have enough degrees- of- control to address scattering in deep tissue.

Progress both in light sculpting and camera technologies dramatically increases both the amount and the rates of optically encoded data that can be exchanged between a computer and biological samples. New algorithms must be developed to control both ends of the acquisition process and obtain the most informative final image within a biologically- defined time window. These algorithms must be capable of managing user input, controlling hardware, and analyzing data simultaneously. By collecting information about the sample in real-time the most efficient frameworks will be able to identify the most informative sampling strategy with instantaneous and partial information.

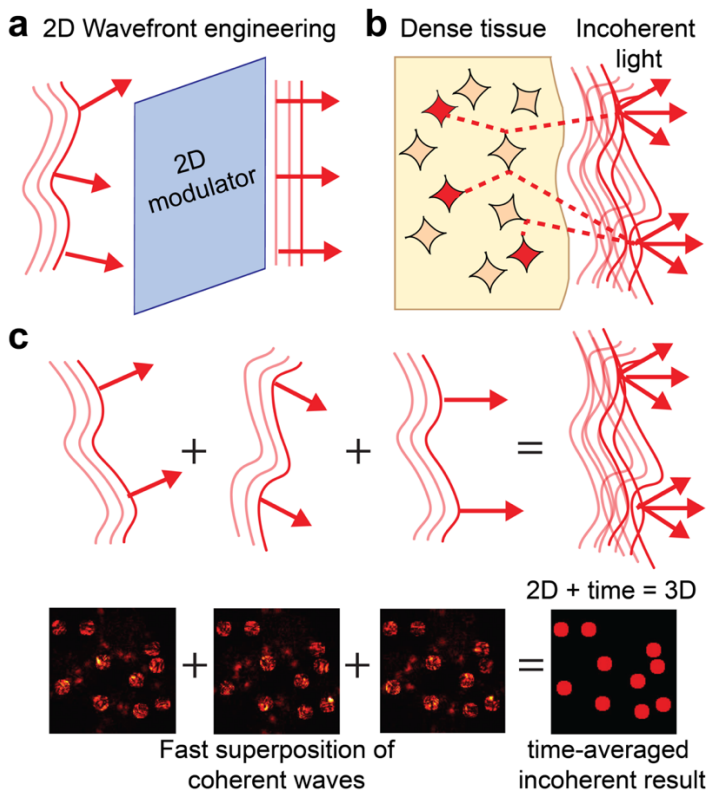

**Figure 1.** (a) A popular approach for deep imaging relies on wavefront engineering methods, either to compensate for optical aberrations between the source and the detector, or to synthesize custom sculpted illumination patterns. (b) Light, scattered or emitted by deep objects in biological tissue is subject to dense aberrations and yields intricate distributions of incoherent light at the surface. (c) Advanced CGH techniques that leverage time as an extra degree of freedom for light sculpting can synthesize incoherent light distributions with greater fidelity, as a rapid superposition of mutually optimized coherent wavefronts.

## Advances in Science and Technology to Meet Challenges

We expect that ongoing and rapid progress in custom, low-volume optoelectronic device manufacturing will open the door to innovative task-driven hardware designs. Custom light modulators tailored for laser beam fast focusing applications have already demonstrated superior capabilities compared to systems designed with commercial, rectangular pixel array modulators [8].

The development of light sculpting technology for deep tissue focusing will require pushing the capabilities of existing hardware and algorithms to obtain the additional DoF needed to engineer light through billions of aberrating features. Perhaps a promising path is to explore the untapped potential of high-speed modulators and leverage time to sculpt light with many more degrees of control than static modulators. Promising results have already been obtained with dynamic computer-generated holography (DCGH) [5] to achieve realistic 3D image renderings in the human eye, by approximating incoherent distributions as a superposition of mutually-optimized coherent waves (Fig. 1c). This methodology has the potential to achieve the necessary gap in performance to focus light precisely through dense biological tissue.

The combination of high-throughput sensing, and light sculpting technologies enables innovative closed-loop imaging modalities for which new algorithms must be developed. The magnitude of data exchange required to view a sample is too large to be stored, making offline processing suboptimal. Real-time data-driven frameworks that can continuously collect, process, and use information gathered from a changing environment to inform and improve the sampling strategy are preferred. For instance, online optical aberration correction algorithms [9] gradually improve image quality and facilitate the long-term monitoring of living tissue. Deep learning models are particularly amenable to handle this type of two-way data stream at high speeds. Their reconfigurable structure can be adjusted on demand with learning algorithms as new data becomes available during operation. Deep learning models have already dramatically accelerated CGH [10] and are now routinely used in image acquisition and processing. By leveraging deep learning methods to simultaneously take control of illumination and acquisition hardware [7], self-adjusting imaging systems eliminate sources of human bias and pave the way for more stable and reproducible imaging experiments.

## Concluding Remarks

The joint development of high-performance sculpted illumination and smart algorithms for online data analysis is critical to enable progress in deep tissue imaging technology. Although this strategy may bring deep imaging capabilities closer to the theoretical limits of accessible depth in biological tissue, many technological and scientific challenges remain to be addressed. Open research avenues include high-performance optoelectronic hardware, optical instrumentation design, and new algorithms for fast, sample-driven acquisition and processing of optically encoded information. In this interdisciplinary area, collaborative projects that explore innovation along multiple directions in parallel are expected to achieve the greatest breakthroughs in performance.

## Acknowledgements

The authors acknowledge financial support from the Burroughs Welcome fund (2018 CASI to NCP), and from the Arnold and Mabel Beckman Foundation (2021 BYI to NCP).

## Section 18 - Multimode fibre based holographic endo-microscopy

David B. Phillips, University of Exeter, UK.
Tomáš Čižmár, Leibniz Institute of Photonic Technology, Germany & Institute of Scientific Instruments of CAS, Czechia

**Status**

Optical imaging through scattering media, such as living tissue, is a grand challenge in biophotonics. Such a capability promises visualisation of structures deep inside the body using non-ionising light. In this chapter we focus on techniques to image through hair-thin strands of multimode optical fibre (MMF) enabling their deployment as endo-microscopes capable of conveying high-resolution images and video from the tip of a needle.

Recovery of images from light signals which have been randomised by propagation through a MMF was first demonstrated in 1967 using analogue holography [1]. Despite this achievement, it took the next four decades for our understanding of light scattering in complex media and the development of digital wavefront manipulation techniques to became sufficiently mature to target real applications. Modern MMF-based endoscopes utilise high-fidelity spatial control over the amplitude, phase and polarisation of light in order to achieve close-to-perfect generation of the desired optical fields at the distal facet (i.e. far end) of a MMF [2]. In imaging applications, these fields most commonly take the form of diffraction-limited foci which are used to scan the scene point-by-point. Specialised techniques are already able to use MMFs to funnel various forms of modern microscopy techniques, as well as spectroscopy and the methods of optical manipulation, into locations with restricted access [3]. In particular, holographic endoscopes have become an exciting technological candidate for *in-vivo* neuroscience, promising micrometre resolution observations of fluorescently labelled neurones residing deep within the living brain [4, 5], as shown in Fig. 1. To date, these systems have been tested on mice models, but they are readily scalable to the size of non-human primates and, ultimately, humans.

Based on the geometry of the MMF, the size of the field-of-view can vary between 10s and 100s of $\mu$m (when imaging in the vicinity of the distal facet of the MMF) and the numerical aperture (NA) ranges between 0.1 and 1. With current light modulation technology, imaging at a few frames per second is achievable with these systems. A unique feature of MMF-based holographic endoscopes is their ability to arbitrarily alter their working distance, and observe objects located right on the distal facet, or move the imaging plane away from the end of the fibre as far as the strength of the returning signals allow. In combination with time-of-flight detection, MMF-based endoscopes can nowadays also offer depth-perception in macroscopic three-dimensional scenes. Such far-field holographic endoscopes look set to further expand the range of applications within the biomedical and industrial inspection domains [6].

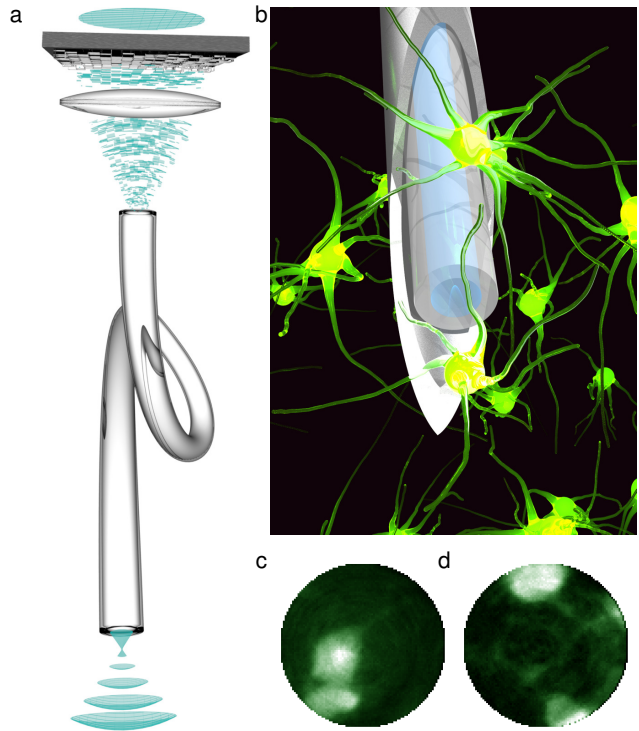

**Figure 1. a,** Formation of a focus at the MMF output using computer holography. An SLM modulates the phase profile of a light field incident from the top such that it is transformed into a diffraction limited focus at the output of the fibre (bottom). **b,** A conceptual schematic of an MMF endoscope applied in neuronal tissue. **c, d,** Neuronal somata recorded by a holographic endo-microscope. Here the field-of-view is 50μm in diameter. Images c, and d, adapted from [5].

## Current and Future Challenges

Scanning imaging through a MMF requires the acquisition of the fibre's transmission matrix (TM), which describes how light fields at either side of any linearly scattering medium are connected [7]. More precisely, the TM of an MMF is a linear matrix operator relating how any input field is transformed via propagation through the fibre. TM acquisition generally requires access to both ends of the MMF: a sequence of known input fields are propagated through the fibre (of a number that should exceed the fibre's mode capacity), and the output fields are holographically measured using an optical set-up with interferometric stability. Once the TM is recovered, it can be used to predict the input field required to generate any desired output (within the spatial bandwidth of the fibre), such as a focussed spot. A key challenge is that the TM of current fibre technology is highly sensitive to mechanical or thermal perturbations – so if the fibre is contorted or changes temperature during use as an endoscope, its TM is altered in an unknown way, and the pre-calibration is no longer valid. This reduces the fidelity of light control at the distal facet, and ultimately disrupts imaging capabilities.

Other challenges pertain to improving the resolution, frame-rate, and signal-to-noise ratio of microscopy techniques it is possible to deliver through optical fibres. Although MMF endoscopy has proven itself capable of adopting numerous scanning-based imaging techniques, it remains highly desirable to extend its portfolio also to prominent wide-field approaches, including super-resolution PALM or STORM, as well as structured-illumination and volumetric (e.g. light-sheet) imaging modalities.

## Advances in Science and Technology to Meet Challenges

There are a range of emerging advances that offer routes to overcome these challenges. Several developments show promise in managing and ultimately alleviating the extreme mechanical sensitivity of holographic endo-microscopy. By virtue of its cylindrical symmetry, the TM of a MMF is not completely random, but contains hidden correlations that become evident when it is represented in a well-chosen basis. In particular, by solving the wave equation in cylindrical coordinates a set of circularly polarised propagation invariant eigenmodes (PIMs) can be derived. These PIMs form a basis in which the experimental TM of a short length of MMF will be sparse and strongly diagonal, as shown in Fig. 2(a-c). Recent work has shown that these features allow prediction of how the TM will be modified when the fibre bends [2], see Fig. 2(d-e). They also constitute prior knowledge which can be used to vastly reduce the number of probe fields required to measure the TM – thus significantly speeding-up the pre-calibration process [8], see Fig. 2(f-g). Furthermore, such correlations also enable an estimate of the TM to be derived with access only to the input end of the fibre by placing a guide-star at the distal facet – meaning re-calibration of a perturbed MMF-based endoscope could be performed *in-situ* [9]. In parallel to these concepts, there is interest in developing new fibres that are more stable to perturbations, with recent work suggesting this may also be possible.

There are two main routes to turning MMF-based endoscopes into single-shot wide-field (scanner-less) imaging devices. The first involves building a mode converter [10] that is able to physically reconstruct an image from MMF-delivered light by unscrambling all modes simultaneously. This method has already been demonstrated using a single diffractive element, however this design suffers from prohibitively low conversion efficiency which is incompatible with fluorescence microscopy [11]. More complicated designs will be necessary to achieve such transformations more efficiently, with an added challenge that they must be able to dynamically adapt to apply new transformations if the fibre itself bends. The second suite of methods rely on computational approaches to recover images from the MMF-randomised signals with no further physical light modulation. However, such an inverse problem is very poorly conditioned, and therefore due to their profound sensitivity to noise computational algorithms can only provide useful results when imaging very sparse scenes [12]. Practical solutions are likely to involve a combination of both physical and computational techniques.

Finally, rapid light modulation devices, such as digital micro-mirror devices (DMDs) and high-speed phase-only spatial light modulators (SLMs) currently underpin holographic endoscopic techniques. DMDs can operate at tens of kHz, but are highly inefficient and are only practical for narrowband sources. SLMs are more efficient yet offer considerably slower modulation rates. New light modulation techniques are needed to overcome these limitations, and one avenue may be the fast piston-based DMD equivalents that are on the horizon.

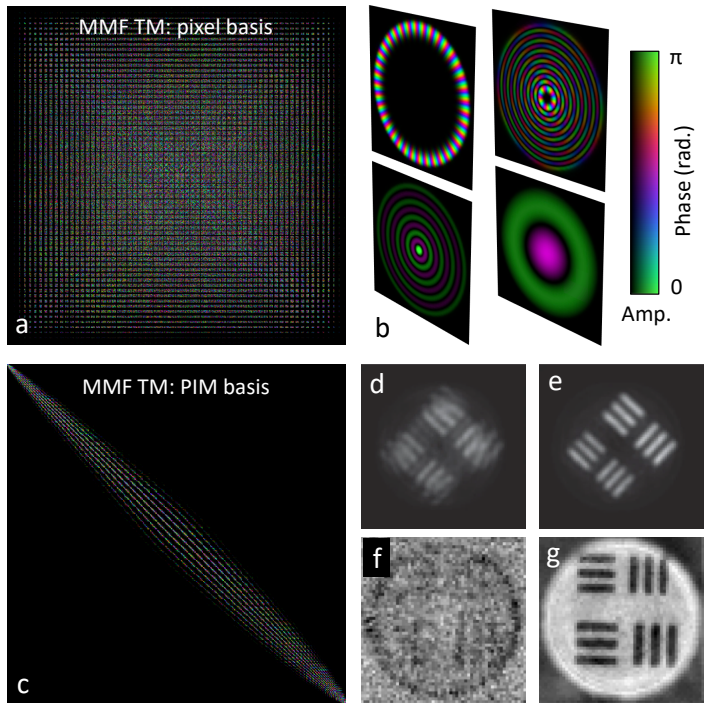

**Figure 2**. **a**, Example of the TM of a MMF represented in the pixel basis. **b**, 4 examples of PIMs that propagate with a constant spatial profile through an ideal MMF. **c**, Example of the same TM as in part a, now represented in the PIM basis. In this case the TM is strongly diagonal, with the remaining spread in the power into off-diagonal elements due to imperfect alignment and limited knowledge of the geometric fibre parameters. **d**, Imaging quality through a MMF that has been bent after calibration. **e**, Image quality restored after prediction of new TM accounting for bending (d, e, adapted from [2]). **f**, Imaging quality through a MMF having only measured 5% of the full TM. **g**, Imaging using the same TM calibration data as in part f, but now incorporating priors to estimate a more faithful TM (f, g, adapted from [8]).

## Concluding Remarks

Over the last decade there has been an explosion of interest in holographic endo-microscopy, driven by key breakthroughs such as the transmission matrix concept and the advent of high-speed digital light shaping techniques. Holographic endoscopes have already been employed to image neurons deep inside living brain tissue, and emerging applications also include depth perception and industrial inspection endoscopy. In parallel, extensive efforts in basic research are focussed on overcoming remaining challenges, particularly the sensitivity of MMF calibration to mechanical fibre deformations and moving from scanning-based to single-shot widefield imaging modalities. Progress so far has impacted numerous domains beyond endoscopic imaging, including high-capacity optical communications, optical computing and quantum optics, to name a few. Looking forward, we foresee a continued expansion of activity in this exciting area of complex media photonics.

## Acknowledgements

D.B.P. thanks the Royal Academy of Engineering, and the European Research Council (ERC starting grant, 804626) for financial support. T.C. acknowledges (ERC consolidator grant, 724530) and the Ministry of Education, Youth and Sport of the Czech Republic (CZ.02.1.01/0.0/0.0/15_003/0000476).

## Section 19 – Ultra-thin imaging endoscope using multi-core fiber

Hervé Rigneault, Aix-Marseille Univ., CNRS, Centrale Marseille, Institut Fresnel, F13013 Marseille, France

Esben Ravn Andresen, Université de Lille, CNRS UMR 8523 – PhLAM – Laboratoire de Physique des Lasers, Atomes et Molécules, F-59000 Lille, France

**Status**

In the same year that a fiber-optic lensless endoscope based on multi-mode fiber was demonstrated [1], a lensless endoscope based on multi-core fiber [Fig. 1 (a1)] was also demonstrated [2]. Both represent the vision of lensless endoscopes: an ultra-thin endoscope whose diameter reduces to the size of the fiber itself (~100 $\mu$m). Having no further distal optics, the lensless endoscope uses a wavefront shaping element on its proximal side to control the phase of the fiber modes at the distal side for imaging. The small diameter makes the fiber a minimally-invasive imaging probe exquisitely suited to acquire images of cells hidden deep in sensitive tissue which must be left to the highest possible extent undisturbed. For instance, multi-mode fiber based lensless endoscopes have already been demonstrated for brain imaging [3-5].

We may take References [3-5] as a starting point to highlight some universal challenges for lensless endoscopes and how multi-core fiber may aid in overcoming them.

*1. Acquisition rate*. In the cited Ref. [3-5] image acquisition is performed by point-scanning, and the wavefront shaping element has to display a specific mask for every pixel of the acquired image. The update rate of the wavefront shaping element is thus the limiting factor for image acquisition rate. Multi-core fibers can overcome this limit under the condition that individual cores do not exchange energy, allowing point-scanning by fast scan mirrors [6] by exploiting the so-called memory effect (the ability to translate the distal wavefront with a simple phase tilt at the fiber input side, see section 13).

*2. Two-photon imaging (and nonlinear imaging in general).* In Ref. [3-5] fluorescence contrast was used to benefit from the specificity of contemporary fluorescent reporter molecules. However, narrowband laser sources were used to overcome the low bandwidth of wavefront shaping in multi-mode fibers and so only one-photon fluorescence contrast was possible. Often two-photon fluorescence contrast is desired for biological imaging allowing deeper penetration and z-sectionning, but this requires illuminating the sample by a train of ultra-short optical pulses which become stretched by modal dispersion in multi-mode fibers. Multi-core fibers are exempt from modal dispersion under the condition that all cores are identical, so virtually undistorted transport of femtosecond pulses and two-photon imaging is possible [7]. Additionally, we note that the multi-core fabrication method allows to include a very high-NA inner cladding which is very efficient at collecting fluorescence light [7] [Fig. 1(a2)].

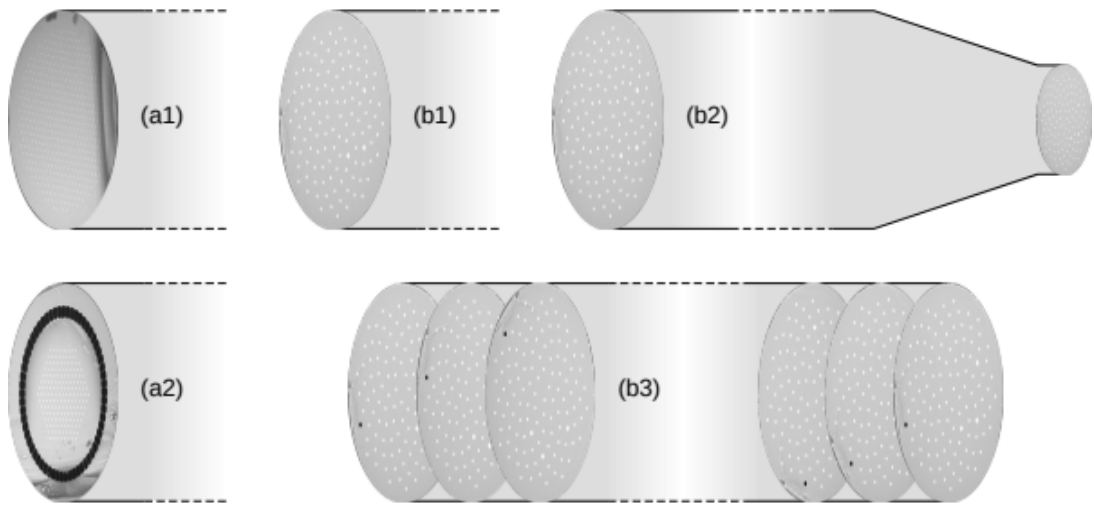

**Figure 1.** Two levels of multi-core fiber design. (a1, a2) "Unmodified" multi-core fiber (a1) without and (a2) with high-NA inner cladding. (b1-b3) "Modified" multi-core fiber with (b1) aperiodic core layout; (b2) a tapered section; and (b3) a helical twist.

## Current and Future Challenges

Recent years have seen the exploration of "Modified" multi-core fibers [Figs. 1(b1)-1(b3)] for lensless endoscopes. During fabrication, such modifications can be made before; after; or during the drawing process. Modified multi-core fibers can help address some of the thornier challenges facing multi-core fiber-based lensless endoscopes.

*3. Multiply-peaked point spread function (PSF).* Periodicity in the core layout of an unmodified multi-core fiber [Fig. 1 (a1)] has the unwanted consequence of additional intensity maxima in the PSF leading to overlapping replica images. Multi-core fiber with completely aperiodic core structure—for instance a Fermat's golden spiral layout [Fig. 1(b1)]—can be made whose resulting PSF is singly-peaked [8].

*4. Strehl ratio.* The Fermat's golden spiral multi-core fiber however does nothing to resolve another shortcoming of multi-core fibers: the low intensity delivered into the PSF. Indeed, only a small portion (known as the "Strehl ratio") of the light emitted from the multi-core fiber can be concentrated in a focus by the wave front shaping element. This is an intrinsic consequence of the low surface coverage of the cores [9], which in turn is a consequence of the large core-to-core distance dictated by the need for low energy exchange between cores. A modification post-drawing of the multi-core fiber tip—a tapering or a homothetic transverse downscaling over a few cm [Fig. 1(b2)]—can dramatically increase the PSF intensity while introducing virtually no additional energy exchange and leaving the memory effect intact [10].

*5. Bend resilience.* In Ref. [3-5] the endoscope fiber was a few cm long multi-mode fiber which is short enough that it remains rigid at all times. A more ambitious vision of the lensless endoscope calls for a longer fiber that can be free to flex, as this would allow to fix it onto a freely-moving animal. However, the transmission matrix changes with the fiber conformation, generally rendering the masks displayed on the wave front shaping element invalid. Multi-core fiber can overcome this challenge in the case where the individual cores follow corkscrew trajectories, i.e. the multi-core fiber is "twisted" [11] [Fig. 1(b3)]. Such a twisted multi-core fiber remains invariant to conformation as long as the twist period remains small compared to the rate of change of radius of curvature along the fiber. Another strategy has been to monitor in real time the multi-core conformation using double path Mach-Zehnder interferometry and to correct for deformation-induced shifts in focal spot position during raster-scanning [12].

## Advances in Science and Technology to Meet Challenges

Recent progress in additive manufacturing open new possibilities to "functionalize" or "augment" optical fiber properties by 3D printing microstructures at the fiber tip [13]. The coming years could see these new opportunities explored in ultra-thin fiber-optic endoscopes. For instance, we recently explored the possibility to overcome the low Strehl ratio (~0.01) achievable by multi-core fibers (point 4) by designing a miniature beam combiner that has the property to artificially increase the surface coverage of the cores. Using a combination of microlens and a 'top lens' the fabricated miniature beam combiner was shown to achieve a Strehl ration of 0.35 suitable for two-photon imaging (Fig. 2).[14]. This 3D micro-printing technique is applicable to any type of fiber and so may well have potential to improve also lensless endoscopes based on multi-mode fiber. Other improvements that will bring ultra-thin fiber-optic endoscope closer to applications are expected such as biocompatible and biodegradable material or their association with smart materials allowing their remote conformation control [15].

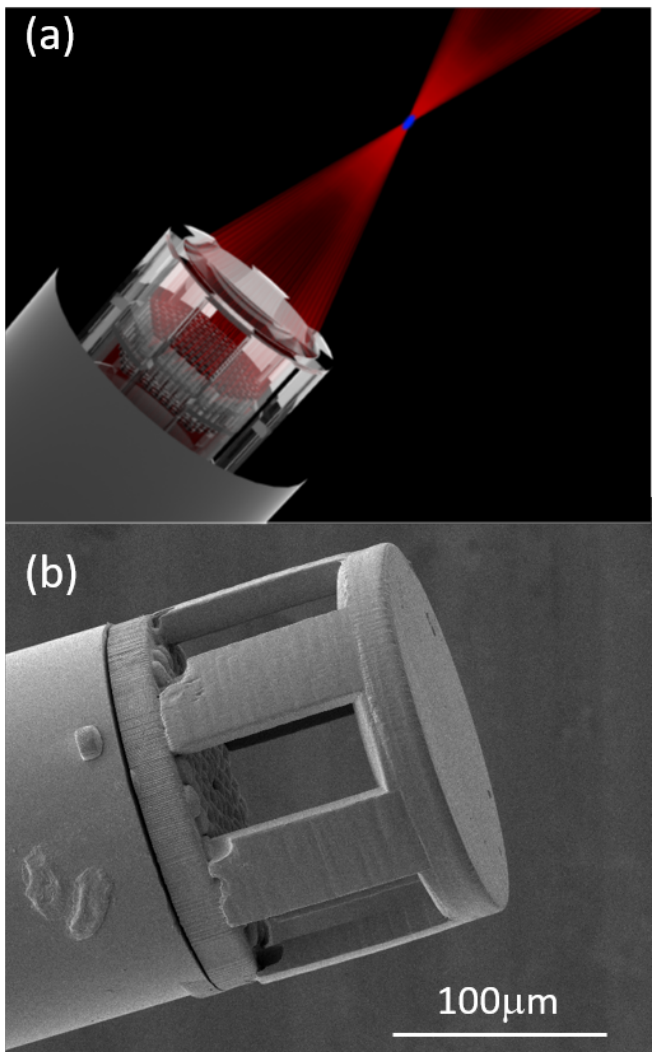

Figure 2. Hyper-Telescope micro-printed at the multi-core fiber end and improving the delivered power into the focus spot by x35. (a) artist view of the hyper-telescope beam combiner, (b) actual fabrication using micro-printed polymer.

## Concluding Remarks

We have seen that most of the challenges brought by ultra-thin fiber-optic endoscope based on multi-core fiber have solutions that can been realized (inner cladding, aperiodicity, twist, taper or micro-printed beam combiner). Now the challenge is to combine all these innovations together into a system that is sufficiently reliable to perform in vivo experiments. Forthcoming innovations brought by additive manufacturing or novel material are expected to simplify this challenging task.

## Acknowledgements

Agence Nationale de la Recherche (ANR-14-CE17-0004-01 "LENIMBRA"); (ANR-20-CE19-0028 "NAIMA"); ANR-16-IDEX-0004 ULNE, LABEX CEMPI (ANR-11-LABX-0007), Equipex Flux (ANR-11-EQPX-0017), Turing Centre for Living systems (ANR-16-CONV-0001), Ministry of Higher Education and Research, Hauts de France council, European Regional Development Fund (CPER Photonics for Society) P4S), FiberTechLille Technology Platform (linky)., NIH R21 EY029406-01, Aix Marseille University (A-M-AAP-ID-17-13-170228-15.22).

## Section 20 − **Learning to Image, sense and control with Multimode fibers**

[1]Babak Rahmani, [1]Christophe Moser, [2]Demetri Psaltis

1 Laboratory of Applied Photonics Devices, School of Engineering, Ecole Polytechnique Fédérale de Lausanne, 1015 Lausanne, Switzerland

2 Laboratory of Optics, School of Engineering, Ecole Polytechnique Fédérale de Lausanne, 1015 Lausanne, Switzerland

As shown in previous sections, conventional methods of imaging through MMFs, such as the gold standard transmission matrix method, require full-field measurements (phase and amplitude) of the fiber's output to construct a mapping between the input and output of the system. The dependence on the phase information makes the system vulnerable as the phase is sensitive to external perturbations. Data driven methods have been recently proposed to circumvent this problem. These techniques learn statistical characteristics of the light propagation using examples of inputs and intensity-only measurements of the output. Within this framework, learning-based methods for imaging through MMFs either 1- seek to retrieve the input information (usually a 2D image) entering the fiber from intensity-only measurements of the output or 2- seek to obtain the required input pattern that projects a desired target at the distal facet of the fiber. It should be noted that such problems are highly ill-posed as many inputs can result in the same intensity profile at the output of the fiber that only differ in their respective phase information.

The inference of the input of the fiber from intensity-only measurements of its output involves the construction of a backward mapping function that is obtained by minimizing a loss function in the following format:

$$L = \mathrm{M}\big(x, A_\theta(y)\big)$$

where $x$ and $y$ are the input and output of the system. We note that $x$ in general is complex, whereas $y$ is always a positive real number. $\hat{x}$ is the solution of the this optimization problem. The operator M represents a metric between the predicted output of the deep neural network (DNN) $A_\theta$ that is parametrized by $\theta$. The loss function is optimized by taking gradients with respect to the learnable parameters of the mapping function $A$, i.e. $\theta$, in a process known as gradient descent. Upon convergence, the mapping function is an estimator of the backward mapping of the MMF system that predicts the input patterns of the system from the corresponding outputs.

Using convolutional networks, Rahmani *et.al*., [1] Borhani *et. al.* [2] and Kakkava et.al. [3] were able to reconstruct sparse-like input images that were scrambled upon propagation through various MMF's lengths up to 1 km. Fig 9. (a) shows the optical setup that is used in data collection step. Fig 9 (b) depicts an example of a speckle pattern which is fed to the DNN together with its reconstruction. Other authors showed the same performance of the DNNs (2D Pearson correlation around 95 percent) with more complex input images [4].

Authors in [5] used a simpler DNNs, such as a fully connected single layer for reconstruction of the images scrambled through MMF with comparable fidelity as that of the complex architecture DNNs.

Learning-based methods for projection through MMFs seek to find the correct input pattern that upon propagation through the fiber produces a desired image at its distal side [6, 7]. It is again assumed that the fiber system is characterized without resorting to holographic measurements. Using a more complex training procedure, authors in [6] were able to project arbitrary images through MMFs for various wavelengths. The training algorithm therein involves the construction of a mapping mimicking the forward propagation of the light from the input to the output (Model) followed by learning the backward mapping of the system (Actor). Some sample images projected through the experimental system using the Actor-Model algorithm are plotted in Fig. 10 (b).

(a)

(b)

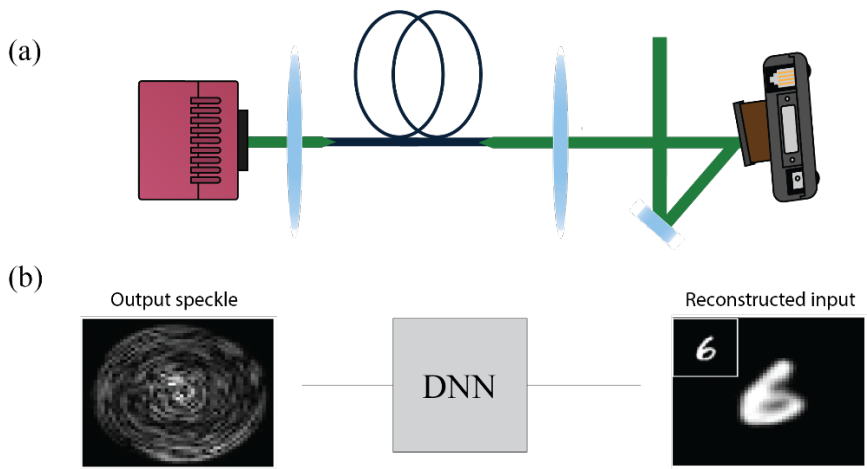

Output speckle        DNN        Reconstructed input

Fig. 9. (a) Experimental setup including spatial light modulator (SLM), MMF and a camera. (b) Inference pipeline with the fiber output speckle as input to the DNN and the output of the DNN. The inset shows the groundtruth originally sent through the fiber.

The end-to-end characterization of learning-based methods allows for inherent learning of perturbations incurring on the imaging system. Despite severe decorrelation of the imaging system due to external sources of perturbations such as thermal variability and mechanical misalignment, the authors in [6] showed a stable imaging fidelity over several hours.

Another source of perturbation is the drift in the wavelength of the laser source that decorrelates the output intensity with time. In a study conducted by Kakkava *et.al.* [8-10], it was shown that the DNNs can correct for the decorrelation rendered by the wavelength change of the fiber with an extended bandwidth.

Other researchers investigated the performance of the DNNs under severe mechanical perturbations [11-13]. Specifically, the MMF was positioned in different configurations while examples of inputs and outputs being collected. It was shown that DNNs were able to reconstruct the input information sent through the fiber when the DNN has been given examples of input-output for the entire positional configurations. Other lines of work use

DNNs to characterize perturbed MMFs for sensing applications such as temperature [14] and mechanical sensors [15, 16]. In addition to sensing in the linear domain, DNNs have started to be used for the characterization of nonlinear dynamics in MMFs. The results of such studies open up novel perspective for the use of machine learning in multimode fibers [17-20].

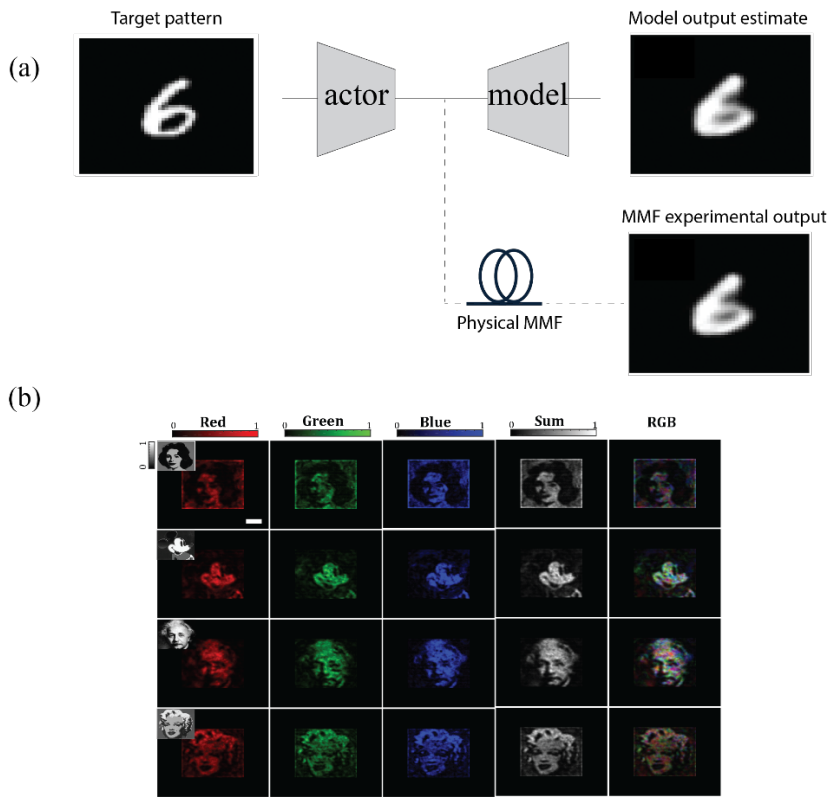

Fig. 10. (a) Projection pipeline with the target image as input of the DNN. The Actor predicts the SLM pattern with the help of the Model. Upon convergence, the predicted SLM pattern is also sent through the physical MMF. (b) Example of images projected experimentally through the MMF for three wavelengths corresponding to three colors.

# Section 21- Transmission eigenchannels in diffusive random media

Hui Cao, Dept. of Applied Physics, Yale University, USA.
Alexey Yamilov, Dept. of Physics, Missouri University of Science and Technology, USA.
Hasan Yılmaz, National Nanotechnology Research Center (UNAM), Bilkent University, Turkey.

**Status**

Coherent control of wave transport by spatial wavefront shaping has overcome the limitations imposed by incoherent diffusion. In a linear scattering system with static disorder, the mapping from the incident to the transmitted waves is deterministic, and fully described by the field transmission matrix $t$. The eigenvectors of $t^\dagger t$ provide the input wavefronts for a set of transmission eigenchannels. Any incoming wave can be decomposed into a linear superposition of these eigenchannels, each propagating independently through the system with a transmittance equal to the corresponding eigenvalue $\tau$. In a lossless diffusive system of average transmittance $\langle \tau \rangle \ll 1$, the transmission eigenvalues $\tau$ range from 1 (open channels) to 0 (closed channels). Therefore, selective excitation of individual eigenchannels leads to diverse non-diffusive behaviors.

Transmission eigenchannels not only have very different transmittance, but also feature distinct energy distributions inside a diffusive system. In contrast to a linear decrease of energy density with depth in a diffusive system, an open channel reaches the energy maximum near the center of depth, while a closed channel exhibits an exponential decay [1-3]. Therefore, an open channel has energy built-up deep inside a diffusive system, and enhances light-matter interaction.

The depth profile of an open channel can be modified by modulating the width of a diffusive waveguide with reflecting sidewalls (Fig. 1a). This enables inverse design of an eigenchannel profile [4]. In a wide diffusive slab with open boundaries (Fig. 1b), all transmission eigenchannels are localized in the transverse direction (parallel to the slab surface), and their lateral size is much smaller than the slab width [5]. None of them expands laterally while propagating through the slab, in sharp contrast to transverse diffusion of a narrow beam with identical width but arbitrary wavefront.

Open channels exist at any frequency, although their input wavefronts are frequency specific. Even for a broadband input light, it is possible to find an incident wavefront to enhance transmittance at all frequencies involved [6].

**(a) Longitudinal profile of open channels**

**(b) Transverse localization of transmission eigenchannels**

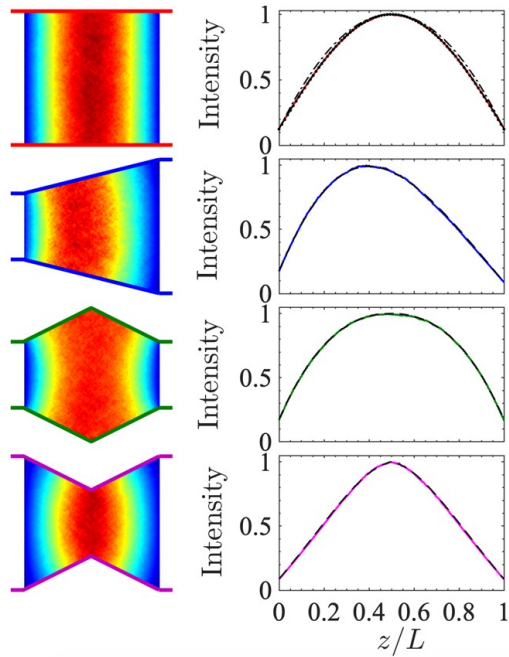

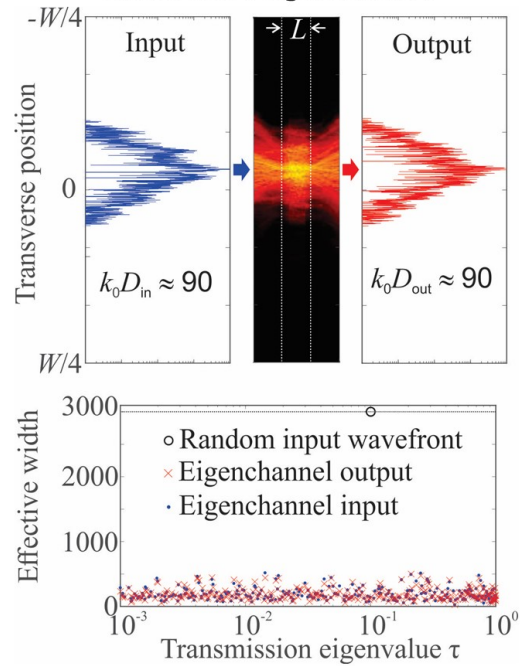

**Figure 1.** (a) Engineering the depth profile of open channels by modulating the width of a diffusive waveguide [4]. (b) Transverse localization of transmission eigenchannels in a wide diffusive slab [5].

## Current and Future Challenges

While there have been extensive theoretical and numerical studies on transmission eigenchannels, experimentally it is hard to observe them directly. Typical scattering samples are wide slabs with open boundaries, and it is impossible to control incident light in all angles. Such incomplete channel control (ICC) greatly reduces the transmittance of open channels and their penetration depth [7].

The main challenge is to achieve complete control of incident fields in all spatial modes. To observe the internal structures of individual eigenchannels, it is necessary to probe energy density everywhere inside a disordered system. Experimentally fluorescent beads are embedded inside three-dimensional scattering samples to extract information about internal light distribution from their fluorescence [8]. However, it is difficult to control exact locations of these beads as well as to separate the individual fluorescence for a large number of them. Numerically it is a daunting computational task to simulate wave propagation in a three-dimensional diffusive system of practical dimensions. Finally, the ability to precisely tailor sample geometry, dimension and scattering parameters is essential to uncover their effects on the transmission eigenchannels.

The transmission matrix is specific to the disorder realization, so are its eigenchannels. In theoretical and numerical studies, the spatial profiles of transmission eigenchannels are averaged over large ensembles, whereas experiments are conducted on a single sample and the eigenchannel profile may deviate from the average. Knowledge of the extent of this realization-to-realization deviation is critical to practical applications.

Absorption is ubiquitous in optical systems, and its effect on coherent wave transport differs from that of decoherence on mesoscopic electron transport. Moreover, the spatial distribution of absorption can be homogeneous or inhomogeneous, which will impact the transmission eigenchannels in different ways [9].

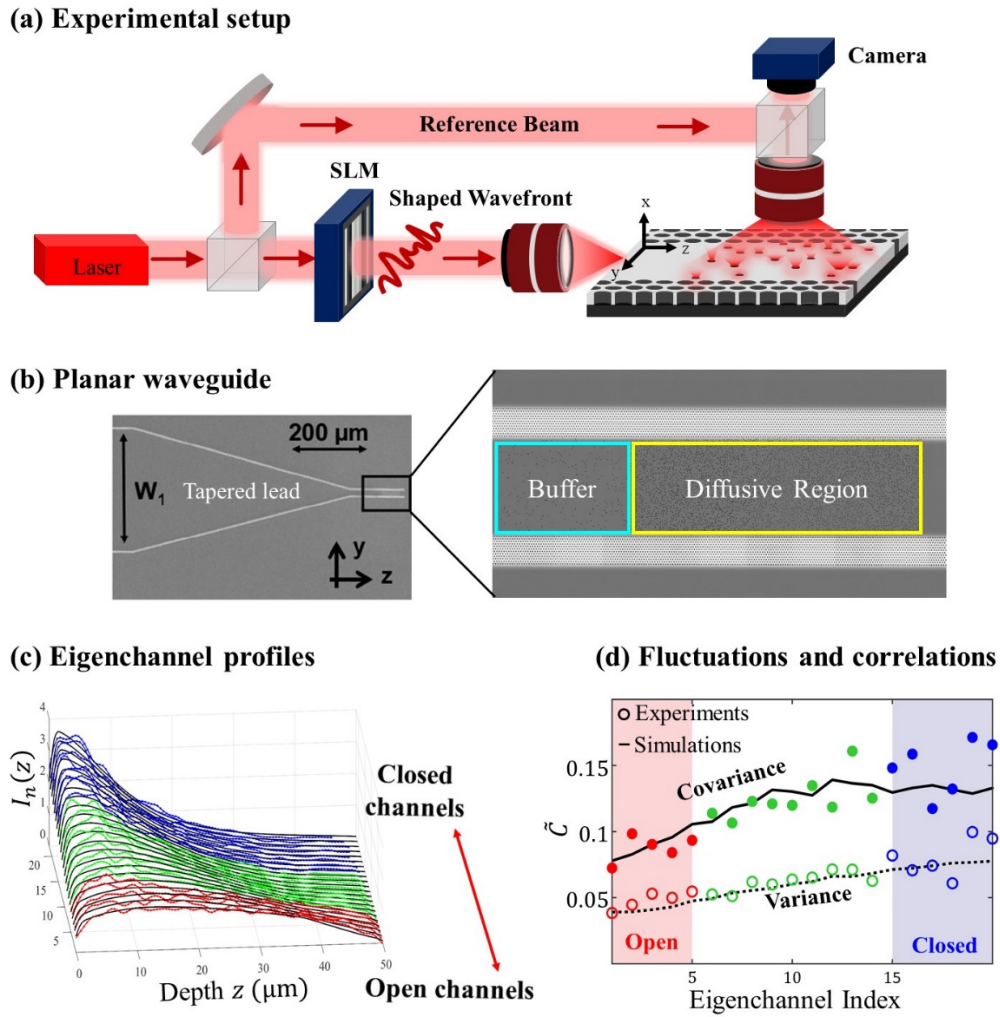

**(a) Experimental setup**

**(b) Planar waveguide**

**(c) Eigenchannel profiles**

**(d) Fluctuations and correlations**

**Figure 2.** (a) Schematic of experimental platform for excitation and observation of individual transmission eigenchannels in a planar silicon waveguide. (b) Scanning electron micrograph showing the tapered lead (left), buffer and diffusive region (right) of the waveguide. (c) Distinct depth profiles of open and closed channels. (d) Realization-to-realization fluctuations (variance) and correlations (covariance) of eigenchannel profiles [10].

**Advances in Science and Technology to Meet Challenges**

A unique experimental platform [10] has been built in the two-dimensional optical waveguide geometry (Fig. 2a). The waveguides, fabricated in silicon-on-insulator wafers, have highly reflecting sidewalls composed of photonic crystal layers (Fig. 2b). Air holes are randomly distributed inside a section of the waveguide. Their size and location can be precisely controlled to tune the transport mean free path $\ell$. The length of the disordered section is much larger than the transport mean free path but smaller than the localization length, to ensure diffusive transport. The waveguide width determines the number of spatial modes. A monochromatic laser beam is wavefront-shaped by a spatial light modulator (SLM), and then injected via the edge of the wafer into a ridge waveguide. The waveguide width is adiabatically tapered down to couple light into all waveguide modes incident on the disordered region. While light is predominantly scattered by air holes in the waveguide plane, a small amount is scattered out of plane. It is collected and measured with an interferometer to recover the field distribution inside the disordered region. In order to access the field incident on the disordered section of the waveguide, a weakly-scattering zone (buffer) is added, and light escaping from it to the third dimension is detected.

This experimental platform enables complete channel control and measurement of the full transmission matrix *t* [9]. From *t*, transmission eigenchannels are found and individually excited by wavefront shaping. The spatial structures of both open and closed channels inside a planar waveguide are observed directly from the vertical dimension (Fig. 2c). The depth profiles of higher-transmission channels exhibit smaller fluctuations from one disorder realization to another, compared to lower-transmission eigenchannels and random input wavefronts. Realization-to-realization fluctuations of the depth-profiles of different eigenchannels exhibit correlations. These correlations are weaker for higher-transmission eigenchannels, indicating they are more isolated than lower-transmission eigenchannels (Fig. 2d). Open channels have robust depth profiles, which are consistent between disorder configurations, allowing reliable energy delivery deep inside diffusive system. This platform also enables experimental study of energy deposition to a region of arbitrary size and shape anywhere inside the scattering system.

**Concluding Remarks**

The experimental and numerical studies on transmission eigenchannels represent the first steps towards physical understanding of how open and closed channels are formed inside diffusive systems, and eventually the development of a comprehensive theory. The next step is to extend these studies to wide slabs with open boundaries and to volumetric diffusive systems.

Another direction is to explore and exploit the remarkable properties of transmission eigenchannels such as, for example, the enhanced range of the angular memory effect for open channels. Finally, it will be interesting to explore the eigenchannels of other operators, e.g., for time delay of a pulse propagating through a diffusive medium, or energy deposition deep inside a scattering system.

**Acknowledgments**
The authors thank their coworkers and collaborators who have contributed to the works described in this contribution. They also acknowledge financial support from National Science Foundation grants DMR-1905442, DMR-1905465, and from Office of Naval Reseeach grant N00014-20-1-2197.

### Section 22 - Optimal information extraction from scattering systems

Dorian Bouchet[1], Allard P. Mosk[2] and Stefan Rotter[3], [1]Université Grenoble Alpes, CNRS, LIPhy, Grenoble, France, [2]Debye Institute for Nanomaterials Science, Utrecht University, Utrecht, The Netherlands, [3]Institute for Theoretical Physics, TU Wien, Vienna, Austria.

**Status**

In free space, measuring the field scattered by an object allows one to directly estimate the value of observables characterizing this object, such as its position, its size or its shape. However, in the case of an object embedded in a complex scattering environment, the situation is more involved due to the influence of all absorption and scattering processes acting on the incident light. To characterize all these processes, a convenient tool is given by the system's scattering matrix, which connects incident to outgoing far-field modes (see sections 9 and 10). In the last decade, the possibility to optically measure significant fractions of scattering matrices has enabled a multitude of new approaches to image hidden objects in complex scattering environments [1].

A promising way to analyze such scattering experiments is to use tools borrowed from information theory. Notably, if the object can be described by a set of parameters, the Fisher information can be employed to quantitatively assess how well the object can be reconstructed [2]. It then turns out that, from the knowledge of the system's scattering matrix (which is now a function of these parameters), one can not only predict the achievable precision in the estimation of these parameters, but also design incident coherent fields that optimally probe the system [3,4] (Fig. 1a). These "maximum information states" are specifically tailored to precisely characterize an object of interest within its complex environment and could thus serve to improve the quality of imaging techniques in scattering environments, such as semiconductor nanostructures or biological tissues. Remarkably, these customized light fields are not only optimal for extracting the maximum amount of Fisher information on a target, but also for manipulating this target with light [5] (Fig. 1b)—a feature, which is rooted in a fundamental connection between information and measurement back-action.

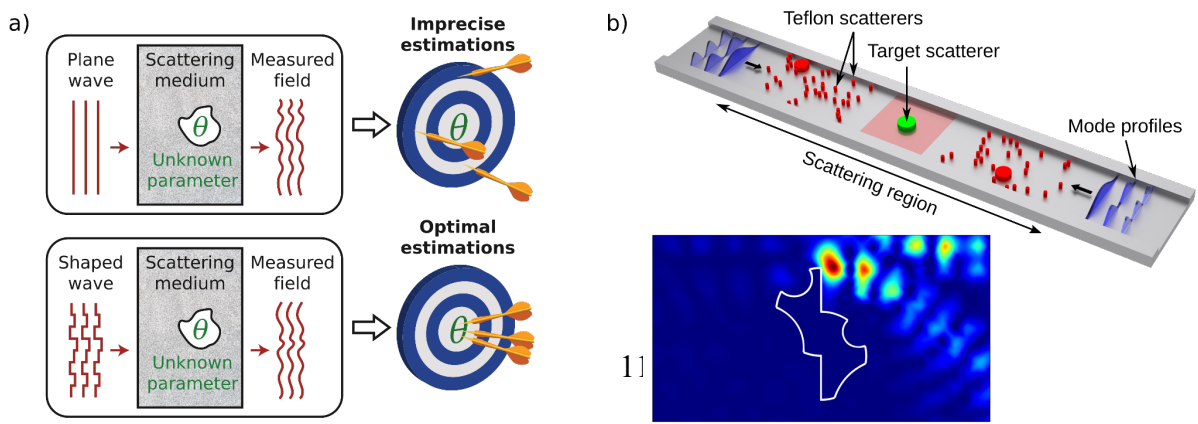

**re 1.** a) Principle of an optimal field specifically shaped for the estimation of a parameter θ. Plane wave illumination leads, in many cases, to imprecise estimations (top), while the optimal field enables one to reach the best possible estimation precision (bottom). Figure adapted from [4]. b) Top: sketch of a scattering system consisting of a two-dimensional waveguide within which scatterers (red cylinders) are randomly distributed. The object of interest is represented by a green cylinder. Bottom: Intensity distribution experimentally measured around the object, which maximizes the Fisher information accessible in the far field about the object orientation, and simultaneously maximizes the transfer of angular momentum between the field and the object. Figure adapted from [5].

## Current and Future Challenges

Employing the Fisher information to improve imaging techniques entails describing entire images as ensembles of parameters. The large number of such parameters that need to be introduced in this case, however, easily makes a corresponding procedure prohibitively difficult. Indeed, calculating the achievable precision in the estimation of p parameters requires the inversion of a $p \times p$ Fisher information matrix, which becomes numerically expensive when $p$ exceeds around 10,000. While the available computational power still gradually increases over the years, a suitable alternative could be to find suitable sparse representations to describe images of interest [6].

Different options are available to experimentally determine optimal input fields that maximize the Fisher information, such as brute-force optimization, iterative phase conjugation, or scattering matrix measurements. At this stage, however, all these procedures rely on a deterministic control over the parameters to be estimated, therefore strongly limiting the applicability of the approach. Developing experimental procedures to identify optimal fields without such a high level of control would thus be of great interest. Alternatively, in the case of engineered nanostructured samples, numerical models of scattering experiments could potentially be employed to determine optimal fields purely numerically [7]; it remains to be assessed, however, if such models are sufficiently accurate for these fields to be usable in actual experiments (see section 13).

In theory, one can also expect to improve the achievable precision by simultaneously taking advantage of both the spatial and quantum degrees of freedom of the incident light. In quantum metrology, general procedures already exist to identify such optimal input states that maximize the Fisher information [8]. Nevertheless, it remains unclear how such procedures can be employed in the case of complex scattering systems, as the evolution of any input state is then typically described by an unknown sub-unitary operator. Moreover, while experiments that involve spatially-shaped quantum states of light start to emerge (see section 23), most quantum states remain experimentally difficult to generate, restricting the practical application of quantum metrology and imaging protocols.

## Advances in Science and Technology to Meet Challenges

For experiments in which optimal input fields cannot be easily identified using information theory, the association of wavefront shaping devices with deep learning algorithms constitutes a possible alternative for the development of imaging applications based on task-specific input fields (see section 12). Indeed, deep learning algorithms offer the possibility to implicitly measure scattering matrices using training data sets. Integrating light-shaping devices (such as spatial light modulators, digital micro-mirror devices or tunable metasurfaces) as trainable layers in artificial neural networks offers a practical approach to identify input fields that are specifically tailored for a given imaging task [9], with no guarantee, however, that the true optimal solution will be reached.

It also often happens that one cannot rely on scattering matrices to describe the influence of absorption and scattering processes upon the incident light, for instance in the case of media (such as biological tissue) that feature an unknown time-variation. Reconstructing an image through such media requires the use of statistical correlations—rather than deterministic relations—between fields at the detector and fields in the object plane (see section 15). Interestingly, such statistical approaches can also benefit from quantitative analyses based on information theory [10], in order to guide the experimental development of imaging techniques at large depths.

## Concluding Remarks

The tools of information theory offer a wide range of possibilities to quantitatively assess and optimize the capabilities of imaging techniques in scattering environments. We expect that information-driven approaches will be key to the development and refinement of new computational microscopy methods, and notably to identify task-specific illumination schemes that optimally probe complex scattering media (see also section 17). In addition, an information-driven paradigm may provide us with a new level of understanding of light-matter interactions in complex systems.

## Acknowledgements

The authors thank M. Kühmayer for his help with editing the figures. A.P.M. acknowledges support from the Nederlandse Organisatie voor Wetenschappelijk Onderzoek NWO (Vici 68047618), and S.R. acknowledges support by the Austrian Science Fund (FWF) under project number P32300 (WAVELAND).

# Section 23 - Imaging in complex media with quantum states of light

Yaron Bromberg, Racah Institute of Physics, The Hebrew University of Jerusalem

Hugo Defienne, School of Physics and Astronomy, University of Glasgow

**Status**

Quantum states of light play a key role in basic research on quantum mechanics, as well as in emerging quantum technologies. Early studies on the propagation of quantum states of light in complex media focused on fundamental questions such as the survival of quantum features of light in random media [1]. Following recent rapid developments in photonic quantum technologies, it became critical to understand the effect of scattering of quantum states of light also from a technological perspective. For example, implementation of free-space quantum communication in real-life settings, such as ground-satellite links, requires dealing with scattering and aberrations induced by atmospheric turbulence [2]. In quantum computation, quantum supremacy was recently demonstrated by mixing squeezed states occupying one hundred modes using a multiport interferometer with random mode couplings [3]. In this chapter, we discuss recent advancements and future challenges towards utilizing quantum resources for imaging in complex media.

Over the past few decades, researchers have considered various approaches for utilizing quantum properties of light to surpass classical bounds in optical imaging. These include enhanced sensitivity and noise rejection, utilizing anti-bunching for super-resolution imaging [4] and imaging with undetected photons [5]. Motivated by the potential of quantum imaging to lift some of these limits, in recent years we are witnessing a growing interest in going one step further, and utilize quantum states of light for imaging through turbid media.

**Current and Future Challenges**

Two general approaches can be envisaged to harness quantum states for imaging in scattering media.

First, the advantages that these states provide in terms of imaging performance (e.g. resolution and sensitivity) can directly benefit existing imaging methods in scattering media. In this case, the main challenge is to adapt the corresponding optical tools, such as adaptive optics and wavefront shaping, to quantum sources. Although imaging photon pair correlations through thin static diffusers by wavefront correction was recently achieved [6], such a task remains extremely challenging in real-world situations because of the high complexity of natural scattering media and the very low intensity of quantum light sources. To advance these issues, a promising avenue is to use a hybrid approach in which quantum light is guided through scattering using an intense classical light as a beacon [7]. Figure 1a illustrates this concept with a beam of entangled photons that is re-focused after a turbid medium by implementing real-time aberration correction using scattered light from the classical pump laser as a feedback signal. Such a hybrid approach enables the incorporation of classical wavefront correction algorithms into any quantum imaging system. Furthermore, it can lead to the development of a novel multimodal imaging framework in which quantum and classical light operate in parallel.

The second approach consists of harnessing properties that are unique to quantum light, such as entanglement, to develop new protocols for imaging inside scattering media. In this case, the main current challenge is to determine which states and specific quantum resources could potentially be used to improve such a task. In this respect, an interesting scheme was recently proposed in the context of quantum communication through multimode optical fibers [8]. Adapted to the imaging problem, its concept is illustrated in Figure 1b. One photon from an entangled pair illuminates an object hidden behind or inside a disordered medium, while its twin photon is sent towards a light shaping system (e.g. spatial light modulators). Unlike classical approaches, the shaped light does not penetrate the scattering medium and no information can be retrieved from intensity signals at the output. However, if one programs the shaping system so that it mimics the optical disorder, object information can be retrieved via intensity correlation measurements between the two distant cameras. In essence, the scattering medium is rendered transparent by carefully 'scrambling' the photons that did not enter it, rather than unscrambling the photons that did. Even if the benefits of this non-local shaping scheme compared to classical methods are not yet established, it raises intriguing questions regarding the use of genuine quantum resources for imaging inside scattering samples, and invites us to explore them.

Interestingly, other recently developed quantum imaging approaches can indirectly contribute to advance the problem of imaging through scattering media. For example, quantum imaging schemes with undetected photons [5] allow imaging with mid-infrared light, a wavelength range that is naturally less sensitive to scattering in biological tissues. Furthermore, quantum-dots-based super-resolution approaches [4] have also the potential to access deeper layers inside disordered media by using these emitters as guidestars for wavefront shaping.

**Advances in Science and Technology to Meet Challenges**

One of the main challenges of quantum imaging is the low light levels associated with quantum light sources. This challenge becomes particularly severe in imaging through scattering media, as scattering spreads the photons over multiple spatial modes, making it extremely difficult to efficiently collect the scattered photons. Furthermore, to truly exploit quantum features of light, it is generally needed to probe temporal and spatial correlations between the detected photons. This requires multi-channel detection at the single photon level, with excellent spatial and temporal resolutions. While impressive progress has been made in measuring spatial correlations using EMCCD and ICCD cameras [9], the inherently slow frame rates of such cameras limit the applicability for imaging in natural scattering media such as biological tissue, and prevents implementation of feedback-based wavefront shaping. The most promising technology for meeting the strict requirements for quantum imaging through scattering media is SPAD arrays which combine the multi-channel detection of cameras with the sensitivity of single photon detectors. More importantly, it provides high-resolution timestamps to photon-detection events at each pixel. SPAD array technology is rapidly progressing over the past decade, demonstrating arrays with pixel numbers approaching a few thousand and photon detection efficiencies as high as 60% [10]. Yet for meeting the demanding needs of deep quantum imaging, further advancements are required in terms of the number of pixels, array fill-factor, low cross-talk, and on-chip multi-channel coincidence registration.

The vast amount of spatial-temporal information made available by SPAD arrays, sets new challenges in terms of data readout, transfer, processing, and management. To fully exploit the information carried by high-order correlations between multiple photon detection events, smart on-chip and FPGA-based processing protocols will have to be developed. Efficient on-the-fly processing of the registered correlations will open the door for new deep-imaging strategies. For example, in the photon-starved regime, wavefront optimization schemes based on high-order spatial-temporal correlations may turn out to be more efficient than intensity-based schemes, as more information is extracted per photon detection event. Similarly, the development of new algorithms based on high-order spatiotemporal photon correlations may initiate a new paradigm in computational imaging in scattering samples.

The key to the success of quantum imaging in scattering media depends on whether the quantum advantages provided by non-classical states of light can balance the cost of using weak light sources, which reflects in low signal-to-noise ratios compared to imaging with classical light sources. The answer to this question depends on technological and scientific breakthroughs, which at this point are difficult to predict. Nonetheless, we strongly believe that research on quantum imaging through scattering will also lead to the development of quantum-inspired imaging schemes that will utilize high-order correlations using classical bright sources.

**Concluding Remarks**

Up to today, only very few research groups have investigated the propagation of quantum optical states in scattering media, with nearly all studies focused on fundamental aspects and communication applications. Regarding imaging in scattering media, nearly everything remains to be done. Whether it is by adapting already existing deep-tissue imaging methods to quantum imaging approaches, or by truly harnessing the unique properties of quantum light to see deeper, this research field has a strong potential for developing new imaging methods that can complement or even surpass classical approaches. In the coming years, this research will also benefit from the rapid development of novel imaging sensors, such as SPAD cameras, allowing detection of single photons across multiple channels in parallel at unprecedented temporal resolution and speed.

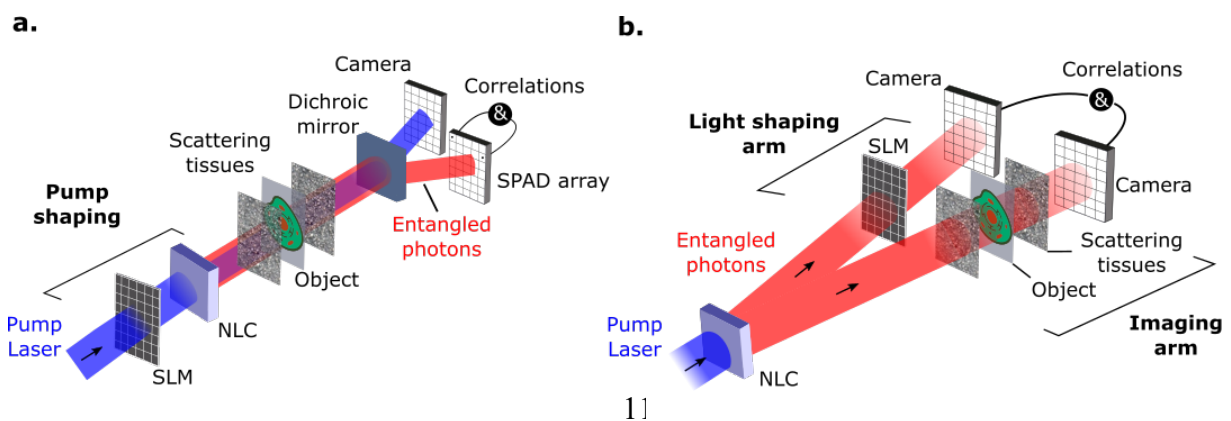

**Figure 1.** Conceptual illustrations of imaging through scattering media using quantum light. **a.** By applying classical wavefront shaping to the bright pump beam that stimulates the generation of entangled photons in a nonlinear crystal (NLC), scattering of entangled photons can be compensated in real-time. **b.** Entangled photon pairs are used to implement non-local wavefront shaping. If the propagation path of the photon that does not penetrate the medium is manipulated so that it mimics the scattering experienced by its twin photon, an aberration-free image can be retrieved by measuring photon correlations between the two cameras. Concepts inspired from [7,8].

## Acknowledgments

Y. B. is supported by the Zuckerman STEM Leadership Program. H.D. acknowledges funding from the European Union's Horizon 2020 research and innovation programme under the Marie Skłodowska-Curie grant no. 840958.